\documentclass[12pt]{article}
\pdfoutput=1
\usepackage[a4paper]{geometry}
 
\usepackage[T1]{fontenc} 
\usepackage{comment}

\usepackage{jheppub,hyperref,float,array,adjustbox,mathtools,physics,xcolor}
\usepackage{lipsum}
\usepackage{booktabs}
\usepackage{pdflscape}
\usepackage{geometry}
\usepackage{fancyhdr}
\usepackage{changepage}
\usepackage[english]{babel}

\usepackage{xcolor,tikz,pgfplots,amsmath,amsfonts}
\usepackage{graphicx}
\usepackage[mathscr]{euscript}
\usetikzlibrary{matrix,calc,positioning,decorations.markings,decorations.pathmorphing,decorations.pathreplacing}
\usetikzlibrary{arrows,cd}
\usetikzlibrary{shapes.misc}
\usetikzlibrary {shapes.geometric}
\usetikzlibrary {shapes.multipart}
\usetikzlibrary{decorations.markings}
\usetikzlibrary{backgrounds}
\usetikzlibrary{shapes.gates.logic.US}
\usetikzlibrary{circuits.ee.IEC}
\usepackage{bbding}

\definecolor{terradisiena}{RGB}{233,116,81}
\definecolor{strisciadipietro}{RGB}{229,204,255}
\definecolor{verdepetrolio}{RGB}{33,100,119}

\usepackage{tikz}
\usetikzlibrary{automata,positioning,calc}
\usetikzlibrary{decorations.markings}
\tikzset{->-/.style={decoration={markings, mark=at position #1 with {\arrow{>}}},postaction={decorate}}}
\tikzset{-<-/.style={decoration={markings, mark=at position #1 with {\arrow{<}}},postaction={decorate}}}
\tikzset{auto shift/.style={auto=right,->, to path={ let \p1=(\tikztostart),\p2=(\tikztotarget), \n1={atan2(\y2-\y1,\x2-\x1)},\n2={\n1+180} in ($(\tikztostart.{\n1})!1mm!270:(\tikztotarget.{\n2})$) -- ($(\tikztotarget.{\n2})!1mm!90:(\tikztostart.{\n1})$) \tikztonodes}}}

\usetikzlibrary {shapes.geometric}

\DeclareMathOperator{\SO}{SO}
\DeclareMathOperator{\SU}{SU}
\DeclareMathOperator{\UU}{U}

\newcommand{\ee}{\mathrm{e}}
\newcommand{\mi}{\mathrm{i}}

\definecolor{USPcol}{rgb}{0.94, 0.73, 0.80}
\definecolor{SUcol}{rgb}{0.894, 0.902, 0.976}
\definecolor{SOcol}{rgb}{0.5, 0.85, 0.8}

\title{
\begin{center}
Overlooking 3d dualities \\from  mezzanines and  balconies
\end{center}
}

\author[a]{Antonio Amariti,}	
\author[a,b]{Chiara Mascherpa,}

\affiliation[a]{INFN, Sezione di Milano, Via Celoria 16, I-20133 Milano, Italy}
\affiliation[b]{Dipartimento di Fisica, Università degli studi di Milano, Via Celoria 16, I-20133}

\emailAdd{antonio.amariti@mi.infn.it}
\emailAdd{chiara.mascherpa@mi.infn.it}

\abstract{
We derive new 3d $\mathcal{N}=2$ dualities with $\mathrm{U}(N)$ gauge groups involving pairs of two-index tensors interacting through a quartic superpotential, (anti)-fundamentals and possibly an adjoint.
A strong hint for their validity follows from  T-duality on brane setups with O6 planes and  stacks of multiple NS fivebranes. These setups engineer two families of 4d models known in the literature as mezzanine and balcony  of type $A_k$ and $D_{k+2}$. In 4d these models admit a generalization of Seiberg duality, tested also at the brane level.
We study the reduction of such dualities from both the brane and the field theory perspective and we establish the matching of the three-sphere partition functions, assuming the validity of the 4d superconformal indices. The latter is achieved through a double scaling limit, whose structure is dictated by the brane picture, and  it requires the validity of new 3d confining dualities for quivers with two gauge nodes. 
Moreover we provide a proof of these confining dualities via the tensor deconfinement technique.
}

\begin{document}
\maketitle
\flushbottom
\allowdisplaybreaks


\section{Introduction}
\label{sec:intro}

The study of the IR behavior of quantum field theories requires sophisticated  techniques because in general well defined UV descriptions flow to strong coupling and a non-perturbative analysis is necessary.
Supersymmetry offers powerful tools to tackle such a problem, for example the fact that strongly coupled descriptions can be reformulated in terms of weakly coupled ones thanks to the existence of an IR dual description.  
Supersymmetric dualities are quite ubiquitous, and their analysis has been a fertile field of research over the last decades.
Focusing on the case of 4d supersymmetric gauge theories, the prototypical example of an IR duality was originally worked out in \cite{Seiberg:1994pq} and it is referred to as Seiberg duality.
It was originally proposed for $\SU(N)$ SQCD, but it was soon realized that it could be generalized to SQCD with a different gauge group \cite{Intriligator:1995ne,Intriligator:1995id}.
Another possible generalization was related to the presence of extra charged matter fields, such as adjoints and symmetric and antisymmetric two-index tensors.
The case with an adjoint and polynomial superpotential was proposed in \cite{Kutasov:1995ve,Kutasov:1995np,Kutasov:1995ss} and various generalizations were then found in \cite{Brodie:1996vx,Intriligator:1995ff,Leigh:1995qp,Intriligator:1995ax,Brodie:1996xm}.
Focusing on these cases, the Arnold ADE singularity classification emerges, see \cite{Intriligator:2003mi} for an explicit 
discussion in the case of adjoint matter.
In the presence of two-index symmetric and/or antisymmetric tensors a similar classification for the $A_k$ and $D_{k+2}$ case was worked out in \cite{Intriligator:1995ax} and \cite{Brodie:1996xm} respectively. These dualities are the starting point of our discussion.

Our goal consists of finding an analogous 3d $\mathcal{N}=2$ classification for such dualities, generalizing the results obtained in \cite{Amariti:2025gca} for the $A_1$ case.
Avatars of such dualities in three dimensions have been proposed in \cite{Kapustin:2011vz} in the presence of CS coupling for some of the $A_k$ cases and in \cite{Amariti:2015mva} from the brane engineering of  the 4d dualities.
This last approach is one of the main tools that  motivates our analysis.
Indeed, starting from the analysis of ARSW \cite{Aharony:2013dha} it became clear that many 3d $\mathcal{N}=2$ theories could be obtained by dimensionally reducing 4d dualities, with an appropriate prescription (for example the originally 3d Aharony  \cite{Aharony:1997gp} and Giveon-Kutasov \cite{Giveon:2008zn} dualities were derived along these lines in \cite{Aharony:2013dha}). 
This explained the similarities already observed between 4d parent dualities and 3d ones.
In many cases the ARSW prescription has been successfully applied at the level of localization, reducing integral identities matching the superconformal index across dual phases to integral identities matching the squashed three sphere partition functions.
However this last procedure is not \emph{a priori} well defined if not all of the Coulomb branch get lifted on $S^1$ \cite{Aharony:2013kma}.
A way out consists of studying the dimensional reduction through a double scaling limit, keeping the radius small and taking some large scalars in the dynamical or background vector multiplets. 

Such a double scaling was engineered at the level of localization in \cite{Amariti:2024bdd}, by elaborating on the approach of \cite{Choi:2018hmj} to the study of gravitational setups using the index.
In short, while the ARSW prescription corresponds to matching the leading saddle of an integral identity, the double scaling approach allows to match subleading saddles, and such  a matching strongly suggests the existence of a 3d $\mathcal{N}=2$ duality.

A tricky aspect of this analysis consists of finding the correct backgrounds for the scalars in the vector multiplet. Indeed one starts by considering a real mass assignment on the electric side, through the study of a  large vev limit for the scalars in the weakly gauged flavor symmetries. The duality dictionary automatically assigns the corresponding vevs on the dual side. On the other hand finding the  assignment of the  backgrounds  for the scalars in the dynamical vector multiplets is not straightforward and  a judicious choice is necessary in order to cancel the divergent contributions from the gravitational anomalies across the dual phase.

A powerful tool to address this problem is given by the brane engineering of the 4d duality. A general prescription to work out the 4d/3d reduction in this geometric setting was proposed in \cite{Amariti:2015yea,Amariti:2015mva} through T-duality on the compact direction. The key-point is that this  setup embodies the double scaling prescription. Indeed the real mass flow is engineered in this case by moving the flavor branes along the compact T-dual direction and on the other hand the backgrounds  for the scalars in the dynamical vector multiplets that preserve the duality on the circle are found by the 
Hanany-Witten (HW) transition \cite{Hanany:1996ie}.

The brane setup for the $A_k$ and $D_{k+2}$ dualities discussed in this paper have been obtained in \cite{Brunner:1998jr} and here we  study the reduction at the geometric level and then examine brane configurations that allow the study of the integral identities at the level of localization.
We assume the  validity of the 4d identity\footnote{This is in general a rather strong assumption that needs a separate analysis, we will be back on this issue in the conclusions} and we then  look for a 3d identity by scanning over the sub-leading saddles, which allows us to propose a purely 3d duality.
Such a scan requires to isolate  gauge sectors in the 3d case that we can locally confine, providing a final identity between two gauge theories with a single gauge group. The confining sectors obtained here are actually new, and consist of theories with two gauge groups  connected by bifundamental matter in addition to fundamentals and possibly adjoints. We have proven that these theories are confining through  a separate analysis, by generalizing  the argument of \cite{Benvenuti:2024glr,Hwang:2024hhy}.
A consistent procedure that always produces the expected identity was found in  \cite{Amariti:2025gca}  and it corresponds to finding, after  confinements,  an identity of the type 
\begin{equation}
\label{sqeq}
Z_{ele}^2(\vec m) = Z_{mag}^2(\vec m) \, ,
\end{equation}
 between two perfect square partition functions, where the (complex)  parameters $m$  satisfy  a constraint inherited from the cancellation of the 4d anomalies. The pure 3d limit is then taken by lifting this constraint through an appropriate large mass limit. 
Here we provide a generic treatment of the $A_k$ and $D_{k+2}$  scaling through this procedure, and we further   
discuss the cases where the canonical ARSW prescription can be applied as well, finding a perfect agreement.

The paper is organized as follows. In Section \ref{branepicture} we give a brief overview, based on the results of \cite{Brunner:1998jr}, of the brane setups associated with the 4d duality that we aim to reduce. This allows us to define, borrowing the terminology of \cite{Brodie:1996xm},  the Mezzanine and Balcony dualities for the $A_k$ and $D_{k+2}$ case.
Then in Section \ref{doublescaling} we review the strategy adopted in the paper for the reduction of  4d dualities to 3d through a  double scaling prescription, by focusing on the brane interpretation of this reduction and summarizing the mathematical formulation of the reduction of the 4d superconformal index to the 3d squashed three sphere partition function.
The last auxiliary results that are necessary to our analysis are provided in  Section \ref{newconf}, where we derive two new confining dualities for quiver gauge theories with two gauge groups. This derivation is mostly based on the results of \cite{Benvenuti:2024glr,Hwang:2024hhy} for the deconfinement in 3d  of an adjoint of an $\UU(N)$ gauge group with power law superpotential.
Remarkably, these quivers represent new examples of confining dynamics in 3d $\mathcal{N}=2$ theories, extending the known landscape beyond single-node gauge groups.
The core of the paper consists of the analysis performed in  Section \ref{secAd} and Section
\ref{secDd}. In these sections we consider each 4d duality, reviewing the field content, the superpotential and summarizing the table of charges that can be used to built the 4d identity between the superconformal index.
Then we reduce the identity to 3d, using the double scaling approach, by considering a non trivial background for the weakly gauged global symmetries. We found by inspection that there is a consistent prescription that produces a well defined relation of the type \eqref{sqeq}. We then read the final 3d duality by extracting the square root from this relation and interpreting the identity properly. When possible we compare our results with other approach (e.g. the ARSW prescription in the Mezzanine models with antisymmetric tensors) finding a perfect agreement.
A further step of our analysis regards the construction of dualities with CS terms or with $\SU(N)$ gauge groups. These possibilities are discussed in Section \ref{CSSU}. The dualities with CS terms are built from the ones derived above by triggering real mass flows for the fundamental flavors. In these cases we find agreement with other results derived in the literature. The dualities with $\SU(N)$ factors on the other hand are derived by a standard gauging of the topological symmetry.
We conclude the analysis in Section \ref{sec:conc} by proposing various open questions and possible extensions of our results.

%
%
%
%
%
%
\section{The brane picture for the mezzanine and balcony dualities}
\label{branepicture}
%
%
%
%
%

\begin{figure}[ht] \begin{center}
\includegraphics[width=15cm]{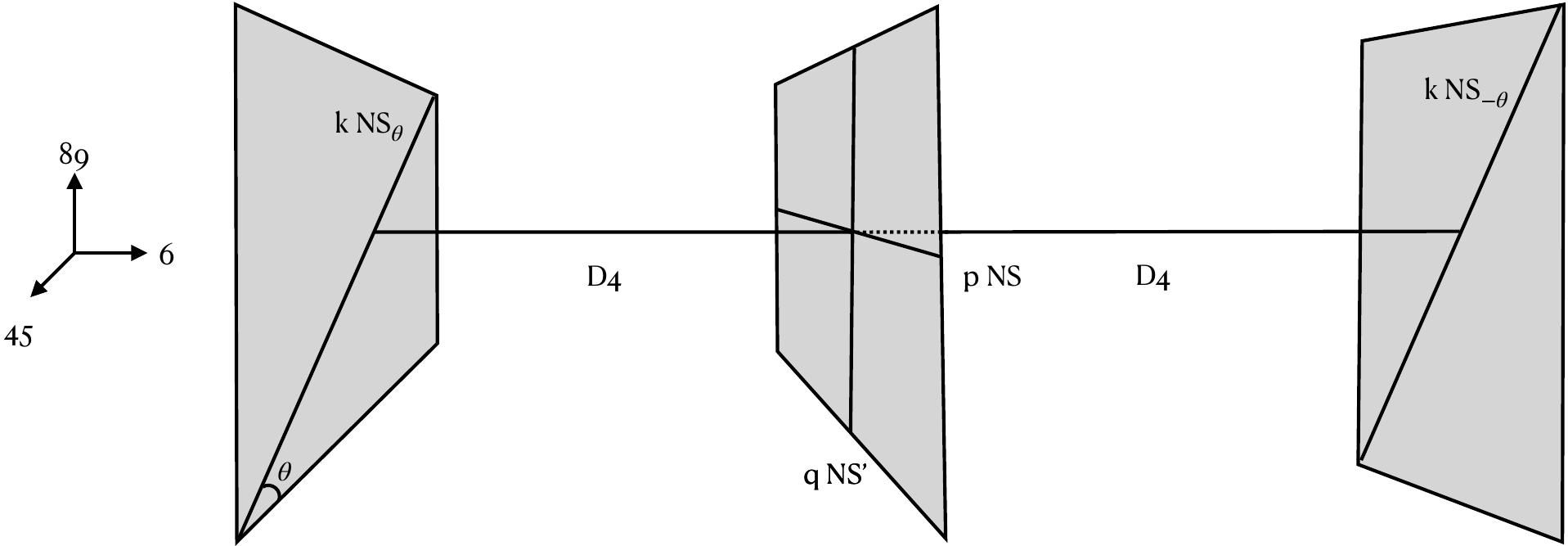}
\caption{In this figure we survey the various brane setups considered in the paper. The mezzanine models correspond to $(p,q) = (1,0)$ and $(p,q)=(k,0)$ for $A_k$ and $D_{k+2}$ respectively.
The balcony models correspond to $(p,q) = (0,1)$ and $(p,q)=(0,k)$ for $A_k$ and $D_{k+2}$ respectively.
}
\label{fig:branefig}
\end{center} \end{figure}

The brane engineering of the dualities studied in this paper has been discussed in \cite{Brunner:1998jr} through an HW construction \cite{Hanany:1996ie} (see also 
\cite{Brodie:1997sz,Elitzur:1997fh,Elitzur:1997hc,Landsteiner:1997vd,Landsteiner:1998gh,Giveon:1998sr} for related works).
Here we will be sketchy, and we refer the reader to our previous paper for more details.
The brane setup is realized by considering NS, D4 and D6 branes in addition to an O6$^\pm$ plane. 
The NS  are all extended along $x_{0123}$ and in addition in two extra directions, corresponding to $x_{45}$ for an NS fivebrane and $x_{89}$ for an NS$'$ fivebrane.
Here we deal with setups preserving four supercharges, which are compatible with a further rotation of an NS
brane  in  $x_{89}$  with respect to  $x_{45}$  by an angle $\theta$. In this case we have an 
NS$_{\theta}$ fivebrane.
The 4d gauge theories live on the  D4 branes, which are  stretched between pairs of NS branes, and placed at finite distance along  $x_6$.
The D6 branes are on the other hand extended along $x_{0123789}$, while the D6' are extended along $x_{0123457}$. In addition configurations with rotated D6$_\theta$ branes can be considered
\footnote{In principle one can also consider  D6$_{\pm w}$ branes with $w \neq \theta$, inducing a coupling 
between the fundamentals and the adjoint. We do not consider this configuration below, but we observe that
our analysis applies to this case straightforwardly. This is because the cubic coupling induced by the deformation   modifies the dual picture, Higgsing the dual gauge group and the HW transition is modified accordingly.}.
The models considered in 
\cite{Amariti:2025gca} correspond to the case with $N_c$ D4 branes extended along $x_6$ between an NS$_{\theta}$ and an NS$_{-\theta}$ with a further NS or NS'  fivebrane placed symmetrically on $x_6$ (say $x_6=0$)  with respect to the NS$_{\pm \theta}$ fivebranes. In addition we consider O6$^{\pm}$ extended along the directions  $x_{0123789}$  that occupies on $x_6$ the same position of the NS or NS' fivebrane.
We have summarized below the brane setup and the direction of each brane
\begin{equation}
\setlength{\arraycolsep}{6pt} 
\begin{array}{l|cccccccccc}
              & 0 & 1 & 2 & 3 & 4 & 5 & 6 & 7 & 8 & 9 \\
\hline
\text{D4} & \text{X} & \text{X} & \text{X} & \text{X} & \cdot & \cdot & \text{X} & \cdot & \cdot & \cdot \\
\text{NS} & \text{X} & \text{X} & \text{X} & \text{X} & \text{X} & \text{X} & \cdot & \cdot & \cdot & \cdot \\
\text{NS}' & \text{X} & \text{X} & \text{X} & \text{X} & \cdot & \cdot & \cdot & \cdot & \text{X} & \text{X} \\
\text{NS}_{\theta}
& \text{X} & \text{X} & \text{X} & \text{X}
& \multicolumn{2}{c}{(\text{X})_{\hat\theta}}
& \cdot & \cdot
& \multicolumn{2}{c}{(\text{X})_{\theta}} \\
\text{D6} & \text{X} & \text{X} & \text{X} & \text{X} & \cdot & \cdot & \cdot & \text{X} & \text{X} & \text{X} \\
\text{D6}' & \text{X} & \text{X} & \text{X} & \text{X} & \text{X} & \text{X} & \cdot & \text{X} & \cdot & \cdot \\
\text{D6}_{\theta}
& \text{X} & \text{X} & \text{X} & \text{X}
& \multicolumn{2}{c}{(\text{X})_{\hat\theta}}
& \cdot & \text{X}
& \multicolumn{2}{c}{(\text{X})_{\theta}} \\
\text{O6} & \text{X} & \text{X} & \text{X} & \text{X} & \cdot & \cdot & \cdot & \text{X} & \text{X} & \text{X}\rlap{\quad ,}
\end{array}
\label{eq:brane_config}
\end{equation}
where we have denoted by $\hat\theta$ the angle $\frac{\pi}{2}-\theta$.

In the first case, where an NS is considered, we can either choose the O6$^+$ or the O6$^-$
plane, and it corresponds to a conjugated pair of $\SU(N_c)$ two-index either symmetric or antisymmetric tensors. There is a third possibility, where we have an NS' fivebrane and a configuration with an O6$^+$ extended along $x_7>0$ and half O6$^-$ extended along $x_7<0$  must be considered.  In this case we need for consistency also eight semi-infinite D6 branes on the O6$^-$ plane.
This configuration corresponds to the pair of antisymmetric and conjugated symmetric tensors.
Here we modify this picture as explained in \cite{Brunner:1998jr} by considering stacks of  multiple NS branes.
We consider $k$ pairs of NS$_{\pm \theta}$ branes. Furthermore we distinguish two cases at $x_6=0$. In the first case we have $1$ NS or NS' fivebrane, while in the second case we have 
$k$ NS or NS' fivebranes (see Figure \ref{fig:branefig}). 
The various choices discussed in this paper are summarized in (\ref{tabella}). 
\begin{equation}
\label{tabella}
\begin{array}{|l|lcc|}\hline
                    \text{Model}                       &       \text{Fivebrane(s)}    &      \text{O}6     &  \text{Section}  \\
                                           \hline 
\text{Mezzanine } A_k^{(A)}      &    \hspace{.5cm}      1 \text{ NS}             &       -          & (\ref{MAA3d})     \\
\text{Mezzanine } A_k^{(S)}      &    \hspace{.5cm}        1 \text{ NS}                 &       +        & (\ref{MSS3d})     \\
\text{Balcony } A_{k}      &        \hspace{.5cm}        1 \text{ NS}  '             &      \pm    & (\ref{BA3d})     \\
\text{Mezzanine } D_{k+2}^{(A)}      &         \hspace{.5cm}     k \text{ NS}               & $-$   & (\ref{MDA3d})     \\
\text{Mezzanine }  D_{k+2}^{(S)}      &        \hspace{.5cm}      k \text{ NS}           &       +    & (\ref{MDS3d})     \\
\text{Balcony }  D_{k+2}      &      \hspace{.5cm}          k \text{ NS}  '          &      \pm    & (\ref{BD3d})     \\
\hline
\end{array}
\end{equation} 
In the first column, we specify the model, with the terminology used above. 
We referred in general to the $A_k$ and $D_{k+2}$ cases without specifying whether $k$ is generic, odd or even. Such details depend on the structure of the chiral ring truncation and on other issues associated with the validity of the duality. 
In the second column, we indicate the type and number of fivebranes placed at $x_6=0$.
The sign of the O6, either O6$^+$ or O6$^-$ or a pair of O6$^\pm$ appears on the third column. The last  column refers to the section where we reviewed the 4d duality and derived the 3d one.
Observe that the balcony models studied from the brane perspective differ from those introduced at the field theory level, because there is an extra superpotential interaction involving eight fundamentals (arising from the eight semi-infinite D6 branes) and the conjugated symmetric tensor, of the form $\Delta W_{balcony} = \tilde S T^2$. This term triggers an RG flow on the electric side and induces an Higgsing on the magnetic side, thus modifying the duality. It is also possible to reverse the flow by adding a mass term for eight pairs of fundamentals and antifundamentals in this model. 
For this reason we keep referring to these models as balcony models, keeping the same terminology as in the field theory description.

%
%
%
%
%
%
%
%
\section{Double scaling and 4d/3d reduction}
\label{doublescaling}
%
%
%
%
%
%

In this section, we summarize some general aspects of the reduction of 4d $\mathcal{N}=1$ dualities to 3d by making use of a modified version of the ARSW prescription.
This corresponds to  the double scaling prescription  proposed in 
\cite{Amariti:2024bdd} (see \cite{ArabiArdehali:2015ybk,Hwang:2018riu} for previous discussions in the same direction and \cite{ArabiArdehali:2026kvt} for recent applications) and then used in \cite{Amariti:2025gca} in order to reduce to 3d the limiting mezzanine and balcony $A_1$ dualities.

At the field theory level we focus on 4d systems with a space-like direction compactified on a circle.
In the 3d effective description one assigns non-vanishing real masses to some matter fields. At the same time one can consider a non vanishing background value for some of the real scalars in the vector multiplet. 
The combined action generates, for the theory on $\mathbb{R}^{1,2} \times S^1$,  new gauge sectors with, depending on the mass assignment, massless charged matter fields. 
The gauge symmetry is broken by a Higgs mechanism, and  the instantons associated with this breaking are  encoded in superpotential monopole interactions  of  AHW-type.
Preserving a duality on the circle then requires finding the corresponding Higgsing on the dual gauge group (while the duality dictionary enforces the non-zero real masses).
The double scaling prescription considered in \cite{Amariti:2024bdd,Amariti:2025gca} and used below corresponds to assigning the real masses and the backgrounds for the scalars proportionally to the inverse compactification radius, 
such that, when the radius becomes small, we are dealing with theories at large distances on the compact direction $S^1$.
In the  3d limit, one deals with gauge theories coupled through monopole superpotentials. 
Observe that depending on the value of the background scalars there can be points of enhanced symmetry on $S^1$, at which enhancements of massless chiral and vector multiplets typically take place.
Various known pure 3d results are then recovered from this analysis, in principle by performing further RG flows and local dualities, where by local duality we mean that in the small radius limit, where the theories are at infinite 
distance on $S^1$, one can dualize only a single gauge sector.

The structure just outlined has a counterpart at the level of the brane engineering, where one space-like direction (for example $x_3$) is compact. In such  a case the type IIA setup can be T-dualized along the compact direction 
and a IIB setup emerges at large T-dual radius. The NS fivebranes are compact along the $x_3$ direction, while the D4 and the D6 branes become D3 and D5 branes respectively and they occupy the position $x_3=0$. The O6$^+$ planes 
split into pairs of O5$^\pm$ planes\footnote{Here we do not consider the case of O3 planes, see \cite{Amariti:2015mva} for a discussions on such setups}, one at $x_3=0$ and one at the
point symmetric on the  T-dual circle with respect to the origin, denoted in \cite{Amariti:2015mva} as the mirror point.
Observe that the picture does not in general guarantee that the configuration is stable. For example, when we consider the semi-infinite eight half D6 branes on the  O6$^\pm$ configuration, charge conservation imposes that 
that four semi-infinite half D5 branes are located at $x_3=0$ and four half semi-infinite half D5 branes are located at the mirror point.
Various configurations on the compact direction can be considered, corresponding to the assignment of  non-vanishing backgrounds to the scalars in the vector multiplets of the original gauge group. We can also assign a background to the scalars in the vector multiplets of the weakly gauged global symmetries. Such backgrounds are encoded in the brane picture by displacing the D3 and D5 branes at different positions along the compact $x_3$ direction. Observe that in this way Euclidean D1 branes, extended along $x_6$ and $x_3$, encode the AHW-type interactions from the brane perspective.
The presence of points of enhanced symmetry emerges also in the brane setup. For example, a two-index tensor in the original description is lifted at generic positions of the D3 on the circle, except when the D3 are placed at the mirror point, where the they cross an O5 plane. Furthermore if pairs of D3 branes move symmetrically with respect to $x_3=0$, they give rise to pairs of unitary gauge groups connected by massless pairs of bifundamentals.
The crucial aspect of the brane engineering emerges in the dual phase, where the choice of background scalars in the vector multiplets that preserve the duality is automatically encoded in the HW transition in the presence of the compact direction.

There is a natural way to study the double scaling just reviewed on the field theory side and in the brane picture when considering the reduction of the 4d superconformal index to the 3d partition function. 
The index is a matrix integral over the holonomy of the gauge group 
$
I(\vec v) = \int [dz] \mathcal{I}(\vec z,\vec v)
$, where $
    z_j = e^{2 \pi \mi \zeta_j}$ and  $v_k = e^{2\pi \mi m_k}
$.
The integrand  $\mathcal{I}$ corresponds to a product of one-loop determinant, consisting of elliptic Gamma functions, having as arguments the fugacities of the gauge and flavor symmetries, $z_j $ and $v_k$, with the appropriate weights of their representation.

In order to reduce the index to the squashed three-sphere partition function $Z_{S_b^3}$  we  define a basis for the fugacities of each field  as,
\begin{equation}
    (pq)^{R_a/2} \prod_{k} v_i^{e_k^a} \equiv y_a \equiv \ee^{2\pi \mi \Delta_a}.
\end{equation}

In the  double scaling limit,  the real masses are taken to be large with a $1/r$ scaling in the  $r \rightarrow 0$ limit of the radius  of the circle. 
Therefore, we parametrize the 4d fugacities as
\begin{equation}
\label{pars}
    \zeta_i = \sigma_i^* + \sigma_i {r}\,,  \qquad \Delta_k = \mu_k^* + \mu_k {r}\,,
\end{equation}
where we assign fixed values $\sigma_i^*$ and $ \mu_k^*$ in addition to a term of order $\mathcal{O}(r)$.

The elliptic Gamma function  can be reformulated as an infinite product of 3d hyperbolic Gamma functions
\cite{NARUKAWA2004247} (see also \cite{ArabiArdehali:2019tdm}). 
A hyperbolic gamma function corresponds to the one-loop determinant for a 3d multiplet, and the presence of an infinite number of these multiplets for each field signals the presence of a tower of KK modes. The KK tower is indeed labeled by the inverse radius, and integrating out the massive contributions gives a Gaussian factor to the integrand, associated with the gravitational anomaly of the 4d system.
Depending  on the choice of the fixed values in (\ref{pars}) some modes, coming from the matter and gauge fields, can be massless in the KK tower.
Restricting to a single gauge and flavor holonomy and denoting  $k = \sigma^* + \mu^*$ and $x r = (\sigma + \mu)r$, we have
\begin{equation}
    \Gamma_e(k + x r) \underset{r \to 0}{\sim}
    \begin{cases}
        \ee^{-\frac{\mi \pi (x - \omega) }{ 6 r_1 \omega_1\omega_2}} \; \Gamma_h(x) \quad \quad & k \in \mathbb{Z} \\
        \ee^{\mi \pi Q(k + x r)} & k \notin \mathbb{Z}\,,
    \end{cases}
\end{equation}
with
\begin{equation}
  \begin{aligned}
    Q(k + xr) = 
    - \frac{1}{\omega_1 \omega_2} &\Bigg(
    \frac{ k (2k - 1)(k - 1)}{6 r^2} + 
    \frac{(x - \omega)\left( 6k (k - 1) + 1 \right)}{6r} + \\ & +
    \frac{(2k - 1)(6x^2 + \omega_1^2 + \omega_2^2 + 3 \omega_1\omega_2 - 12x\omega )}{12} \Bigg) + \mathcal{O}(r)\,.
  \end{aligned}
\end{equation}

When one considers a 4d duality encoded in an integral identity the reduction procedure just spelled out gives rise to a 3d identity, corresponding to the matching of the leading saddle-point approximation. This is the procedure adopted in \cite{Aharony:2013dha} in order to reduce 4d dualities to 3d at the level of the three sphere partition function. 
This procedure is well defined if the 3d partition function is not divergent, and it requires the lifting of the whole Coulomb branch for the theory in the circle.
However, such a condition is not automatic even in simple cases. For example, this is not the case for $SO(N)$ SQCD \cite{Aharony:2013kma}.
A more refined analysis at the level of the saddle point is then necessary in order to produce well defined 3d identities starting from 4d ones. This is where the double scaling prescription comes into play.
In the 4d/3d reduction of the index the double scaling corresponds to looking for subleading saddle-points of the 4d index. Such saddles are found by exploring the background for the gauge holonomy,
accompanied by different scalings for the flavor holonomies.
Such scalings produce a different spectrum of massless vector and chiral matter fields, and they also modify the divergent contributions factored out from the partition function. 
Matching the divergent prefactor is then necessary  in order to preserve the duality at the level of the integral identity.
The final identities obtained using this prescription correspond to the expected ones from the field theory and the brane analysis discussed above. We refer the reader to \cite{Amariti:2024bdd} for a more rigorous and mathematical approach, and in the following we  adopt the same conventions in order to study the reduction of the balcony and mezzanine dualities reviewed above.

%
%
%
%
%
%
%
%
%
%
\section{New confining sectors}
\label{newconf}
%
%
%
%
%
%
%
%
%
%
In this section we discuss two 3d $\mathcal{N }=2$ confining quiver gauge theories that will be useful in the analysis below.
It is indeed quite common that when one studies the reduction of a 4d duality to 3d the presence of finite size effects forces one to
consider effective descriptions in which monopole superpotentials are introduced in order to constraint the generation of topological 
and axial symmetries that usually arise in 3d.  These finite size effects are usually removed by considering further real mass deformations.
These deformations do not in general preserve the effective duality when trivial backgrounds are chosen for the vector multiplet on both the electric and the magnetic side.
The typical situation is that on one side one can consider a vanishing background while on the dual side a non trivial gauge sector at large 
distance in the Coulomb branch emerges. This is in general a 3d confining gauge theory, and in the IR it is described by gauge singlets, 
corresponding to  mesons (and possible baryons) and  monopoles.
In our analysis below we have found two models of this type, that are expected to confine (the brane picture confirms this expectation) but that have not been discussed in the literature.
Here we provide a proof of this fact, by exploiting the results of \cite{Benvenuti:2024glr,Hwang:2024hhy}, i.e. we prove that these sectors are confining through tensor deconfinement.
We refer the reader to 
\cite{Pasquetti:2019uop,Benvenuti:2020wpc,Etxebarria:2021lmq,Benvenuti:2021nwt,Bottini:2022vpy,Bajeot:2022lah,Bajeot:2022wmu,Amariti:2022wae,Amariti:2023wts,Amariti:2024sde,Jiang:2024ifv,Amariti:2024gco,Benvenuti:2024glr,Hwang:2024hhy,Amariti:2025jvi,Amariti:2025lem,Jia:2025koz} 
for recent applications of tensor deconfinement techniques in supersymmetric dualities.

In this section we provide first some salient features of the results of \cite{Benvenuti:2024glr,Hwang:2024hhy} and then we study the cases of interest.

\subsection{Deconfining an $\UU(N)$ adjoint with power law superpotential}
\label{decsarasec}

In this subsection, we review the deconfinement of a $\UU(N)$ adjoint chiral
multiplet $X$ with  superpotential $X^{k+1}$, following
\cite{Benvenuti:2024glr,Hwang:2024hhy}.  The basic idea is that such a theory is
equivalent, in the infrared, to a linear quiver with $k-1$ gauge nodes, each one with an adjoint and connected by 
 pairs of conjugated bifundamentals. Depending on the rank of the gauge group there will also be a non trivial monopole superpotential.
  We  focus on two choices, i.e. $N=2k+1$ and $N=3k$, which correspond to the cases studied below.
 
\begin{figure}[ht] \begin{center}
\includegraphics[width=7cm]{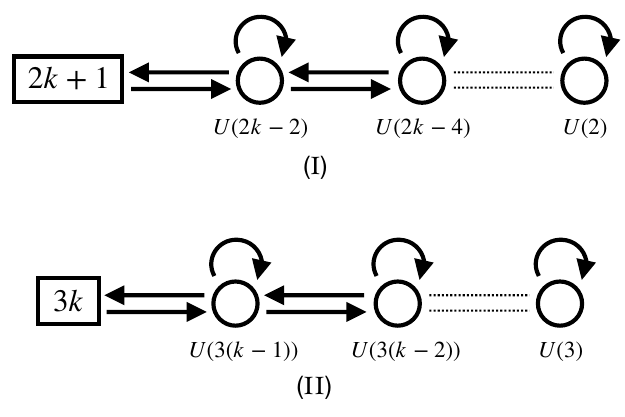}
\caption{
In these figures we depict  two confining 3d 
$\mathcal{N}=2$ quivers, studied in \cite{Benvenuti:2024glr,Hwang:2024hhy} and reviewed in this section. 
We represented through  dotted lines whole tails with unitary gauge groups. In the first case the gauge groups  decrease along the tail as  $\UU(2(k-j))$, while in the second case they decrease as  $\UU(3(k-j))$. }
\label{fig:ADk}
\end{center} \end{figure}
 
 \subsubsection*{The case of $\UU(2k+1)$}
 
 We start by considering the confining duality involving the linear quiver  in figure \ref{fig:ADk} (I). 
 This theory is confining when we consider the  superpotential  
 \begin{equation}
 W = W_{\mathcal{N}=4} + \sum_{j=1}^{k-1} \mathcal{M}^{\dots \bullet_j \dots} + \mathcal{M}^{--\dots--} \; ,
 \end{equation}
where $\mathcal{M}^{\dots \bullet_j \dots} $ correspond to the flux $+1$ monopole for the $j$-th gauge group
$\UU(2j)$ in the linear quiver and $\mathcal{M}^{--\dots--}$  has flux $-1$ for every gauge group in the linear quiver.
The model confines, as shown in \cite{Benvenuti:2024glr,Hwang:2024hhy}, and the dual WZ model has a single adjoint $\Phi$ (mapped to the adjoint operator of the tail as explained in \cite{Benvenuti:2024glr})
with superpotential $W=\Phi^{k+1}$.
Below, we  use this confining duality to deconfine an $\UU(2k+1)$ adjoint\footnote{While the original derivation regarded an $\SU(N)$ adjoint,  here we always focus on the $\UU(N)$ case. Such a distinction was explicitly considered also in  \cite{Benvenuti:2024glr,Hwang:2024hhy}.}.

At the level of the three sphere partition function this deconfinement translates into the identity 
\begin{eqnarray}
\label{idgrak}
&&
\prod_{a,b} \Gamma_h(\mu_a + \nu_b)=
\int \prod_{j=1}^{k-1}\Bigg( \prod_{i=1}^{2(k-j)} d\sigma_i^{(j)}
\!\!\!\!\!\!
\prod_{1\leq i<\ell\leq 2(k-j)} 
\!\!\frac{\Gamma_h(\pm (\sigma_{i}^{(j)} - \sigma_{\ell}^{(j)}) + (k-1) \tau_k)}
{\Gamma_h(\pm (\sigma_{i}^{(j)} - \sigma_{\ell}^{(j)})+ (k-1) \tau_k)}\Bigg)
\nonumber \\
\times
&&\prod_{j=1}^{k-2}\prod_{i=1}^{2(k-j)} \prod_{\ell=1}^{2(k-j-1)}
\Gamma_h(\pm (\sigma_{i}^{(j)} - \sigma_\ell^{(j+1)} )+  \tau_k)
\prod_{i=1}^{2(k-1)} \prod_{a=1}^{2k+1} 
\Gamma_h(\sigma_i^{(k-1)} + \mu_a,-\sigma_i^{(k-1)}+ \nu_a) 
\nonumber \\
\times
&&
e^{\sum_{i=1}^{2(k-1)} 2 \pi (2-k) \tau_k \sigma_{i}^{(k-1)} + \sum_{j=2}^{k-1} \sum_{i=1}^{2(k-j)} 2 \pi \tau_k \sigma_{i}^{(j)}} \; .
\end{eqnarray}

\begin{figure}[ht] \begin{center}
\includegraphics[width=15cm]{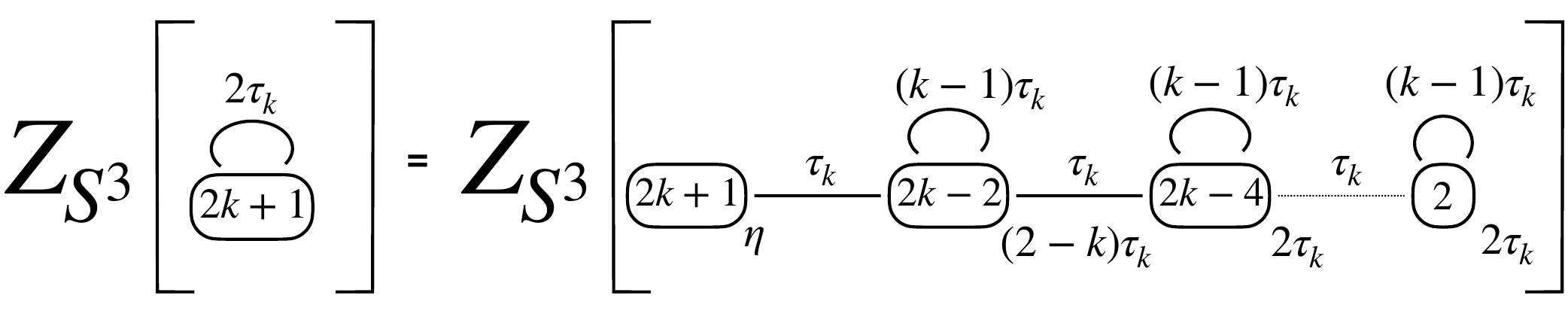}
\caption{Graphical representation for the deconfinement of an $\UU(2k+1)$ adjoint $\Phi$ with superpotential $W = \Phi^{k+1}$. The various gauge nodes are all unitary and they are identified with their rank.}
\label{fig:Akgr}
\end{center} 
\end{figure}

To avoid clutter, we can reformulate the identity just discussed at the graphical level. 
We refer to Figure \ref{fig:Akgr} for a graphical representation of the formula given above.
In the figure, we have also gauged the flavor symmetry, treating it as $\UU(2k+1)$, which corresponds to the case studied below.
The parameters appearing on the bifundamentals and on the adjoints correspond to the parameters in the hyperbolic gamma functions in formula 
(\ref{idgrak}) while the subscripts on the gauge groups are the FI parameters dictated by the monopole superpotential.
We have also added an FI for the $\UU(2k+1)$ gauge symmetry, that is associated with an unconstrained parameter $\eta$.
The lines connecting the gauge groups correspond to the pairs of bifundamentals and antibifundamentals appearing in Figure \ref{fig:ADk}.

 \subsubsection*{The case of $\UU(3k)$}
 
 The second confining duality used below involves the linear quiver in figure \ref{fig:ADk} (II). 
This theory is confining when we consider the  superpotential  
 \begin{equation}
 W = W_{\mathcal{N}=4} + \sum_{j=1}^{k-1} \mathcal{M}^{\dots \bullet_j \dots} + \mathcal{M}_A^{--\dots--} \; ,
 \end{equation}
where $\mathcal{M}^{\dots \bullet_j \dots} $ correspond to the flux $+1$ monopole for the $j$-th gauge group
$\UU(3j)$ in the linear quiver and $\mathcal{M}^{--\dots--}_A$  has flux $-1$ for every gauge group in the linear quiver dressed by one of the adjoints (any dressing is equivalent because of the chiral ring constraints).
Again, the model confines, as shown in \cite{Benvenuti:2024glr,Hwang:2024hhy}, and the dual WZ model has a single adjoint $\Phi=...$ with superpotential $W=\Phi^{k+1}$.

We use  this confining duality below to deconfine an $\UU(3k)$ adjoint.
The identity at the level of the partition function for the deconfinement of an $\UU(3k)$ adjoint $\Phi$ with superpotential $W = \Phi^{k+1}$ can be represented graphically as in Figure  \ref{fig:ADk}.
\begin{figure}[ht] \begin{center}
\includegraphics[width=15cm]{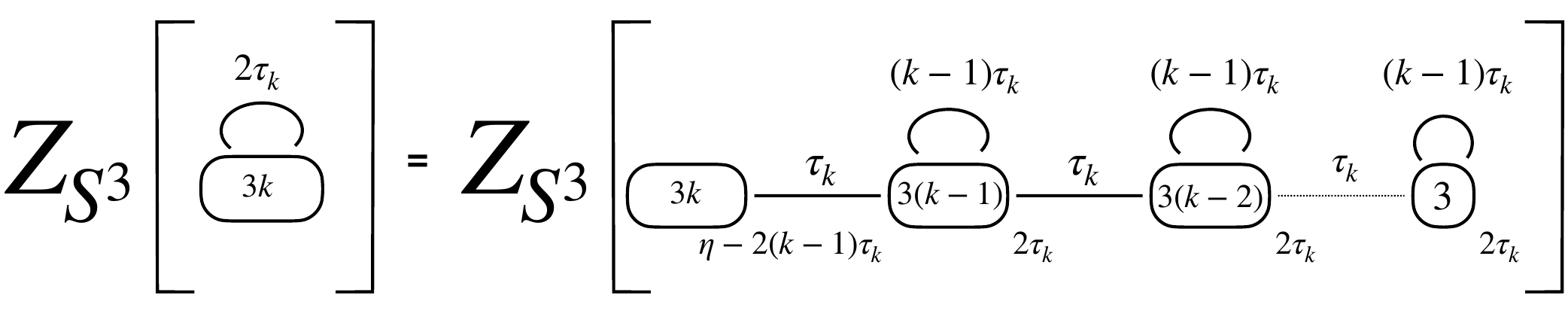}
\caption{Graphical representation for the deconfinement of an $\UU(3K)$ adjoint $\Phi$ with superpotential $W = \Phi^{k+1}$. The various gauge nodes are all unitary and they are identified with their rank.}
\label{fig:Dkgr}
\end{center} 
\end{figure}

\subsection{Confining $A_k$ duality with two gauge groups}
\label{sec:Aconf}

Here we prove a confining duality for a 3d $\mathcal{N}=2$ quiver that plays a crucial role in the 4d/3d reduction of the mezzanine dualities. 
The duality corresponds to a 3d version of the one studied in \cite{Brodie:1997sz} for a product of two $\SU(N _{1,2})$ gauge groups connected by a pair of bifundamentals, interacting through the superpotential 
\begin{equation}
W = (f \tilde f)^{k+1}\; ,
\end{equation}
and each one with 
additional flavors, $F_1$ and $F_2$ respectively. 
\begin{figure}[H] \begin{center}
\includegraphics[width=7cm]{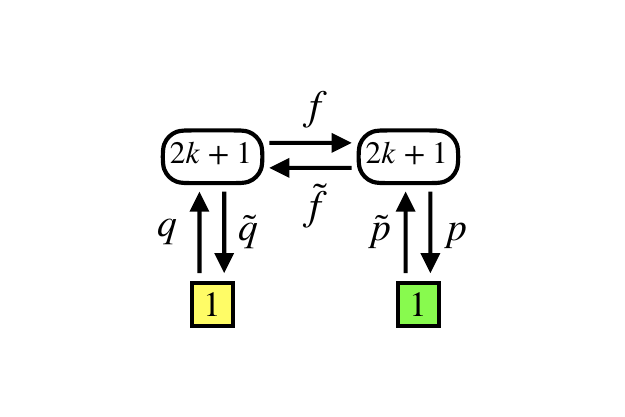}
\caption{Quiver representation of the confining $\UU(2k+1)^2$ gauge theory with two fundamentals and a pair of conjugated bifundamentals.}
\label{fig:eleAk}
\end{center} \end{figure}
Here we restrict to $N_1=N_2=2k+1$, where $k+1$ is the power of the superpotential involving the bifundamentals, and to $F_1=F_2=1$.
We represent the field content with the help of the quiver in Figure \ref{fig:eleAk}.
\begin{figure}[ht] \begin{center}
\includegraphics[width=15cm]{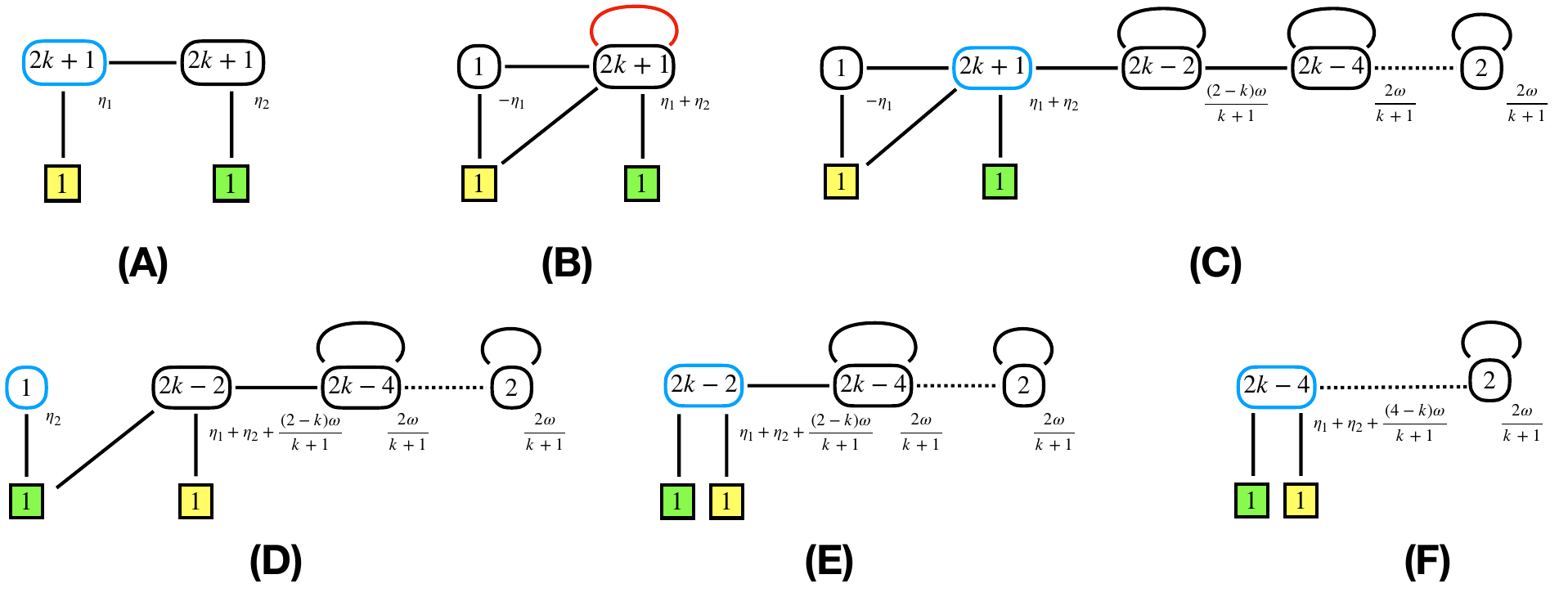}
\caption{Quiver representation of the iterative steps that we used in order to prove the confinement of the model. Observe that in this case we adopted a different notation for the bifundamentals in the quiver, by associating to each line connecting a pair of gauge nodes a conjugated pair of bifundamental and antibifundamental. We adopted the same notation for the fundamentals connected to the green and the yellow box, associated with two abelian flavor symmetries. The nodes represented in blue are the ones that undergo an IR duality in the subsequent step. In the second quiver (figure {\bf(B)}) we denoted the adjoint with a red color, this is because in that case we deconfine it by using the confining duality reviewed above. 
We also added on each gauge node the relative FI term, denoting the presence of monopole superpotentials in the action. Observe that the FI terms in these figures are the ones appearing in the partition functions.}
\label{DecAk}
\end{center} \end{figure}

In order to keep a compact notation in the body of the paper we report the explicit formula for the 
squashed three sphere partition function of this quiver 
\begin{eqnarray}
\label{genpfakconf}
Z_{\UU^2(2k+1)}(\vec \mu;\vec \nu;\tau_+,\tau_{-};\vec \eta)
&=&\frac{1}{((2k+1)!)^2}\!\!
 \int \prod_{i=1}^{2k+1} \prod_{r=1}^2 \ee^{-\mi \pi \eta_r \sigma_i^{(r)}}\!\!  \dd{\sigma_i^{(r)}}
  \Gamma_h(  \mu_{r}\!-\!\sigma_i^{(r)}\!, \nu_{r}\!+\!\sigma_i^{(r)})
\nonumber  \\
&\times&
\frac{\prod_{i,\ell=1}^{2k+1} \Gamma_h( \tau_\pm \pm (\sigma_i^{(1)}-\sigma_\ell^{(2)}))}
{\prod_{1\leq i<\ell \leq 2k+1}\prod_{r=1}^2
\Gamma_h(\pm(\sigma_i^{(r)}-\sigma_\ell^{(r)}))}\,,
\end{eqnarray}
with $m_A = \mu_1+\nu_1$ and $m_{A'} = \mu_2+\nu_2$.

In order to prove the fact that such a model is confining we follow the following iterative steps, which are summarized graphically in Figure \ref{DecAk}, where for simplicity each pair of conjugated field is associated with an un-oriented segment.
\begin{itemize}
\item[{\bf(A)}] We start by considering the $\UU(2k+1) \times \UU(2k+1)$ quiver with $W=(f \tilde f)^{k+1}$. In addition, here we consider all the possible mesons as flipped (indeed the model that we need to confine is obtained after a sequence of HW transitions at the brane level, where all the mesonic flippers  are turned on. However, the procedure that we describe here can be generalized to the case where these flippers are not turned on. The difference is that the final structure of the confining superpotential is more  complicated, while in the case at hand in each step some of the flippers are massive and they are 
integrated out at zero vev.
In the figure, we have highlighted a node in the quiver in blue, indicating that this node undergoes a duality (in this case, Aharony duality). 
At this step we consider generic FI terms $\eta_1$ and $\eta_2$ for the two gauge groups. We add the FI as a suffix to each node in the Figure. 
We refer to the value of these FI as the parameters appearing in the three sphere partition function, since our goal here is to derive an exact relation 
to plug in the formulas obtained from the double scaling analysis in the body of the paper.
We also associate an axial real mass $m_A$ to the fundamentals on the LHS of the quiver and $m_{A'}$ to the fundamentals on the RHS.
\item[(B)]
The second quiver in the Figure is obtained after the application of Aharony duality and in this case an adjoint field appears at the second gauge node, with a
power law superpotential of order $k+1$. This adjoint in depicted in red in the figure because we are going to deconfine it in favour of a tail of gauge groups. Such a tail was found in 
\cite{Benvenuti:2024glr,Hwang:2024hhy}, where it was named $\mathcal{C}_p(\SU(N))$ theory.
The duality in this case generates a flipper with real mass parameter $2m_A$ and two monopole operators that contribute to the partition function as
\begin{equation}
\label{mon1}
\Gamma_h\left( \pm \frac{\eta_1}{2}-m_A+\frac{\omega}{k+1} \right)\; .
\end{equation}
While the mesonic flipper simplifies in the low energy spectrum, the monopole operators act as  singlets and we need to keep them at low energy.
\item[(C)]
The third quiver represents the deconfinement of the adjoint with power law superpotential. There is an $\mathcal{N}=4$ tail with the addition of the monopole superpotential. This  is reflected in the structure of the FI terms explicitly represented in the quiver.  
\item[(D)] The fourth quiver is obtained by Aharony duality on the blue node of the previous step. The flippers that are generated at this step have mass parameters
$m_{A'}+m_A+\frac{\omega}{k+1}$,$\frac{2\omega}{k+1}+2m_A$ and $2m_{A'}+\frac{2k\omega}{k+1}$ and they are all integrated out. In addition a pair of fundamental fields 
charged under the first gauge group are integrated out, corresponding to presence of a  "triangle" superpotential at the previous step. 
Observe that at this step the colors of the flavor nodes are inverted accordingly.
Furthermore, two monopole operators, that contribute to the partition function as
\begin{equation}
\label{mon2}
\Gamma_h\left( \pm \frac{ \eta _1+\eta _2}{2}-m_{A'}-m_A-\frac{2(k-1) \omega }{k+1}\right)\; ,
\end{equation}
are generated.
\item[(E)] The fifth quiver is obtained by confining the $\UU(1)$ node represented in blue at the previous step. The massive flipper generated at this step has real mass parameter 
$2m_{A'}+\frac{2k\omega}{k+1}$ and the two monopole operators generated at this step  contribute to the partition function as
\begin{equation}
\label{mon3}
\Gamma_h\left( 
\pm \frac{\eta_2}{2}-m_{A'}+\frac{1}{k+1}\omega\right)\; .
\end{equation}
The quiver obtained so far can be reconfined sequentially. In order to explain this reconfinement, we first explicitly dualize the $\UU(2k-2)$ node and then extract the general result.
\item[(F)] The sixth quiver is obtained by Aharony duality on $\UU(2k-2)$. The massive flippers generated at this step have mass parameter
$\frac{3\omega}{k+1}+m_A+m_{A'}$, $2m_{A'}+\frac{2\omega}{k+1}$ and $\frac{4\omega}{k+1}+2m_A$ while the two monopoles contribute to the partition function as
\begin{equation}
\label{mon4}
\Gamma_h \left(
\pm \left(\frac{1}{2} \left(\eta _1+\eta _2\right)+\frac{ 2-k}{k+1} \omega\right)-\frac{k-2 }{k+1}\omega-m_{A'}-m_A
\right).
\end{equation}
We can iterate the procedure and then organize the various monopoles and flippers.
Summarizing we obtained three types of mesons, 
$\frac{(2\ell+1)\omega}{k+1}+m_A+m_{A'}$,
$2m_{A'}+\frac{2j\omega}{(k+1)}$ and $2m_A+\frac{2j\omega}{k+1}$ with  $ \ell=0,\dots,k-1$ and 
$j=0,\dots,k$. Observe that the first types of mesons (the ones labeled by $\ell$ ) have multiplicity $2$, because they are always generated in pairs.
Such flippers can be associated with the mesonic combinations of the electric theory, by representing the fields as in Figure \ref{fig:eleAk}. They are 
$M_{q p}^{(\ell)} \equiv q f(\tilde f f)^\ell  p$, 
$M_{\tilde q \tilde p}^{(\ell)} \equiv \tilde p \tilde f(f \tilde f )^\ell  \tilde q$
$M_{q \tilde q}^{(j)} \equiv q (f \tilde f )^j  \tilde q$ and 
$M_{p \tilde p}^{(j)} \equiv \tilde p (\tilde f f)^j p$ respectively, where $\ell$ and $j$ vary over the  range of values given above.
The difference in such ranges is consistent with the chiral ring truncation imposed by the $(f \tilde f)^{k+1}$ superpotential.

On the other hand the contribution to the three sphere partition function of the monopoles generated in the sequence of Aharony duality is
\begin{equation}
\label{monj}
\Gamma_h \left(
\pm \left(\frac{1}{2} \left(\eta _1+\eta _2\right)+\frac{\omega  (2 j-k+2)}{k+1}\right)+\frac{(2- k) \omega }{k+1}-m_{A'}-m_A
\right).
\end{equation}
where  $j=0,…,k-2$. The case $j=0$ corresponds to the monopoles in formula \eqref{mon4} and the case $j=k-2$ correspond to the monopoles of the last $\UU(2)$ gauge group in the sequence. Observe that there is a symmetry $j \rightarrow k-2-j$ in the spectrum of (\ref{monj}), that allows to reorganize the spectrum (adding also the contribution of the monopoles in formula (\ref{mon2}) as 
\begin{equation}
\label{mon12}
\Gamma_h \left(
\pm \frac{\eta _1+\eta _2}{2} -m_{A'}-m_A-\frac{2 j \omega}{k+1} 
\right),
\end{equation}
with $j=0,\dots,k-1$.

Summarizing we have three type of monopoles, that contribute to the partition function as (\ref{mon1}), (\ref{mon3}) and (\ref{mon12}) respectively and they correspond to the monopoles 
$\mathcal{M}_{\pm,\bullet}$,  $\mathcal{M}_{\bullet,\pm}$, $\mathcal{M}_{+,+}^{(j)}\equiv \mathcal{M}_{+,+} \text{tr} (f \tilde f)^{(j)} $  and  
$\mathcal{M}_{-,-}^{(j)}\equiv \mathcal{M}_{-,-} \text{tr} (f \tilde f)^{(j)} $ respectively. 
\end{itemize}

\subsection{Confining $D_k$ duality with two gauge groups}
\label{sec:Dconf}
Here we prove the  second confining duality relevant for our analysis,  corresponding to a 3d $\mathcal{N}=2$ quiver.
\begin{figure}[ht] \begin{center}
\includegraphics[width=7cm]{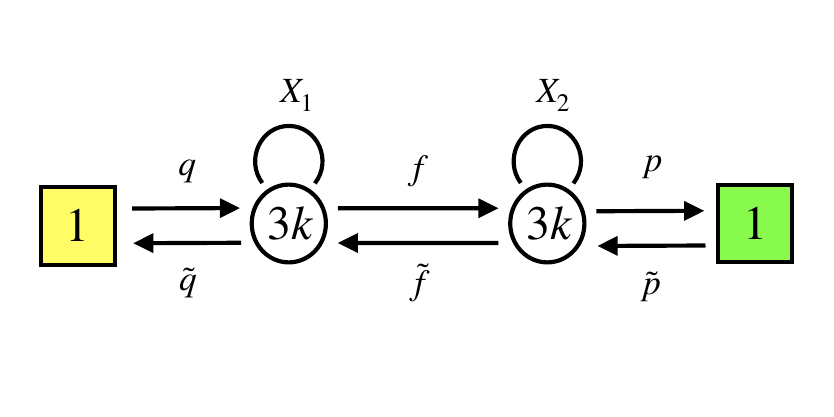}
\caption{Quiver representation of the confining $\UU(3k)^2$ gauge theory with two fundamentals and a pair of conjugated bifundamentals. In this case there are also two adjoints $X_{1,2}$ represented by curved lines without any arrow.}
\label{fig:eleDk}
\end{center} \end{figure}
The duality is  a 3d version of the one studied in \cite{Brodie:1996vx} for a product of two $\SU(N _{1,2})$ gauge groups, each one with an adjoint and connected by a pair of bifundamentals, interacting through the superpotential\footnote{The $(-1)^{k+1}$ was added in \cite{Brodie:1996vx} to guarantee the chiral ring truncation. In this paper we do not analyze this issue, and refer the reader to \cite{Brodie:1996vx}  and \cite{
Brodie:1996xm} for proper discussion.} 
\begin{equation}\label{spotappD}
W = \tr X_1^{k+1} + (-1)^{k+1} \tr X_2^{k+1} + f X_1 \tilde f +\tilde f X_2 f \; ,
\end{equation}
and each one with 
additional flavors, $F_1$ and $F_2$ respectively. 
Here, we restrict to $N_1=N_2=3k$ and to $F_1=F_2=1$.
We represent the field content with the help of the quiver in Figure \ref{fig:eleDk}.

Again, we report the explicit formula for the 
squashed three sphere partition function of this quiver 
\begin{eqnarray}
\label{genpfdkconf}
Z_{\UU^2(3k)}(\vec \mu;\vec \nu;\tau_{\vec X}\!\!\!&\!\!;\!\!&\!\!\!\tau_+,\tau_{-};\vec \eta)
=
\prod_{r=1}^{2} \frac{\Gamma_h^{3k}(\tau_{X_r})}{(3k)!}\!
\int \prod_{i=1}^{3k}  
 \prod_{r=1}^2 \ee^{-\mi \pi \eta_r \sigma_i^{(r)}} \!\!  \dd{\sigma_i^{(r)}}
\Gamma_h(  \mu_{r}\!-\!\sigma_i^{(r)}\!\!, \nu_{r}\!+\!\sigma_i^{(r)})\,
\nonumber  \\
&\times&
\frac{\prod_{i,\ell=1}^{3k} \Gamma_h( \tau_\pm \pm (\sigma_i^{(1)}-\sigma_\ell^{(2)}))
\prod_{r=1}^2
\Gamma_h(\pm(\sigma_i^{(r)}-\sigma_\ell^{(r)}+\tau_{X_r}))
}
{\prod_{1\leq i<\ell \leq 3k}\prod_{r=1}^2
\Gamma_h(\pm(\sigma_i^{(r)}-\sigma_\ell^{(r)}))}\,,
\end{eqnarray}
with $m_A = \mu_1+\nu_1$ and $m_{A'} = \mu_2+\nu_2$.

\begin{figure}[ht] \begin{center}
\includegraphics[width=10cm]{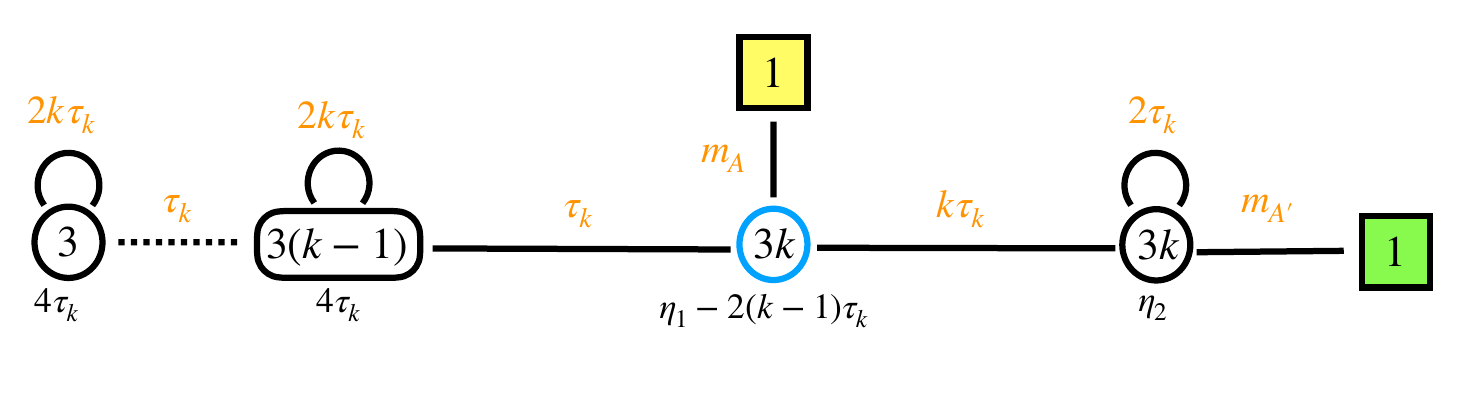}
\caption{Quiver obtained by deconfining the adjoint $X_1$. Similarly to the quivers in Figure \ref{DecAk},
the conjugated (bi)fundamentals are represented as single  lines.
Again  the FI terms are explicitly given, and in addition here we also represented, in orange,  the real mass parameters that enter in the partition function.}
\label{fig:eleDkdec}
\end{center} \end{figure}

Again, we inductively prove that this model is confining  by first deconfining an adjoint (here $X_1$) and then by acting sequentially with Aharony and confining dualities.
In the following, similarly to the $A_k$ case, we represent each pair of conjugated field with an un-oriented segment.
We start by considering the original model and referring to the axial masses as $m_A$ and $m_{A'}$ for the pairs $(q,\tilde q)$ and $(p,\tilde p)$.
Deconfining the adjoint, we obtain the quiver in Figure \ref{fig:eleDkdec}.
This quiver is the starting point of our analysis. We have represented the real mass parameters that enter in the partition function in orange and we added the effective FI terms. Differently to the case studied above, here the original FI is shifted after deconfining the adjoint. We have furthermore defined the quantity $\tau_k=\frac{\omega}{k+1}$.

\begin{figure}[ht] \begin{center}
\includegraphics[width=14cm]{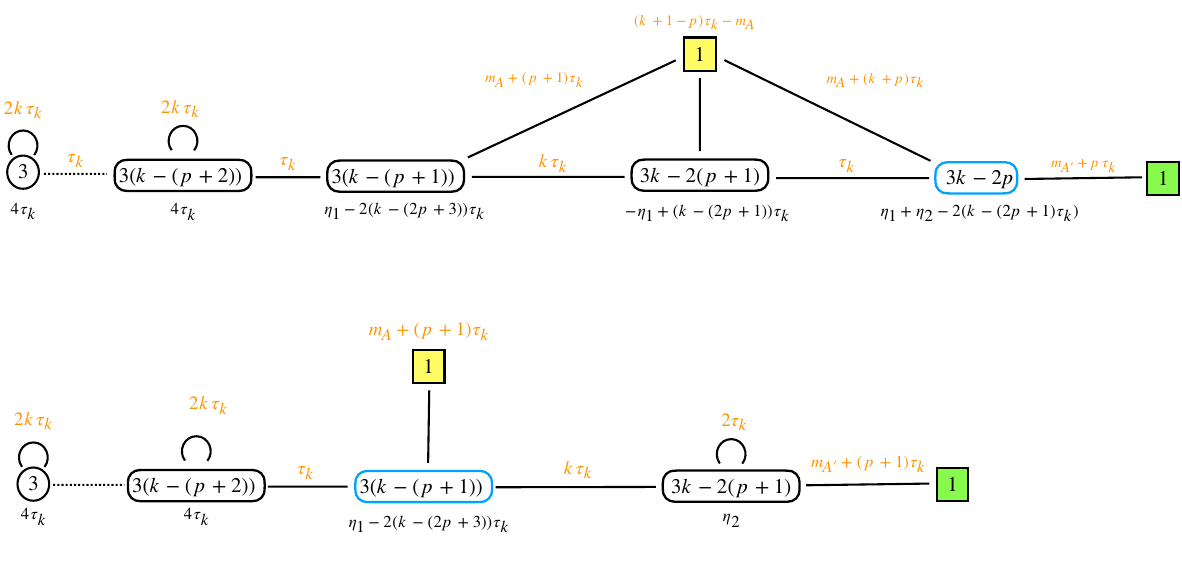}
\caption{
Iterative steps used in the proof  of confinement. The gauge groups represented in blue undergo an IR duality.
In the figure on the top the $\UU(3k-2p)$ nodes confines in the IR, then in the second figure the 
 $\UU(3(k-(p+1))$ nodes undergoes an IR Aharony duality and it becomes
  $\UU(3k-2(p+2))$. The quiver after this step comes back to be the one in the first figure on the top with decreased gauge ranks. By iterating this procedure one can prove the confining duality.
}
\label{fig:Dkpgen}
\end{center} \end{figure}

Next, we start by sequentially acting with duality (on the blue node) and then we confine the rightmost gauge node.
These two steps are iterated until we reach an $\UU(k)$ gauge theory with a single adjoint.
We can summarize these two steps by the two quivers in Figure \ref{fig:Dkpgen}
where $p=0,\dots,k-1$. Observe that the case $p=-1$ for the second quiver is represented in Figure \ref{fig:eleDk}.

The contribution to the three sphere partition function of the mesons and of the monopoles arising from  confinement of the $\UU(3k-2p)$ node in the first quiver can be collectively organized as 
\begin{equation}\
\label{mesp}
\prod_{j=0}^{k-1} \Gamma_h(2m_{A'}+2j \tau_k) \Gamma_h(2m_A+2(k+j) \tau_k) \Gamma_h(m_A+m_{A'}+(k+2j) \tau_k)^2
\end{equation}
for the mesons and 
\begin{equation}
\label{monp}
\prod_{j=0}^{k-1}
\Gamma_h \left(\pm \frac{\eta_1+\eta_2}{2}-2 (j+k-1)\tau_k -m_A-m_{A'}
\right)
\end{equation}
for the monopoles.
The same argument can be applied to the second quiver, where the $\UU(3(k-(p+1))$ node undergoes Aharony duality for $p=1,\dots,k-2$ (the case $p=k-1$ deserves a separate analysis).
In this case, the contribution of the mesons and of the monopoles can be summarized as 
\begin{equation}
\label{mesmonp}
\prod_{j=0}^{k-1} \Gamma_h\left(\pm \frac{\eta _1}{2}-m_A - (2j-1) \tau _k,2m_A+2 j \tau_k\right).
\end{equation}
The process ends when $p=k-1$, when we are left with a single adjoint with mass parameter $2 \tau_k$ and one flavor. The model has a power law superpotential of order $k+1$ for the adjoint and it corresponds to the confining limit of the Kim-Park \cite{Kim:2013cma} duality. The mesons and the monopoles of this confining duality are 
\begin{equation}
\label{mesmonppk}
\prod_{j=0}^{k-1} \Gamma_h\left(2 m_{A'}+2 (k+j) \tau_k ,\pm \frac{\eta_2}{2} -m_{A'}-(2j-1)\tau_k  \right).
\end{equation}

We can summarize the mesonic contributions in terms of the original fields as 
$M_{p \tilde p}^{(\ell)} =\tilde p (X_2)^\ell p$, $N_{q\tilde q}^{(\ell)}= q (f \tilde f)X_1^\ell \tilde q$, $M_{q p}^{(\ell)} = q f(X_1)^\ell  p$ and
$M_{\tilde q \tilde p}^{(\ell)} = \tilde p \tilde f(X_2)^\ell \tilde q$ appearing in  \eqref{mesp}, 
$M_{q \tilde q}^{(\ell)} = q (X_1 )^\ell  \tilde q$ appearing in  \eqref{mesmonp}, 
and 
$N_{p\tilde p}^{(\ell)}= p (f \tilde f)X_2^\ell \tilde p$
appearing in  \eqref{mesmonppk}.
On the other hand the monopoles 
$\mathcal{M}_{\pm,\bullet}$,  $\mathcal{M}_{\bullet,\pm}$ and $\mathcal{M}_{\pm ,\pm }$ appear in 
\eqref{monp}, \eqref{mesmonp} and \eqref{mesmonppk} dressed by the adjoints.

%
%
%
%
%
%
%
%
%
%
\section{4d/3d reduction for the $A_k$ cases}
\label{secAd}
%
%
%
%
%
%
%
%
%
%
In this section, we  apply the double scaling prescription to the reduction to 3d of the 4d $A_k$ dualities proposed in  \cite{Intriligator:1995ax}.
In a previous paper we have studied the limiting cases of such dualities, corresponding to $k=1$, while here we  generalize the construction to generic $k$.
Here we first briefly summarize the salient features of  the 4d dualities under investigation, referring the reader to the original references for further details.
In order to simplify the reading, we keep the 4d terminology of \cite{Brodie:1996xm}, referring to the models as \emph{mezzanine} and \emph{balcony}.
We  then study the reduction of these dualities by first gauging the baryonic symmetry, such that the resulting models involve a $\UU (N_c)$ gauge theory with fundamental and antifundamental flavors. 

The models have a D-brane interpretation, as reviewed in section \ref{branepicture},  corresponding to taking $k$ left and $k$ right NS$_{\pm \theta}$  fivebranes, while keeping a single central NS or NS' fivebrane in Figure \ref{fig:branefig}.
In the first case we have also $N_c$ D4 branes and $2N_f$ D6$_{\pm \theta }$ branes. In addition, we consider either an O6$^-$ or an O6$^+$ plane. We refer to the first case as the mezzanine $A_k^{(A)}$ duality, while the second one is denoted as mezzanine $A_k^{(S)}$ duality.

If, on the other hand, we consider an NS' brane, there is an additional stack of eight semi-infinte D6 branes, and this case corresponds to a deformation of the balcony $A_k$ duality. The deformation indeed triggers an RG flow that Higgses the original magnetic phase and the duality is modified accordingly. In the following, we discuss the reduction of the duality in the deformed case and we  comment on the undeformed case as well.
%
%
%
%
%
%
%
%
\subsection{Mezzanine $A_{k}^{(A)}$}
\label{MAA3d}
%
%
%
%
%
%
%
%

This model was first discussed in {\bf Section 2.6 of \cite{Intriligator:1995ax}}. The field content consists of a pair of conjugated antisymmetric two-index tensors $A$ and $\tilde A$ in addition to $N_f$ pairs of fundamentals and antifundamentals $Q$ and $\tilde Q$.
In the  following we refer to this model as mezzanine $A_k^{(A)}$.
On the electric side the superpotential is
\begin{equation}
\label{eq:WeleAkAAt}
W = (A \tilde A)^{k+1}.
\end{equation}
This model is dual to an $\SU(\tilde N_c = (2k+1) N_f -N_c-4k)$ gauge theory with a pair of conjugated antisymmetric two-index tensors $a$ and $\tilde a$ in addition to $N_f$ pairs of fundamentals and antifundamentals $q$ and $\tilde q$.
In this case, a large number of mesons, acting as singlets in this dual phase, must be introduced to flip certain operators in order to match the chiral ring.
There are bifundamental mesons  $M_j = Q (A \tilde A)^j \tilde Q$ with $j=0,\dots,k$ and antisymmetric mesons 
$P_\ell = Q (\tilde A A) ^\ell \tilde A Q$ and $\tilde P_\ell = \tilde Q (A \tilde A)^\ell  A \tilde Q$ with $\ell=0,\dots,k-1$. The bounds on the labels $j$ and $\ell$ follow from the constraints on the chiral ring imposed by \eqref{eq:WeleAkAAt} and the dual superpotential is
\begin{equation}
\label{eq:WmacAkAAt}
W = (a \tilde a )^{k+1}
+
\sum_{j=0}^{k} M_j q (\tilde a a)^{k-j} \tilde q
+
\sum_{\ell=0}^{k-1} \left( P_{\ell}  q ( \tilde a a)^{k-\ell-1} \tilde a  q+ \tilde P_{\ell}  \tilde q ( a\tilde  a)^{k-\ell-1}  a \tilde q\right).
 \end{equation}

The reduction to 3d of the $\UU(N_c)$ version of this duality was studied in \cite{Amariti:2025gca} for the case $k=1$, while here we consider  the $k>1$ case.

In order to study the reduction, we summarize 
the global charges of the electric side 
\begin{equation}
\label{gloele}
\begin{array}{c|c|ccccc}
               &\SU(N_c)&\SU(N_f)   & \SU(N_f) & \UU(1)_B & \UU(1)_X & \UU(1)_R\\
               \hline
Q            & \square &  \square &1 &   \frac{1}{N_c}&0 &1- \frac{N_c+2k}{(k+1)N_f} \\
\tilde Q    & \overline \square & 1  &  \square &\! \! \! \! \! -\frac{1}{N_c} &0  & 1- \frac{N_c+2k}{(k+1)N_f} \\
\tilde A    & \begin{array}{c} \overline \square \vspace{-2.9mm} \\  \square  \end{array}  &1   & 1&\! \! \! \! \!   -\frac{2}{N_c}&\! \! \! \! \! -1  &\frac{1}{k+1} \\
A            & \begin{array}{c} \square \vspace{-2.9mm} \\  \square  \end{array} &1    & 1&\frac{2}{N_c} & 1 &\frac{1}{k+1}  \\
\end{array}
\end{equation}
and of the magnetic side
\begin{equation}
\label{dualfuge}
\begin{array}{c|c|cccccc}
&\SU(\tilde N_c)&\SU(N_f)   & \SU(N_f) & \UU(1)_B & \UU(1)_X & \UU(1)_R\\
\hline
q            & \square &  \overline \square &1 &   \frac{1}{\tilde N_c}& \frac{k(N_f-2)}{\tilde N_c} &1- \frac{\tilde N_c+2k}{(k+1)N_f} \\
\tilde q    & \overline \square & 1  &  \overline\square&\! \! \! \! \! -\frac{1}{\tilde N_c} &\! \! \! \! \!   -\frac{k(N_f-2)}{\tilde N_c} & 1- \frac{\tilde N_c+2k}{(k+1)N_f} \\
\tilde a    &\begin{array}{c} \overline \square \vspace{-2.9mm} \\  \square  \end{array}  & 1& 1&\! \! \! \! \!  -\frac{2}{\tilde N_c}&\! \! \! \! \!  - \frac{N_c-N_f}{\tilde N_c}&\frac{1}{k+1} \\
a            & \begin{array}{c} \square \vspace{-2.9mm} \\  \square  \end{array}    &1 & 1&\frac{2}{\tilde N_c} &  \frac{N_c-N_f}{\tilde N_c}&\frac{1}{k+1}  \\
\hline 
P_\ell             & 1  & \begin{array}{c} \square \vspace{-2.9mm} \\  \square  \end{array}  & 1& 0  &\! \! \! \! \! -1 &\frac{\tilde N_c-N_c+2(\ell+1)N_f}{N_f (k+1)}\\
\tilde P_\ell    & 1  & 1  & \begin{array}{c} \square \vspace{-2.9mm} \\  \square  \end{array}  &0   &1 &\frac{\tilde N_c-N_c+2(\ell+1)N_f}{N_f (k+1)} \\
M_j            & 1  & \square  &\square  & 0 & 0& \frac{\tilde N_c-N_c+(2 j+1)N_f}{N_f (k+1)}\\
\end{array}
\end{equation}
The integral formulas matching the superconformal index of the electric and of the dual phase for the $\UU(N)$ duality are obtained by integrating over the fugacities of the 
 the $\UU(1)_B$ symmetry from the formulas given in  {\bf Section  9.3.1 of \cite{Spiridonov:2009za}}. We refer the reader to \cite{Amariti:2025gca} for the detailed results for the case $k=1$. The case of higher $k$  is obtained by modifying the contributions of the $\UU(1)_X$ and $\UU(1)_R$-charges as read from the tables of charges above.
 
 For completeness, in this case we report the formulas for the electric and the magnetic indeces explicitly, while in the following examples we omit them, leaving the details to the reader.
 
 On the electric side, after gauging the baryonic symmetry, we have
\begin{equation}
\label{indexeleAA}
I_{ele.} = I_{\UU(N_c)}^{[F\square;F \overline \square; 1  \overline{A};1A]} 
(\vec u;\vec v; t_{\tilde A};t_A)\,,
\end{equation}
where, using the conventions spelled out in \cite{Amariti:2025gca}  the arguments separated by a semicolon transform under  different representations. Such  representations are specified in the square brackets in the index  and 
the parametrization of the fugacities can be read from  \eqref{gloele}. We have
\begin{equation}
\label{parele0}
\begin{array}{lll}
u_b = m_b (pq)^{\frac{1}{2} \left(1-\frac{N_c+2k}{(k+1) N_f}\right)}, & \quad b=1,\dots,N_f\,, &\quad
\text{with} \quad \prod_{b=1}^{N_f} m_b=1\,,\\
v_b = n_b (pq)^{\frac{1}{2} \left(1-\frac{N_c+2k}{(k+1)N_f}\right)}, &\quad b=1,\dots,N_f\,, &\quad
\text{with} \quad \prod_{b=1}^{N_f} n_b=1\,,\\
t_{\tilde A} = x^{-1} (pq)^{\frac{1}{2(k+1)}}, \\
t_A = x (pq)^{\frac{1}{2(k+1)}} .\\
\end{array}
\end{equation}
There is a further constraint on the fugacities, denoted as balancing condition in the mathematical literature, that removes the anomalous axial symmetry. In this case the  balancing condition is \cite{Spiridonov:2009za}
\begin{equation}
\label{bcaa}
\prod_{b=1}^{N_f} u_b \, v_b
=
(pq)^{N_f-\frac{N_c+2}{k+1}},
\end{equation} 
and it is  satisfied by the parametrization \eqref{parele0}.
The magnetic index is
\begin{eqnarray}
\label{ima}
I_{mag.} &=& \prod_{\ell=0}^{k-1}
\prod_{1 \leq b < c \leq N_f} \Gamma_e( (t_A t_{\tilde A})^\ell t_{\tilde A} u_b u_c,
(t_A t_{\tilde A})^\ell t_A v_b v_c)\nonumber \\
&\times & \prod_{j=0}^k
\prod_{b,c=1}^{N_f} \Gamma_e((t_A t_{\tilde A})^j u_b v_c)\;
I_{\UU(\tilde N_c)}^{[N_f\square;N_f \overline \square; 
1 \overline A;1 A]} 
(\vec {\tilde u};\vec {\tilde v};  \tilde t_{\tilde a};\tilde t_a),
\end{eqnarray}
where the duality dictionary translates into the 
parametrization of the dual fugacities in the argument of $I_{\UU(\tilde N_c)}$, that can be read analogously from \eqref{dualfuge}.

\begin{figure}[ht] \begin{center}
\includegraphics[width=5cm]{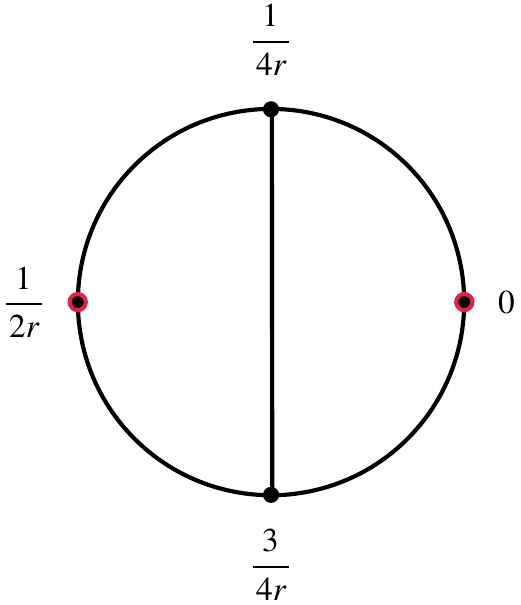}
\caption{
Pictorial representation of the circle direction, referred to as $x_3$.
 Here  $\frac{1}{r}$ is
  the periodicity of the compact scalar $\sigma$  on the compact direction.
The red circles  correspond to  the positions of the orientifolds in the 
T-dual geometric setup.
}
\label{fig:eles1}
\end{center} \end{figure}

To reduce the theory to three dimensions, we  make use of the double scaling limit reviewed in Section \ref{doublescaling}. This is done by considering in the associated brane picture 
a configuration symmetric with respect to the vertical axis in Figure \ref{fig:eles1}.
We take $N_c=2N$ and $N_f=2F+2$ and displace $F$ D6 at $x_3=0$ and, symmetrically to the vertical axes, $F$ at $x_3=1/2 r$; we then displace the other two symmetrically at $x_3=1/4 r$ and $x_3=3/4 r$. 
We also displace $N$ D3 at the origin and $N$ D3 at $x_3=1/2 r$.
The dual gauge group $\UU(\tilde N_c \equiv 2 \tilde N  = 4kF-2N+4)$ is then broken into two $\UU((2k+1)F-N-2k)$, at the origin and at $1/2 r$ and two $\UU(2k+1)$, at $1/4 r$ and at $3/4 r$. 
Each $\UU(2k+1)$ has a fundamental and antifundamental flavor that interact with each other through a pair of conjugated bifundamental fields, arising from the antisymmetric fields. Such fields also inherit the  superpotential \eqref{eq:WeleAkAAt}.
These two $\UU(2k+1)$ have opposite  FI, dictated by the presence of the orientifold.

Our analysis is devoted to  obtaining this configuration, which is read from the brane picture, by reducing  the identity between the electric index in \eqref{indexeleAA} and the magnetic index in \eqref{ima}.
By applying the double scaling prescription, 
reviewed in Section \ref{doublescaling}, we aim to 
obtain an identity that is a perfect square aside from the two  $\UU(2k+1)$ factors contributions. 
This requires to further identify the mass parameters at the origin with the ones at $1/(2r)$.
Then, using the results of  Section \ref{sec:Aconf}, where we show that these sectors are confining, we  show
that their contribution simplifies to  a perfect square as well.
This gives rise to an identity such as
\begin{eqnarray}
\label{quasiquadratoperfetto}
&&
\left(
Z_{\UU(N)}^{(F\square; F \overline \square; 1 \overline{A};1 A)}(\vec \mu;\vec \nu; \tau_{\tilde A};\tau_A;\Lambda) \right)^2
=\left(Z_{\UU((2k+1)F-N-2k)}^{(F \square; F \overline \square; 1 \overline{A} ;1 A)}
\left(\vec {\tilde \mu};\vec {\tilde \nu}; \tilde  \tau_{\tilde a};\tilde  \tau_a;-\Lambda
\right)\right)^2
 \nonumber\\
  &&  \;
 \prod_{j=0}^{k}\prod_{1\leq b,c \leq F}\!\!\!
\Gamma_h^2(\mu_b+ \nu_c+2 j\tau_k)  \,\,\,\,
\prod_{\ell=0}^{k-1}\prod_{1\leq b<c \leq F} \!\!\!\!\!
\Gamma_h^2(\tau_{\tilde A} \!+\!\mu_b \!+\!\mu_c\!+\!2 \ell \tau_k,\tau_A \!+\!\nu_b \!+\!\nu_c\!+\!2\ell\tau_k) \nonumber
\\
 &&  \;
 {\color{red} Z_{\UU(2k+1)^2} (\tilde \mu_{F+1},\tilde \nu_{F+2}; \tilde \nu_{F+1},\tilde \mu_{F+2};\tilde \tau_a,\tilde \tau_{\tilde a};\Lambda;-\Lambda )}  \,\,
 {\color{red} \prod_{j=0}^{k}\prod_{b=F+1}^{F+2}\!\! \Gamma_h(\mu_b\!+\! \nu_b+ 2 j \tau_k)} \nonumber
 \\
&&  \;
\textcolor{red}{ \prod_{\ell=0}^{k-1}\Gamma_h(\tau_{\tilde A}\!+\!\mu_{F+1}\!+\!\mu_{F+2} +2\ell\tau_k,\tau_A\!+\!\nu_{F+1}\!+\!\nu_{F+2}+2\ell\tau_k) }
\;,
\end{eqnarray}
where the red factors reconstruct the $\UU(2k+1)\times \UU(2k+1)$ sector. The notation for  the matrix integral part of this sector has been shown in  \eqref{genpfakconf}.

The arguments appearing in the dual partition function are related to the ones on the electric side by the dictionary
\begin{equation}
\label{dictionary}
\begin{aligned}
\tilde{\mu}_{b} &= \tau_k - \mu_b + k \frac{F}{\tilde N} x,
&\qquad
\tilde{\nu}_{b} &= \tau_k - \nu_b - k \frac{F}{\tilde N} x,
\\[6pt]
\tilde{\tau}_a &= \tau_k + \frac{N-F-1}{\tilde N} x,
&\qquad
\tilde{\tau}_{\tilde a} &= \tau_k - \frac{N-F-1}{\tilde N} x,
\end{aligned}
\end{equation}
where we have introduced for shortness $\tau_k=\frac{\omega}{k+1}$. 

Furthermore, the electric fugacities are constrained by a balancing condition, inherited from the 4d one.
In the 3d theory on $S^1$, it translates into a constraint imposed by the KK monopole superpotential. 
The constraint is obtained from \eqref{bcaa} and it reads: 
\begin{equation}
\label{bc1}
\sum_{b=1}^F(\mu_b+\nu_b)+\frac{1}{2}(\mu_{F+1}+\mu_{F+2}+\nu_{F+1}+\nu_{F+2})=2(F\omega-\tau_k (N-1))\,.
\end{equation} 
The sector $\UU(2k+1)\times\UU(2k+1)$ is represented in Section \ref{sec:Aconf} in Figure \ref{fig:eleAk}. It can be simplified using the prescription outlined in Section \ref{sec:Aconf}. Indeed, we can study the $\UU(2k+1)\times \UU(2k+1)$ along the lines of Section \ref{sec:Aconf}, upon setting $\eta_1=\eta_2$ and $\mu_{F+1}+\nu_{F+1}=\mu_{F+2}+\nu_{F+2}$ for symmetry reasons. It confines giving rise to two towers of monopoles, $\mathcal{M}_{+,+}^{(j)}$ and $\mathcal{M},_{-,-}^{(j)}$ and two additional monopoles $\mathcal{M}_{\pm,\bullet}$ and  $\mathcal{M}_{\bullet,\pm}$. 
The notation used for the monopole operators has been already discussed in subsection \ref{decsarasec}.
Observe that in this confining sector there are also towers of mesons that appear in the IR. Such mesons are flipped with the singlets represented in red outside the integral in formula (\ref{quasiquadratoperfetto}).

Effectively, all the contributions in \eqref{quasiquadratoperfetto} coming from the monopole  sector of the confining duality simplify to 
\begin{equation}
\Gamma_h^2\left(\pm \frac{\eta}{2}-\frac{1}{2}(\tilde{\mu}_{F+1}+\tilde{\nu}_{F+1})+\tau_k\right)\prod_{\ell=0}^{k-1}\Gamma_h^2\left(-(\tilde{\mu}_{F+1}+\tilde{\nu}_{F+1})-\frac{2 \omega \ell}{k+1}\right).
\end{equation}
If we now plug this back into \eqref{quasiquadratoperfetto}, we observe that the identity is of the form of \eqref{sqeq}. After extracting the square root and substituting the balancing condition \eqref{bc1}, we obtain the final identity
 \begin{equation}
\label{antisymmetricidons1}
\begin{aligned}
&
Z_{\UU(N)}^{(F\square; F \overline \square;1  \overline{A};1 A)}(\vec \mu;\vec \nu; \tau_{\tilde A};\tau_A;\Lambda) \;=\;Z_{\UU((2k+1)F-N-2k)}^{(F \square; F \overline \square; 1 \overline{A} ;1 A)}
\left(\vec {\tilde \mu};\vec {\tilde \nu}; \tilde  \tau_{\tilde a};\tilde  \tau_a;-\Lambda\right)\times
\\
&\quad\times\prod_{\ell=0}^{k-1}\prod_{1\leq b< c \leq F}
\Gamma_h(\tau_{\tilde A} +\mu_b +\mu_c+2\ell\tau_k,\tau_A+\nu_b +\nu_c+2\ell\tau_k)
\\
&\quad \times\prod_{j=0}^{k}
\prod_{1\leq b,c \leq F}
\Gamma_h(\mu_b+ \nu_c+2j\tau_k)\;
\prod_{\ell=0}^{k-1}\Gamma_h\left(2\left(F\omega-\tau_k (N+\ell)\right)-\sum_{b=1}^{F} (\mu_b+\nu_b)\right)
\\
&\quad \times\;
\Gamma_h\left(\pm \frac{\eta}{2} +F\omega+\tau_k (1-N)-\frac{1}{2}\sum_{b=1}^{F} (\mu_b+\nu_b)\right)
\,,
\end{aligned}
\end{equation}
which holds in absence of any additional constraints.
The final duality can be summarized as follows
\begin{equation}
\label{MAkApure3d}
\begin{array}{|cc|}
\hline
\UU(N) & \UU((2k+1)F-N-2k) \\
(A,\tilde A) \oplus F(Q,\tilde Q)& (a,\tilde a) \oplus  F (q,\tilde q) \\
W = \eqref{eq:WeleAkAAt} & W = \eqref{eq:WmacAkAAt}+ Flip[\mathcal{M}^{\pm},\mathcal{M}^{(0)}_{\ell} ] \\ 
\hline
\end{array}
\end{equation}
where $\mathcal{M}^{\pm}$ are the flux $\pm 1$ monopole for the dual gauge group  $\UU((2k+1)F-N-2k)$,
 $ \mathcal{M}^{(0)} $ is the  bare  $\SU((2k+1)F-N-2k))$ monopole and 
 $\mathcal{M}^{(0)}_\ell= \mathcal{M}^{(0)} \, Tr(a \tilde a)^{\ell}$  with $\ell=0,\dots,k-1$.

The flippers are the duals of the relative monopole operators acting as singlets in the magnetic phase.

We conclude the analysis by discussing the reduction of this duality from the ARSW prescription. In such a case, it is indeed possible to study the effective duality on $S^1$ by considering on the electric side $N$ D4 branes and $F+1$ pairs of D6 branes.
The real mass flow is triggered on the electric side by assigning a large positive mass to a fundamental and a large negative mass to an antifundamental.
On the dual side one has to consider a further Higgsing of the  $\UU((2k+1)F-N-2k+1)$ gauge group to  $\UU((2k+1)F-N-2k) \times \UU(1)$.
In the  $\UU(1)$ sector there are a fundamental and an antifundamental, while $F$ pairs remain in the $\UU((2k+1)F-N-2k)$ sector.
The mesons $M_j$ split in the two sectors, while the mesons $P_{\ell}$ are massless only in the dual non-abelian sector.
By defining the massless mesons that survive in the $\UU(1)$ dual sector as
${\bf M}_j \equiv Q_{F+1} (A \tilde A)^j \tilde Q_{F+1}$, we observe that only ${\bf M}_{k}$ interacts  with the dual fundamentals through a cubic superpotential.
This $\UU(1)$ sectors can be further confined, because it is a deformation of SQED. After confining this node, we are left with the two monopoles with flux $\pm 1$ while the mesonic direction gets a  mass with the meson  ${\bf M}_{k}$.
Then, we are left with the monopole and antimonopole of this sector that act as singlets in the dual phase, in addition to the mesons ${\bf M}_{0,\dots,k-1}$. 

These fields correspond to the monopoles worked out from the double scaling studied above.
We can provide an explicit derivation of this fact by studying the charges of these field directly focusing on the mass flow at the level of the three sphere partition function.

We start by considering the balancing condition for the theory on the circle:
\begin{equation}
\label{AAtBBC}
\sum_{a=1}^{F+1} (\mu_a + \nu_a)= 2(F-1) \omega -2 \tau_k(N-2)\; .
\end{equation}
Then we consider the $\UU(1)$ sector in the dual phase. The matrix integral for this sector is
\begin{equation}
\label{confs}
\int d\sigma \Gamma_h(\tilde \mu_{F+1}+\sigma, \tilde \nu_{F+1}-\sigma) e^{i \pi \eta \sigma} =
 \Gamma_h \left(\pm\frac{\eta}{2}+\omega-\frac{\tilde \mu_{F+1}+\tilde \nu_{F+1}}{2} \right) 
  \Gamma_h \left( \tilde \mu_{F+1}+\tilde \nu_{F+1} \right) ,
\end{equation}
where 
$ \tilde \mu_a+\tilde \nu_b  = 2\tau_k - \mu_a-\nu_b$ for any $a,b=1,\dots,F+1$.

The tower of mesons ${\bf M}_{j}$ contributes to the partition function as
\begin{equation}
\label{mondoppi}
\prod_{j=0}^{k} \Gamma_h( \mu_{F+1} +\nu_{F+1}  +2 j \tau_k)\; .
\end{equation}
Observe that, upon using the balancing condition (\ref{AAtBBC}) and the inversion relation for the hyperbolic Gamma function,  we obtain that the product of the hyperbolic Gamma function of the meson ${\bf M}_{k}$ and the one corresponding to the  last term in the RHS of \eqref{confs} is equal to $1$.
This  is interpreted as the presence of an holomorphic mass term involving these two singlets in the superpotential.
We are then left with the tower ${\bf M}_{0,\dots,k-1}$ as expected from the field theory analysis. By substituting 
the balancing condition (\ref{AAtBBC})  in the argument of (\ref{mondoppi}),  one obtains, after some rearrangements, the  corresponding monopoles that appear in (\ref{antisymmetricidons1}).
Furthermore, the contribution of the  other monopoles, in the last line of (\ref{antisymmetricidons1}), correspond to the first term in the RHS of
\eqref{confs}.

%
%
%
%
%
%
%
%
\subsection{Mezzanine $A_{k}^{(S)}$}
\label{MSS3d}
%
%
%
%
%
%
%
%

A second  model, discussed in {\bf Section 2.7 of \cite{Intriligator:1995ax}}, consists of a pair of conjugated symmetric two-index tensors $S$ and $\tilde S$ in addition to $N_f$ pairs of fundamentals and antifundamentals $Q$ and $\tilde Q$. In the  following, we refer to this model as mezzanine $A_k^{(S)}$. On the electric side, the superpotential is
\begin{equation}
\label{eq:WeleAkSSt}
W = (S \tilde S)^{k+1}\;.
\end{equation}
This model is dual to an $\SU(\tilde N_c = (2k+1) N_f -N_c+4k)$ gauge theory with pair of conjugated symmetric two-index tensors $s$ and $\tilde s$ in addition to $N_f$ pairs of fundamentals and antifundamentals $q$ and $\tilde q$.
There are bifundamental mesons  $M_j = Q (S \tilde S)^j \tilde Q$ with $j=0,\dots,k$ and symmetric mesons 
$P_\ell = Q (\tilde S S) ^\ell \tilde S Q$ and $\tilde P_\ell = \tilde Q (S \tilde S)^\ell  S \tilde Q$ with $\ell=0,\dots,k-1$. Again, the bounds on the labels $j$ and $\ell$ follow from the constraints on the chiral ring imposed by \eqref{eq:WeleAkSSt} and the dual superpotential is
\begin{equation}
\label{eq:WmacAkSSt}
W = (s \tilde s )^{k+1}
+
\sum_{j=0}^{k} M_j q (\tilde s s)^{k-j} \tilde q
+
\sum_{\ell=0}^{k-1} \left( P_{\ell}  q ( \tilde s s)^{k-\ell-1} \tilde s  q+ \tilde P_{\ell}  \tilde q ( s\tilde  s)^{k-\ell-1}  s \tilde q\right).
 \end{equation}
The reduction to 3d of the $\UU(N_c)$ version of this duality was studied in \cite{Amariti:2025gca} for the case $k=1$, while here we consider  the $k>1$ case.
The global charges of the electric side are
\begin{equation}
\begin{array}{c|c|ccccc}
               &\SU(N_c)&\SU(N_f)   & \SU(N_f) & \UU(1)_B & \UU(1)_X & \UU(1)_R\\
               \hline
Q            & \square &  \square &1 &   \frac{1}{N_c}&0 &1- \frac{N_c-2k}{(k+1)N_f} \\
\tilde Q    & \overline \square & 1  &  \square &\! \! \! \! \! -\frac{1}{N_c} &0  & 1- \frac{N_c-2k}{(k+1)N_f} \\
\tilde S   & \overline{\square \! \square}&1   & 1&\! \! \! \! \!   -\frac{2}{N_c}&\! \! \! \! \! -1  &\frac{1}{k+1} \\
S            & \square \! \square &1    & 1&\frac{2}{N_c} & 1 &\frac{1}{k+1}  \\
\end{array}
\end{equation}
and of the magnetic side are
\begin{equation}
\begin{array}{c|c|cccccc}
&\SU(\tilde N_c)&\SU(N_f)   & \SU(N_f) & \UU(1)_B & \UU(1)_X & \UU(1)_R\\
\hline
q            & \square &  \overline \square &1 &   \frac{1}{\tilde N_c}& \frac{k(N_f+2)}{\tilde N_c} &1- \frac{\tilde N_c-2k}{(k+1)N_f} \\
\tilde q    & \overline \square & 1  &  \overline\square&\! \! \! \! \! -\frac{1}{\tilde N_c} &\! \! \! \! \!   -\frac{k(N_f+2)}{\tilde N_c} & 1- \frac{\tilde N_c-2k}{(k+1)N_f} \\
\tilde s    & \overline{\square \! \square}& 1& 1&\! \! \! \! \!  -\frac{2}{\tilde N_c}&\! \! \! \! \!  - \frac{N_c-N_f}{\tilde N_c}&\frac{1}{k+1} \\
s            & \square \! \square   &1 & 1&\frac{2}{\tilde N_c} &  \frac{N_c-N_f}{\tilde N_c}&\frac{1}{k+1}  \\
\hline 
P_\ell             & 1  & \square \! \square   & 1& 0  &\! \! \! \! \! -1 &\frac{\tilde N_c-N_c+2(\ell+1)N_f}{N_f (k+1)}\\
\tilde P_\ell    & 1  & 1  &\square \! \square   &0   &1 &\frac{\tilde N_c-N_c+2(\ell+1)N_f}{N_f (k+1)} \\
M_j            & 1  & \square  &\square  & 0 & 0& \frac{\tilde N_c-N_c+(2 j+1)N_f}{N_f (k+1)}\\
\end{array}
\end{equation}
Again, the integral formulas matching the superconformal index of the electric and of the dual phase for the $\UU(N)$ duality are obtained by integrating over the fugacities of the 
 the $\UU(1)_B$ symmetry from the formulas given in  {\bf Section  9.3.2 of \cite{Spiridonov:2009za}}. We refer the reader to \cite{Amariti:2025gca} for the detailed results for the case $k=1$. The case of higher $k$ is obtained by modifying the contributions of the $\UU(1)_X$ and $\UU(1)_R$-charges as read from the tables of charges above.

In the following, we study the reduction of this duality to 3d. We will be more schematic in the analysis, since the procedure is the same as in previous section. 
We start by taking again $N_c=2N$ and $N_F=2F+2$, then we displace $F$ flavor fugacities at the origin and at $1/(2r)$ and one at $1/(4r)$ and $3/(4r)$. In this case the dual group is broken into $\UU((2k+1)F-N+2k)\times\UU((2k+1)F-N+2k)\times\UU(2k+1)\times\UU(2k+1)$. We also define the final dual gauge group as $\tilde N =(2k+1)F-N+2k $.

By applying again the double scaling limit, we obtain an identity between partition functions analogous to the one studied above for the mezzanine $A_{k}^{(A)}$ case. We have
\begin{equation}
\label{quasiquadratoperfettoSAk}
\begin{aligned}
&
\left(
Z_{\UU(N)}^{(F\square; F \overline \square; 1 \overline{S};1 S)}(\vec \mu;\vec \nu; \tau_{\tilde S};\tau_S;\Lambda) \right)^2
=\left(Z_{\UU((2k+1)F-N+2k)}^{(F \square; F \overline \square; 1 \overline{S} ;1 S)}
\left(\vec {\tilde \mu};\vec {\tilde \nu}; \tilde  \tau_{\tilde s};\tilde  \tau_s;-\Lambda
\right)\right)^2\times
\\
& 
\prod_{\ell=0}^{k-1}\prod_{1\leq b \leq c \leq F}
\Gamma_h^2(\tau_{\tilde S} \!+\!\mu_b \!+\!\mu_c+2\ell\tau_k,\tau_S \!+\!\nu_b \!+\!\nu_c+2\ell\tau_k)
\prod_{j=0}^{k}\prod_{1\leq b,c \leq F} \Gamma_h^2(\mu_b+ \nu_c+2j\tau_k) 
\\
& {\color{red}
 \prod_{j=0}^{k}\prod_{b=F+1}^{F+2}\!\! \Gamma_h(\mu_b\!+\! \nu_b+2j\tau_k)
 Z_{\UU(2k+1)^2} (\tilde \mu_{F+1},\tilde \nu_{F+2}; \tilde \nu_{F+1},\tilde \mu_{F+2};\tilde \tau_a,\tilde \tau_{\tilde a};\Lambda;-\Lambda )}
 \\
&  {\color{red}\prod_{\ell=0}^{k-1}\Gamma_h(\tau_{\tilde S}\!+\!\mu_{F+1}\!+\!\mu_{F+2} +2\ell\tau_k,\tau_S\!+\!\nu_{F+1}\!+\!\nu_{F+2}+2\ell\tau_k) }\;,
\end{aligned}
\end{equation}
where we have once more highlighted in red the contributions of the $\UU(2k+1)\times\UU(2k+1)$ sector. The magnetic quantities are related to the electric quantities by the following dictionary:
\begin{equation}
\label{dictionarySAk}
\begin{aligned}
\tilde \mu_{b} &= \tau_k -\mu_b +k \frac{F+2}{\tilde N} x, \quad 
\tilde \tau_{\tilde s} =  \tau_k - \frac{N-F-1}{\tilde N} x, \\
\tilde \nu_{b} &= \tau_k -\nu_b - k \frac{F+2}{\tilde N} x,\quad
\tilde \tau_s = \tau_k + \frac{N-F-1}{\tilde N} x.
\end{aligned}
\end{equation}
In this case the mass parameters are constrained by the balancing 
condition
\begin{equation}
\label{bcsymmetricAk}
\sum_{b=1}^F(\mu_b+\nu_b)+\frac{1}{2}(\mu_{F+1}+\mu_{F+2}+\nu_{F+1}+\nu_{F+2})=2 ((F+2)\omega- \tau_k(N+1))\,.
\end{equation}
We can now further identify $\mu_{F+1}+\nu_{F+1}=\mu_{F+2}+\nu_{F+2}$ and the FI. Then, after dualizing the sector $\UU(2k+1)\times\UU(2k+1)$ following again the prescription of Section \ref{sec:Aconf} and extracting the square root from the resulting identity, we find the final identity between partition functions:
 \begin{equation}
\label{symmetricidAk}
\begin{aligned}
&
Z_{\UU(N)}^{(F\square; F \overline \square;1  \overline{S};1 S)}(\vec \mu;\vec \nu; \tau_{\tilde S};\tau_S;\Lambda) \;=\;Z_{\UU((2k+1)F-N+2k)}^{(F \square; F \overline \square; 1 \overline{S} ;1 S)}
\left(\vec {\tilde \mu};\vec {\tilde \nu}; \tilde  \tau_{\tilde s};\tilde  \tau_s;-\Lambda\right)\times
\\
&\quad\times\prod_{\ell=0}^{k-1}\prod_{1\leq b< c \leq F}
\Gamma_h(\tau_{\tilde S} +\mu_b +\mu_c+2\ell\tau_k,\tau_S+\nu_b +\nu_c+2\ell\tau_k)
\\
&\quad \times\prod_{j=0}^{k}
\prod_{1\leq b,c \leq F}
\Gamma_h(\mu_b+ \nu_c+2j\tau_k)\;\Gamma_h\left(\pm \frac{\eta}{2} +(F+2)\omega \!-\!\tau_k (N+1)\!-\!\frac{1}{2}\sum_{b=1}^{F} (\mu_b+\nu_b)\right)
\\
&\quad \times\;\prod_{\ell=0}^{k-1}\Gamma_h\left(2\left((F+2)\omega\!-\!\tau_k (N+\ell+2)\right)-\sum_{b=1}^{F} (\mu_b+\nu_b)\right)
\,.
\end{aligned}
\end{equation}
The final duality can be summarized as follows
\begin{equation}
\label{MAkSpure3d}
\begin{array}{|cc|}
\hline
\UU(N) & \UU((2k+1)F-N+2k) \\
(S,\tilde S) \oplus F(Q,\tilde Q)& (s,\tilde s) \oplus  F (q,\tilde q) \\
W = \eqref{eq:WeleAkSSt} & W = \eqref{eq:WmacAkSSt}+ Flip[\mathcal{M}^{\pm},\mathcal{M}^{(0)}_{\ell} ] \\ 
\hline
\end{array}
\end{equation}
where $\mathcal{M}^{(0)}_\ell= \mathcal{M}^{(0)} \, Tr(s\tilde s)^{\ell}$  with $\ell=0,\dots,k-1$.
The flippers are the duals of the relative monopole operators acting as singlets in the magnetic phase.

We can obtain the same result using a different prescription, spelled out in \cite{Amariti:2025gca} for the limiting case $k=1$. We can consider a theory with $N_c=N+k$ and $N_f=F$ and then displace a $\UU(N)$ at the origin and a $\UU(k)$ at $1/(2r)$. This theory is not symmetric with respect to the vertical line in Figure \ref{fig:eles1}, and we will not need to extract the square root, but, after a shift in the FI, will lead to the same identity \eqref{symmetricidAk}. In the dual, this displacement, it leads to a theory with $\UU(\tilde N)\times \UU(k)$, where $\tilde N=(2k+1)F-N+2k$ once more. After identifying the $\UU(k)$ factors on the electric and magnetic side, we find the following identity between partition functions:
\begin{equation}
\label{quasiquadratoperfettoSAkU1}
\begin{aligned}
&
Z_{\UU(N)}^{(F\square; F \overline \square; 1 \overline{S};1 S)}(\vec \mu;\vec \nu; \tau_{\tilde S};\tau_S;\Lambda)
=Z_{\UU((2k+1)F-N+2k)}^{(F \square; F \overline \square; 1 \overline{S} ;1 S)}
\left(\vec {\tilde \mu};\vec {\tilde \nu}; \tilde  \tau_{\tilde s};\tilde  \tau_s;-\Lambda
\right)\times
\\
&\times\!\!\!\!\prod_{1\leq b,c \leq F}\!\! \Big( \prod_{j=0}^{k}
\Gamma_h(\mu_b+ \nu_c+2j\tau_k) \prod_{\ell=0}^{k-1}
\Gamma_h(\tau_{\tilde S} +\mu_b +\mu_c+2\ell\tau_k,\tau_S +\nu_b +\nu_c+2\ell\tau_k) \Big)\,.
\end{aligned}
\end{equation}
The magnetic and electric quantities are related by the following dictionary:
\begin{equation}
\label{dictionarySAkU1}
\begin{aligned}
\tilde{\mu}_{b} &= \tau_k - \mu_b + \frac{k(F+2)}{\tilde N + k}\, x,&\qquad
\tilde{\nu}_{b} &= \tau_k - \nu_b - \frac{k(F+2)}{\tilde N + k}\, x,\\[6pt]
\tilde{\tau}_s &= \tau_k + \frac{N-F+k}{\tilde N + k}\, x,&\qquad
\tilde{\tau}_{\tilde s} &= \tau_k - \frac{N-F+k}{\tilde N + k}\, x.
\end{aligned}
\end{equation}
The balancing condition, corresponding to a linear monopole constraint \footnote{
Linear monopole superpotential different from the KK monopole play a prominent role in the study of 
3d dualities starting from the analysis of \cite{Benini:2017dud}.}
in the superpotential, reads:
\begin{equation}
\label{constraintSAk}
\sum_{b=1}^F(\mu_b+\nu_b)=2 ((F+1)\omega- \tau_k(N+1))\,.
\end{equation}
We want to obtain the unconstrained 3d theory, to do so we consider again $F\to F+2$ flavors. After this shift we are indeed left with an electric theory $\UU(N)$ dual to a magnetic $\UU(\tilde N)\times\UU(2k+1)\times\UU(2k+1)$. With this displacement we are able to eliminate the constraint \eqref{constraintSAk}. We thus obtain the following identity between partition functions:
\begin{equation}
\label{quasiquadratoperfettoSAkU12F}
\begin{aligned}
&
Z_{\UU(N)}^{(F\square; F \overline \square; 1 \overline{S};1 S)}(\vec \mu;\vec \nu; \tau_{\tilde S};\tau_S;\Lambda)
=Z_{\UU((2k+1)F-N+2k)}^{(F \square; F \overline \square; 1 \overline{S} ;1 S)}
\left(\vec {\tilde \mu};\vec {\tilde \nu}; \tilde  \tau_{\tilde s};\tilde  \tau_s;-\Lambda
\right)\times
\\
&\times \prod_{j=0}^{k}\prod_{1\leq b,c \leq F}
\Gamma_h(\mu_b+ \nu_c+2j\tau_k) \prod_{j=0}^{k}\prod_{b=F+1}^{F+2}\!\! \Gamma_h(\mu_b\!+\! \nu_b+2j\tau_k)\\
&\times\prod_{\ell=0}^{k-1}\prod_{1\leq b \leq c \leq F}
\Gamma_h(\tau_{\tilde S} +\mu_b +\mu_c+2\ell\tau_k,\tau_S +\nu_b +\nu_c+2\ell\tau_k)
 \\
& \times \;
 \prod_{\ell=0}^{k-1}\Gamma_h(\tau_{\tilde A}\!+\!\mu_{F+1}\!+\!\mu_{F+2} +2\ell\tau_k,\tau_S\!+\!\nu_{F+1}\!+\!\nu_{F+2}+2\ell\tau_k) \!\! \\
& \times \;
Z_{\UU(2k+1)^2} (\tilde \mu_{F+1},\tilde \nu_{F+2}; \tilde \nu_{F+1},\tilde \mu_{F+2};\tilde \tau_a,\tilde \tau_{\tilde a};\Lambda_{eff};-\Lambda_{eff} )\; ,
\end{aligned}
\end{equation}
with $\Lambda_{eff} =  \Lambda-\frac{1}{2}(\mu_{F+1}+\mu_{F+2}+\nu_{F+1}+\nu_{F+2})+2\omega$
and 
where the magnetic and electric fugacities are related by:
\begin{equation}
\label{dictionarySAkU11}
\begin{aligned}
\tilde{\mu}_{b} &=\tau_k - \mu_b+ \frac{F+4}{\tilde N + 5k + 2}\, x, &\qquad\tilde{\nu}_{b} &=\tau_k - \nu_b- \frac{F+4}{\tilde N + 5k + 2}\, x,\\[4pt]
\tilde{\tau}_s &= \tau_k+ \frac{N - F + k}{\tilde N + 5k + 2}\, x,&\qquad\tilde{\tau}_{\tilde s} &=\tau_k- \frac{N - F + k}{\tilde N + 5k + 2}\, x .
\end{aligned}
\end{equation}
Here, compared to the dictionary \eqref{dictionarySAkU1} obtained using the first method, we have an additional shift in the $x$ fugacity \footnote{As already noted in \cite{Amariti:2025gca}
such a shift is unphysical because we can always  absorb it by shifting the gauge symmetry without affecting the duality.}.
The balancing condition for the theory at hand reads:
\begin{equation}
\sum_{b=1}^F(\mu_b+\nu_b)+(\mu_{F+1}+\mu_{F+2}+\nu_{F+1}+\nu_{F+2})=2 ((F+3)\omega- \tau_k(N+1))\,.
\end{equation}
If we compare it with the balancing condition \eqref{bcsymmetricAk} obtained using the first method of the symmetric displacement, we notice that they differ by $-\frac{1}{2}(\mu_{F+1}+\mu_{F+2}+\nu_{F+1}+\nu_{F+2})+2\omega$. This is precisely the difference in the FI we have obtained. Thus, after dualizing the $\UU(2k+1)\times\UU(2k+1)$ sector using again the results of Section \ref{sec:Aconf}, we arrive to the same result \eqref{symmetricidAk}.

\subsection{Balcony $A_{k}$ }
\label{BA3d}

We conclude this section with a duality  first introduced in {\bf Section 2.8} of \cite{Intriligator:1995ax} and   denoted  in \cite{Brodie:1996xm} as the $A_k$ balcony duality.
The field content consists of a two-index antisymmetric tensor $A$ and a two-index conjugated symmetric  tensor  $\tilde S$ in addition to $N_f$  fundamentals $Q$  and $N_f-8$ antifundamentals  $\tilde Q$. On the electric side, the superpotential is
\begin{equation}
\label{eq:WeleAkASt}
W = (A \tilde S)^{2(k+1)}\; .
\end{equation}
This model is dual to an $\SU(\tilde N_c = (3k+4) (N_f-4) -N_c)$ gauge theory with 
of a two-index antisymmetric tensor $a$ and a two-index conjugated symmetric  tensor  $\tilde s$ in addition to $N_f$  fundamentals $q$  and $N_f-8$ antifundamentals  $\tilde q$. 

There are bifundamental mesons  $M_j = Q (A \tilde S)^j \tilde Q$ with $j=0,\dots,2k+1$. 
The other mesons 
$P_\ell = Q (\tilde A S) ^\ell \tilde S Q$ and $\tilde P_\ell = \tilde Q A (A \tilde S)^\ell   \tilde Q$ with $\ell=0,\dots,2k$ are in the two-index symmetric or antisymmetric representation of the flavor symmetry group depending on the symmetry of the parity of $\ell$.
If $\ell$ is even, the mesons $P_\ell$ are symmetric and the mesons $\tilde P_\ell$ are antisymmetric.  On the other hand, if 
$\ell$ is odd, the mesons $P_\ell$ are antisymmetric and the mesons $\tilde P_\ell$ are symmetric. 
The dual superpotential is
\begin{equation}
\label{eq:WmacAkASt}
W = (a \tilde s )^{k+1}
+
\sum_{j=0}^{2k+1} M_j q (\tilde a a)^{2k-j+1} \tilde q
+
\sum_{\ell=0}^{2k} \left( P_{\ell}  q ( \tilde s a)^{2k-\ell} \tilde s  q+ \tilde P_{\ell}  \tilde q a (\tilde  s a)^{2k-\ell}   \tilde q\right)\; .
 \end{equation}

The reduction to 3d of the $U(N_c)$ version of this duality was studied in \cite{Amariti:2025gca} for the case $k=1$, while here we consider  the $k>1$ case.

In order to study the reduction, we summarize 
the global charges of the electric side 
\begin{equation}
\begin{array}{c|c|ccccc}
               &\SU(N_c)&\SU(N_f)   & \SU(N_f-8) & \UU(1)_B & \UU(1)_X & \UU(1)_R\\
               \hline
Q            & \square &  \square &1 &   \frac{1}{N_c}&-(2k+1)+\frac{2(4k+3)}{N_f}&1- \frac{N_c+2(4k+3)}{2(k+1)N_f} \\
\tilde Q    & \overline \square & 1  &  \square &\! \! \! \! \! -\frac{1}{N_c} &2k+1+\frac{2(4k+3)}{N_f-8}  & 1- \frac{N_c-2(4k+3)}{2(k+1)(N_f-8)} \\
\tilde S    & \begin{array}{c} \overline{\square\! \square}  \end{array}  &1   & 1&\! \! \! \! \!   -\frac{2}{N_c}&\! \! \! \! \! -1  &\frac{1}{2(k+1)} \\
A            & \begin{array}{c} \square \vspace{-2.9mm} \\  \square  \end{array} &1    & 1&\frac{2}{N_c} & 1 &\frac{1}{2(k+1)}  \\
\end{array}
\end{equation}
and of the magnetic side
\begin{equation}
\begin{array}{c|c|cccccc}
&\SU(\tilde N_c)&\SU(N_f)   & \SU(N_f-8) & \UU(1)_B & \UU(1)_X & \UU(1)_R\\
\hline
q            & \square &  \overline \square &1 &   \frac{1}{\tilde N_c}& 2k+1\!-\!\frac{2(4k+3)}{N_f} &1- \frac{\tilde N_c+2(4k+3)}{2(k+1)N_f} \\
\tilde q    & \overline \square & 1  &  \overline\square&\! \! \! \! \! -\frac{1}{\tilde N_c} &\! \!  \!-2k-1\!-\!\frac{2(4k+3)}{N_f-8}& 1- \frac{\tilde N_c-2(4k+3)}{2(k+1)(N_f-8)} \\
\tilde s    & \overline{\square \! \square}& 1& 1&\! \! \! \! \!  -\frac{2}{\tilde N_c}&1&\frac{1}{2(k+1)} \\
a            & \begin{array}{c} \square \vspace{-2.9mm} \\  \square  \end{array}   &1 & 1&\frac{2}{\tilde N_c} &\! \! \! \! \!  -1&\frac{1}{2(k+1)}  \\
\hline 
P_{2p}             & 1  & \square \! \square   & 1& 0  &\! \! \! \! \! -\frac{(4k+3)(N_f-4)}{N_f} &2- \frac{N_c+2(4k+3)}{(k+1)N_f}+\!\frac{4p+1}{2(k+1)}\\
P_{2q+1}             & 1  & \begin{array}{c} \square \vspace{-2.9mm} \\  \square  \end{array}    & 1& 0  &\! \! \! \! \! -\frac{(4k+3)(N_f-4)}{N_f} &2- \frac{N_c+2(4k+3)}{(k+1)N_f}+\!\frac{4q+3}{2(k+1)}\\
\tilde P_{2p}   & 1  & 1  & \begin{array}{c} \square \vspace{-2.9mm} \\  \square  \end{array}   &0   &\frac{(4k+3)(N_f-4)}{N_f-8}  &2- \frac{N_c-2(4k+3)}{(k+1)(N_f-8)}+\!\frac{4p+1}{2(k+1)} \\
\tilde P_{2q+1}    & 1  & 1  &\square \! \square   &0   &\frac{(4k+3)(N_f-4)}{N_f-8}  &2- \frac{N_c-2(4k+3)}{(k+1)(N_f-8)}+\!\frac{4q+3}{2(k+1)} \\
M_j            & 1  & \square  &\square  & 0 & 
\frac{4 (4k+3) (N_f-4)}{N_f(N_f-8)}
&2\!-\! \frac{(N_f+4)(N_c+2(4k+3))}{(k+1)N_f(N_f-8)}\!+\!\frac{j}{k+1}
\\
\end{array}
\end{equation}
where we have introduced $p=0,\dots,k$ and $q=0,\dots,k-1$.

The integral formulas matching the superconformal index of the electric and of the dual phase for the $\UU(N)$ duality are obtained by integrating over the fugacities of 
 the $\UU(1)_B$ symmetry from the formulas given in {\bf Section  9.3.3 of \cite{Spiridonov:2009za}}. We refer the reader to \cite{Amariti:2025gca} for the detailed results for the case $k=0$. The case of higher $k$  is obtained by modifying the contributions of the $\UU(1)_X$ and $\UU(1)_R$-charges as read from the tables of charges above.
 Observe that in this case, when increasing $k$ we need to consider the different parity for the mesons $P_\ell$ and $\tilde P_\ell$

When reducing this duality to 3d we need to distinguish two slightly different dualities, connected by an RG flow deformation.
The first model under investigation is actually a deformed version of the balcony model, corresponding to adding a cubic superpotential deformation for the conjugated symmetric and eight fundamentals, giving rise to an extra $SO(8)$ global symmetry. The reason is that this is the model that we read from the brane picture, where we can infer the gauge symmetry breaking pattern on the dual sides from the HW transition.
Then, inspired by such a pattern we correctly guess the picture for the underformed case, as we explicitly show in the second part of the analysis.

\subsubsection*{The deformed balcony}

We start by considering the brane picture discussed in \cite{Brunner:1998jr} and reviewed in Section \ref{branepicture}.
In this case, we have $N_f-8=2F+2$ D6$_{\pm \theta}$ and $N_c=2N$ D4 branes and we  
 displace $F$ D6 at $x_3=0$ and, symmetrically to the vertical axes, $F$ at $x_3=1/2 r$; we then displace the other two symmetrically at $x_3=1/4 r$ and $x_3=3/4 r$. 
We also displace $N$ D3 at the origin and $N$ D3 at $x_3=1/2 r$.
Furthermore we have eight D5 branes at the origin and D5 branes at $1/(2r)$ in the T-dual brane picture.
In this case, the dual group is broken into $\UU((4k+3)F-N+2)^2\times \UU(4k+3)^2$. There are $F+4$ fundamentals and $F$ antifundamentals in each $\UU(N)$ gauge sector on the electric side and in each  $\UU((4k+3)F-N+2)$ on the magnetic side.
The presence of the D5 branes gives a cubic interaction between the conjugated symmetric and four fundamentals in the gauge sectors at the origin and at $1/(2r)$.
The two $\UU(4k+3)$ gauge groups in the dual side are connected by a conjugated  pair of  bifundamentals and there is also a pair of conjugated fundamentals for each gauge group. These interact through the  superpotential \eqref{spotappD}.
When we study this configuration at the level of the reduction of the identity between the  superconformal index of the electric and of the magnetic phase to the partition function, we observe that the divergent phase cancels, consistently with the expectations from the brane picture. 
If we further identify the masses at the origin and at $1/(2r)$, we obtain the following  identity between partition function 

\begin{eqnarray}
\label{quasiquadratoperfettobalconyAkbroken}
&&
\left(
Z_{\UU(N)}^{((F+4)\square; F \overline \square; 1 \overline{S};1 A)}(\vec \mu,\vec \zeta;\vec \nu; \tau_{\tilde S};\tau_A;\Lambda) \right)^2
=\left(Z_{\UU((4k+3)F-N+2)}^{((F+4) \square; F \overline \square; 1 \overline{S} ;1 A)}
\left(\vec {\tilde \mu},\vec {\tilde \zeta};\vec {\tilde \nu}; \tilde  \tau_{\tilde s};\tilde  \tau_a;-\Lambda
\right)\right)^2 \nonumber
\\
&&\prod_{\ell=0}^{2k}\prod_{1\leq b<c \leq F} \!\!
\Gamma_h^2(\tau_{\tilde S} +\mu_b +\mu_c+2\ell\tau_k,\tau_A +\nu_b +\nu_c+2\ell\tau_k) 
\prod_{j=0}^{2k+1} \prod_{1\leq b,c \leq F}\!\!
\Gamma_h^2(\mu_b+ \nu_c+2j\tau_k)  \nonumber
 \\
 &&\prod_{p=0}^{k}\prod_{b=1}^F
\Gamma_h^2(\tau_{\tilde S} +2\mu_b +4p\tau_k) \; \prod_{q=0}^{k-1}\prod_{b=1}^F\Gamma_h^2(\tau_A +2\nu_b+2(2q+1)\tau_k) \prod_{b=1}^{F}\prod_{d=1}^4 \Gamma_h^2(\nu_b+ \zeta_d)  \nonumber
\\
&& {\color{red} Z_{\UU(4k+3)^2} (\tilde \mu_{F+1},\tilde \nu_{F+2}; \tilde \nu_{F+1},\tilde \mu_{F+2};\tilde \tau_a,\tilde \tau_{\tilde s};\Lambda;-\Lambda)}
{\color{red} \prod_{j=0}^{2k+1} \prod_{b,c=F+1}^{F+2}\!\! \Gamma_h(\mu_b\!+\! \nu_c+2j\tau_k)} 
 \nonumber
\\
&&  \textcolor{red}{ \prod_{\ell=0}^{2k}\Gamma_h(\tau_{\tilde S}\!+\!\mu_{F+1}\!+\!\mu_{F+2} +2\ell\tau_k,\tau_A\!+\!\nu_{F+1}\!+\!\nu_{F+2}+2\ell\tau_k) } \; ,
\end{eqnarray}
where  we have highlighted in red the contributions of the $\UU(4k+3)^2$ sector and we have introduced $\tau_k=\frac{\omega}{2(k+1)}$. 
The fugacity $\zeta$ are associated with $\SO(8) \rightarrow SO(4) \times \SO(4)$ symmetry breaking pattern, where we properly identify the fugacities at the origin and at $1/(2r)$.  
The explicit  parametrization is  $\zeta_{1} = \xi_1+\frac{3}{4} \omega$,
$\zeta_{2} = -\xi_1-\frac{3}{4} \omega$,
$\zeta_{3} = \xi_2+\frac{3}{4} \omega$ and $\zeta_{4} = -\xi_2+\frac{3}{4} \omega$.
The identity \eqref{quasiquadratoperfettobalconyAkbroken} holds provided we impose the following  balancing condition on the mass parameters
\begin{equation}
\label{bcbalconyAkbroken}
\sum_{b=1}^F(\mu_b+\nu_b)+\frac{1}{2}(\mu_{F+1}+\mu_{F+2}+\nu_{F+1}+\nu_{F+2})=2 ((F+1)\omega- \tau_k(N-1))\,.
\end{equation}
In this case the following dictionary relates the magnetic quantities to the electric quantities:
\begin{equation}
\label{dictionarybalconyAkbroken}
\begin{aligned}
\tilde \mu_{b} &= \tau_k -\mu_b +\frac{(2k+1)(F+1)+1}{\tilde N} x \,, \\
\tilde \nu_{b} &= \tau_k -\nu_b - \frac{(2k+1)(F+1)+1}{\tilde N} x\,, \\
\tilde \zeta_{d} &= \tau_k -\zeta_d +\left( \frac{N-F-1}{2 \tilde N} -\frac{1}{2}\right)x\,, \\
\tilde \tau_a &= \tau_k + \frac{N-F-1}{\tilde N} x\, , \\
\tilde \tau_{\tilde s} &=  \tau_k - \frac{N-F-1}{\tilde N} x\,.
\end{aligned}
\end{equation}

We can now further identify $\mu_{F+1}+\nu_{F+1}=\mu_{F+2}+\nu_{F+2}$ and the FI. Then, after dualizing the sector $\UU(4k+3)^2$ following  the prescription of Section \ref{sec:Aconf} and extracting the square root from the resulting identity, we find the final identity between partition functions:
\begin{equation}
\label{balconyidAkbroken}
\begin{aligned}
&
Z_{\UU(N)}^{(F\square; (F+4) \overline \square;1  \overline{S};1 A)}(\vec \mu;\vec \nu; \tau_{\tilde S};\tau_A;\Lambda) \;=\;Z_{\UU((4k+3)F-N+2)}^{(F \square; (F+4) \overline \square; 1 \overline{S} ;1 A)}
\left(\vec {\tilde \mu};\vec {\tilde \nu}; \tilde  \tau_{\tilde s};\tilde  \tau_a;-\Lambda\right)\times
\\
&\quad\times\prod_{\ell=0}^{2k}\prod_{1\leq b< c \leq F}
\Gamma_h(\tau_{\tilde S} +\mu_b +\mu_c+2\ell\tau_k,\tau_A+\nu_b +\nu_c+2\ell\tau_k)\prod_{b=1}^{F}\prod_{d=1}^4 \Gamma_h^2(\nu_b+ \zeta_d)
\\
 &\quad\times\prod_{p=0}^{k}\prod_{b=1}^F
\Gamma_h^2(\tau_{\tilde S} +2\mu_b +4p\tau_k) \; \prod_{q=0}^{k-1}\prod_{b=1}^F\Gamma_h^2(\tau_A +2\nu_b+2(2q+1)\tau_k) \\
&\quad \times\prod_{j=0}^{2k+1}
\prod_{1\leq b,c \leq F}
\Gamma_h(\mu_b+ \nu_c+2j\tau_k)\;\Gamma_h\Big(\pm \frac{\eta}{2} \!+\!(F\!+\!1)\omega-\tau_k (N\!-\!1)\!-\!\frac{1}{2}\sum_{b=1}^{F} (\mu_b\!+\!\nu_b)\Big)
\\
&\quad \times\;\prod_{\ell=0}^{2k}\Gamma_h\Big(2\left((F+1)\omega-\tau_k (N+\ell)\right)-\sum_{b=1}^{F} (\mu_b+\nu_b)\Big)
\,.
\end{aligned}
\end{equation}

The final duality can be summarized as follows
\begin{equation}
\label{BAkpure3d}
\begin{array}{|cc|}
\hline
\UU(N) & \UU((4k+3)F-N+2) \\
(A,\tilde S) \oplus F(Q,\tilde Q)+4T& (a,\tilde s) \oplus  F (q,\tilde q) \oplus 4 t \\
W = \eqref{eq:WeleAkASt} +\tilde S T^2& W = \eqref{eq:WmacAkAAt} + \tilde s t^2+ Flip[\mathcal{M}^\pm,\mathcal{M}^{(0)}_\ell ] \\ 
\hline
\end{array}
\end{equation}
where $\mathcal{M}^{(0)}_\ell = \mathcal{M}^{(0)} \, Tr(a \tilde s)^{\ell}$  with $\ell=0,\dots,k-1$.
The flippers are the duals of the relative monopole operators acting as singlets in the magnetic phase.

\subsubsection*{The undeformed balcony}

We can also study the undeformed case, that cannot be obtained from the brane picture but that 
corresponds to the original balcony $A_{k}$ duality proposed by 
\cite{Intriligator:1995ax}.
This is the duality reviewed above. In the case at hand, we take $N_c=2N$ and we consider $2F+2$ fundamental flavors and $2F-6$ antifundamental flavors. We consider a background with
$F$ fundamentals and $F-4$ antifundamental flavors both at the origin and at $1/(2r)$ and one fundamental and antifundamental flavor both at $1/(4r)$ and at $3/(4r)$. The dual gauge group then gets broken from $\SU((4k + 3) (2 F - 2) -2N)$ to $\SU((4k+3)(F-2)-N)^2\times\UU(4k+3)^2$.
If we further identify the masses at the origin and at $1/(2r)$, we arrive to the following identity between partition functions:
\begin{eqnarray}
\label{quasiquadratoperfettobalconyAk}
&&
\left(
Z_{\UU(N)}^{(F\square; (F-4) \overline \square; 1 \overline{S};1 A)}(\vec \mu;\vec \nu; \tau_{\tilde S};\tau_A;\Lambda) \right)^2
=\left(Z_{\UU((2k+1)F-N+2)}^{(F \square; (F-4) \overline \square; 1 \overline{S} ;1 A)}
\left(\vec {\tilde \mu};\vec {\tilde \nu}; \tilde  \tau_{\tilde s};\tilde  \tau_a;-\Lambda
\right)\right)^2\times
\nonumber \\
&&\times \prod_{j=0}^{2k+1}\prod_{b=1}^F\prod_{c=1}^{F-4}
\Gamma_h^2(\mu_b+ \nu_c+j(\tau_A +\tau_{\tilde S}))
\times
\prod_{q=0}^{k-1}\prod_{b=1}^{F-4}\Gamma_h^2(\tau_A +2\nu_b+2(2q+1)\tau_k) 
\nonumber  \\
 &&\times\prod_{\ell=0}^{2k} \prod_{1\leq b < c \leq F}
\Gamma_h^2(\tau_{\tilde S} +\mu_b +\mu_c+2\ell\tau_k)
\times\prod_{\ell=0}^{2k}
\prod_{1\leq b<c\leq F-4}\!\!\!
\Gamma_h^2(\tau_A +\nu_b +\nu_c+2\ell\tau_k)
 \nonumber \\
 &&\times\prod_{p=0}^{k}\prod_{b=1}^{F}
\Gamma_h^2(\tau_{\tilde S} +2\mu_b +4p\tau_k) \times 
 \color{red}{ \prod_{j=0}^{k}\prod_{b=F+1}^{F+2}\! \Gamma_h(\mu_b\!+\! \nu_{b-4}+2j\tau_k)}
\nonumber \\
&& \times \;
\color{red}{ \prod_{\ell=0}^{k-1}\Gamma_h(\tau_{\tilde S}\!+\!\mu_{F+1}\!+\!\mu_{F+2} +2\ell\tau_k,\tau_A\!+\!\nu_{F-3}\!+\!\nu_{F-2}+2\ell\tau_k) }\!\!  \nonumber \\
&& \times
{\color{red} Z_{\UU(4k+3)^2} (\tilde \mu_{F+1},\tilde \nu_{F+2}; \tilde \nu_{F+1},\tilde \mu_{F+2};\tilde \tau_a,\tilde \tau_{\tilde s};\Lambda;-\Lambda)}\; , 
\end{eqnarray}
where highlighted in red are the contributions coming from the $\UU(4k+3)^2$ sector and the electric and magnetic fugacities are related by the following dictionary:
\begin{equation}
\label{dictionarybalconyAk}
\tilde \mu_{b} = \tau_k -\mu_b, \quad 
\tilde \nu_{b} = \tau_k -\nu_b, \quad 
\tilde \tau_a = \tau_k - x, \quad 
\tilde \tau_{\tilde s} =  \tau_k+ x\, .
\end{equation}
Notice that the dictionary is different from that of the undeformed case since here the flavor group is not Higgsed. In this case, the balancing condition reads:
\begin{equation}
\label{bcbalconyAk}
\sum_{b=1}^F\mu_b+\sum_{b=1}^{F-4}\nu_b+\frac{1}{2}(\mu_{F+1}+\mu_{F+2}+\nu_{F-3}+\nu_{F-2})=2 ((F-1)\omega- \tau_k N+x)\,.
\end{equation}
With the same procedure of the previous section, we dualize once more the $\UU(4k+3)^2$ sector, which allows us then to obtain a perfect squared identity, which, once taken the square root, is translated into the final identity between partition functions:
\begin{equation}
\label{balconyidAk}
\begin{aligned}
&
Z_{\UU(N)}^{(F\square; (F-4) \overline \square;1  \overline{S};1 A)}(\vec \mu;\vec \nu; \tau_{\tilde S};\tau_A;\Lambda) \;=\;Z_{\UU((4k+3)(F-2)-N+2)}^{(F \square; (F-4) \overline \square; 1 \overline{S} ;1 A)}
\left(\vec {\tilde \mu};\vec {\tilde \nu}; \tilde  \tau_{\tilde s};\tilde  \tau_a;-\Lambda\right)\times
\\
&\quad\times\prod_{\ell=0}^{2k}\Big(\prod_{1\leq b < c \leq F}\!\!\!
\Gamma_h(\tau_{\tilde S} +\mu_b +\mu_c+2\ell\tau_k)
\!\!\!
\prod_{1\leq b < c \leq F-4}
\!\!\!
\Gamma_h(\tau_A +\nu_b +\nu_c+2\ell\tau_k)\Big)\\
 &\quad\times\prod_{p=0}^{k}\prod_{b=1}^F
\Gamma_h(\tau_{\tilde S} +2\mu_b +4p\tau_k) \; \prod_{q=0}^{k-1}\prod_{b=1}^{F-4}\Gamma_h(\tau_A +2\nu_b+2(2q+1)\tau_k) \\
&\quad \times\prod_{j=0}^{2k+1}
\prod_{b=1}^F\prod_{c=1}^{F-4}
\Gamma_h(\mu_b+ \nu_c+2j\tau_k)\;\Gamma_h\Big(\pm \frac{\eta}{2} \!+\!(F\!-\!1)\omega\!-\!\tau_k N\!+\!x\! -\!\frac{1}{2}\sum_{b=1}^{F} (\mu_b\!+\!\nu_b)\Big)
\\
&\quad \times\;\prod_{\ell=0}^{k-1}\Gamma_h\Big(2\left((F-1)\omega-\tau_k (N+\ell+1)+x \right)-\sum_{b=1}^{F} (\mu_b+\nu_b)\Big)
\,.
\end{aligned}
\end{equation}

The final duality can be summarized as follows
\begin{equation}
\label{BAkpure3d2}
\begin{array}{|cc|}
\hline
\UU(N) & \UU((4k+3)(F-2)-N+2) \\
(A,\tilde S) \oplus F Q \oplus (F-4) \tilde Q& (a,\tilde s) \oplus  F q \oplus (F-4) \tilde q  \\
W = \eqref{eq:WeleAkASt} & W = \eqref{eq:WmacAkAAt} + Flip[\mathcal{M}^\pm,\mathcal{M}^{(0)}_\ell ] \\ 
\hline
\end{array}
\end{equation}
where $\mathcal{M}^{(0)}_\ell= \mathcal{M}^{(0)} \, Tr(a \tilde s)^{\ell}$  with $\ell=0,\dots,k-1$.
The flippers are the duals of the relative monopole operators acting as singlets in the magnetic phase.

%
%
%
%
%
%
%
%
%
%
\section{4d/3d reduction for the $D_{k+2}$ cases}
\label{secDd}
%
%
%
%
%
%
%
%
%
%

In this section we  consider the reduction to 3d of the 4d $D_{k+2}$ dualities proposed in  \cite{Brodie:1996xm}.
Differently from the $A_k$ cases, here there is an additional field in the spectrum, corresponding to an adjoint with superpotential of order $k+1$ . This adjoint interacts moreover with the two two-index tensors  through a cubic superpotential.
Also these cases reduce to the one studied in \cite{Amariti:2025gca}, when the adjoint is massive.

By adopting the same the scheme of the $A_k$ case, 
here we first briefly summarize the salient features of  the 4d dualities under investigation, referring the reader to the original references for further details.
Then, we study the reduction of these dualities by first gauging the baryonic symmetry, such that these models involve a $\UU (N_c)$ gauge theory with fundamental and antifundamental flavor. 

From the brane perspective the $D_{k+2}$ models correspond  to choosing $k$  NS or NS' fivebranes at $x_6=0$ in addition to 
 $k$ NS$_\theta$ and $k$ NS$_{-\theta}$ fivebranes.
 We have also $N_c$ D4 branes and $2N_f$ D6$_{\pm \theta}$  branes.  As we did in the $A_k$ cases, we consider the mezzanine and a deformed version of the balcony models.
 The mezzanine models are obtained by adding either an O6$^-$ or an O6$^+$ plane on the NS fivebrane. The first case corresponds to the mezzanine $D_{k+2}^{(A)}$ duality while the second one to the  mezzanine $D_{k+2}^{(S)}$ duality. Then, when we consider the NS' brane, we have a setup with an  O6$^-$ and an O6$^+$ plane, that again implies the presence of a stack of eight semi-infinte D6 branes. This case corresponds to a deformation of the balcony $D_{k+2}$ duality. In the following, similarly to what we did in the balcony $A_k$ case,  we  discuss the reduction of the duality in the deformed case and we  comment on the undeformed case as well.

\subsection{Mezzanine $D_{k+2}^{(A)}$}
\label{MDA3d}

This duality was first discussed in {\bf Section 6.4 of \cite{Brodie:1996xm}}. In addition to the adjoint $Y$, the field content consists of a pair of conjugated antisymmetric two-index tensors $A$ and $\tilde A$ and to $N_f$ pairs of fundamentals and antifundamentals $Q$ and $\tilde Q$. In the  following we refer to this model as mezzanine $D_{k+2}^{(A)}$.
On the electric side the superpotential is
\begin{equation}
\label{eq:WeleDkAAt}
W = Y^{k+1} + A Y \tilde A.
\end{equation}
This model is dual to an $\SU(\tilde N_c =3kN_f -N_c-4)$ gauge theory with a dual adjoint $y$, a
pair of conjugated antisymmetric two-index tensors $a$ and $\tilde a$ and $N_f$ pairs of fundamentals and antifundamentals $q$ and $\tilde q$.
There are bifundamental mesons  $M_j = Q  Y^j A \tilde A \tilde Q$ and $N_j = Q Y^j \tilde Q$
with $j=0,\dots,k-1$ and,  in addition, there are also mesons of the type
$P_j = Q Y^j \tilde A Q$ and $\tilde P_j = \tilde Q Y^j  A \tilde Q$
that are
(anti)-symmetric for (even) odd   $j=0,\dots,k-1$. 
Observe that in this case the chiral ring is truncated classically for odd values of $k$, while for even $k$ a quantum constraint was advocated in \cite{Brodie:1996xm}. We  further comment on this point below in the paper, and here we  proceed by assuming that such a truncation is correct.  
In this case the dual superpotential is
\begin{equation}
\label{eq:WmagDkAAt}
W = y^{k+1} + a y \tilde a
+
\sum_{j=0}^{k-1} (N_j  q y^{k-j-1} \tilde q 
+ M_j q y^{k-j-1} \tilde a a \tilde q
+
 P_{j}  q y^{k-j-1} \tilde a  q+ \tilde P_{j}  \tilde q y^{k-j-1}  a \tilde q).
 \end{equation}
Analogously to the $A_k$ mezzanine case, the duality discussed here reduces to the one studied in \cite{Amariti:2025gca} for the case $k=1$. Indeed, in such a case the adjoint is massive and by integrating it out one obtains the quartic superpotential involving the antisymmetric tensors.

Here, we list the global charges of the electric side
\begin{equation}
\begin{array}{c|c|ccccc}
               &\SU(N_c)&\SU(N_f)   & \SU(N_f) & \UU(1)_B & \UU(1)_X & \UU(1)_R\\
               \hline
Q            & \square &  \square &1 &   \frac{1}{N_c}&0 &1- \frac{N_c+2}{(k+1)N_f} \\
\tilde Q    & \overline \square & 1  &  \square &\! \! \! \! \! -\frac{1}{N_c} &0  & 1- \frac{N_c+2}{(k+1)N_f} \\
Y & Adj.&1&1&0&0&\frac{2}{k+1}\\
\tilde A    & \begin{array}{c} \overline \square \vspace{-2.9mm} \\  \square  \end{array}  &1   & 1&\! \! \! \! \!   -\frac{2}{N_c}&\! \! \! \! \! -1  &\frac{k}{k+1} \\
A            & \begin{array}{c} \square \vspace{-2.9mm} \\  \square  \end{array} &1    & 1&\frac{2}{N_c} & 1 &\frac{k}{k+1}  \\
\end{array}
\end{equation}
and of the magnetic side 
\begin{equation}
\begin{array}{c|c|cccccc}
&\SU(\tilde N_c)&\SU(N_f)   & \SU(N_f) & \UU(1)_B & \UU(1)_X & \UU(1)_R\\
\hline
q            & \square &  \overline \square &1 &   \frac{1}{\tilde N_c}& \frac{kN_f-2}{\tilde N_c} &1- \frac{\tilde N_c+2}{(k+1)N_f} \\
\tilde q    & \overline \square & 1  &  \overline\square&\! \! \! \! \! -\frac{1}{\tilde N_c} &\! \! \! \! \!   -\frac{kN_f-2}{\tilde N_c} & 1- \frac{\tilde N_c+2}{(k+1)N_f} \\
y & Adj.&1&1&0&0&\frac{2}{k+1}\\
\tilde a    &\begin{array}{c} \overline \square \vspace{-2.9mm} \\  \square  \end{array}  & 1& 1&\! \! \! \! \!  -\frac{2}{\tilde N_c}&\! \! \! \! \!  - \frac{N_c-kN_f}{\tilde N_c}&\frac{k}{k+1} \\
a            & \begin{array}{c} \square \vspace{-2.9mm} \\  \square  \end{array}    &1 & 1&\frac{2}{\tilde N_c} &  \frac{N_c-kN_f}{\tilde N_c}&\frac{k}{k+1}  \\
\hline 
P_{2q+1}            & 1  & \square \! \square   &1& 0  & \! \! \! \! \!  -1  &2\frac{2q+k+1}{k+1}+2-2\frac{N_c+2}{(k+1)N_f}\\
P_{2p}            & 1  &  \begin{array}{c} \square \vspace{-2.9mm} \\  \square  \end{array}   &1& 0  & \! \! \! \! \!  -1  &2\frac{2p+k}{k+1}+2-2\frac{N_c+2}{(k+1)N_f}\\
\tilde P_{2q+1}            & 1  &1&{\square \! \square}& 0  & 1  &2\frac{2q+k+1}{k+1}+2-2\frac{N_c+2}{(k+1)N_f}\\
\tilde P_{2p}            & 1  & 1 &\begin{array}{c} \square \vspace{-2.9mm} \\  \square  \end{array}   & 0  & 1  &2\frac{2p+k}{k+1}+2-2\frac{N_c+2}{(k+1)N_f}\\
M_j            & 1  & \square  &\square  & 0 &0&\frac{2(k+j)}{k+1}+2-2\frac{N_c+2}{(k+1)N_f}\\
N_j            & 1  & \square  &\square  & 0 &0&\frac{2j}{k+1}+2-2\frac{N_c+2}{(k+1)N_f}
\end{array}
\end{equation}

The integral formulas matching the superconformal index of the electric and of the dual phase for the $\UU(N)$ duality are obtained by integrating over the fugacities of the 
 the $\UU(1)_B$ symmetry from the formulas given in {\bf Section 9.5 of  \cite{Spiridonov:2009za}}. We refer the reader to \cite{Amariti:2025gca} for the detailed results for the case $k=1$. The case of higher $k$ is obtained by modifying the contributions of the $\UU(1)_X$ and $\UU(1)_R$-charges as read from the tables of charges above.
 Observe that in this case, when increasing $k$ we need to consider the different parity for the mesons $P_\ell$ and $\tilde P_\ell$.

In the following, we study the reduction of this duality to 3d, through the double scaling prescription.
We apply such a prescription by considering 
$N_c=2N$ and $N_f=2F+2$, then we displace $F$ flavor fugacities at the origin and at $1/(2r)$ and one both at $1/(4r)$ and $3/(4r)$. In this case the dual gauge group is broken into $\UU(3kF-N-2)\times\UU(3kF-N-2)\times\UU(3k)\times\UU(3k)$. We also define the final dual gauge group as $\tilde N =3kF-N-2 $. By making  use of the double scaling procedure, we obtain the following identity between partition functions:
\begin{eqnarray}
\label{quasiquadratoperfettoADk}
&&
\left(
Z_{\UU(N)}^{(F\square; F \overline \square; 1 \overline{A};1 A;1 \text{Adj})}(\vec \mu;\vec \nu; \tau_{\tilde A};\tau_A;\tau_Y;\Lambda) \right)^2
\!\!=\!\!
\left(Z_{\UU(3kF-N-2)}^{(F \square; F \overline \square; 1 \overline{A} ;1 A;1 \text{Adj})}
\left(\vec {\tilde \mu};\vec {\tilde \nu}; \tilde  \tau_{\tilde a};\tilde  \tau_a;\tilde\tau_y;-\Lambda
\right)\right)^2 \nonumber
\\
&&\times \prod_{\ell=0}^{k-1}\prod_{1\leq b,c \leq F}\!\!\!\!
\Gamma_h^2(\mu_b\!+\! \nu_c\!+\! \ell\tau_Y,\mu_b\!+\! \nu_c\!+\!2(\omega\!+\!(\ell\!-\!1)\tau_k))  \nonumber \\
&&\times\prod_{\ell=0}^{k-1}\prod_{1\leq b<c \leq F}\!\!\!\!
\Gamma_h^2(\tau_{\tilde A} \!+\!\mu_b \!+\!\mu_c\!+\!\ell\tau_Y,\tau_A \!+\!\nu_b \!+\!\nu_c\!+\!\ell\tau_Y)
\nonumber
 \\
 &&\times\prod_{q=0}^{K-1}\prod_{b=1}^F
\Gamma_h^2(\tau_{\tilde A} +\mu_b +\mu_b+(2q+1)\tau_Y,\tau_A +\nu_b +\nu_b+(2q+1)\tau_Y)
 \nonumber
 \\
  &&\times \color{red}{ \prod_{\ell=0}^{k-1}\prod_{b=F+1}^{F+2}\!\! \Gamma_h(\mu_b\!+\! \nu_b+\ell\tau_Y,\mu_b\!+\! \nu_b\!+\!
2k \tau_k \!+\!\ell\tau_Y)} \nonumber\\
&& \times 
{\color{red} Z_{\UU^2(3k)}( \tilde \mu_{F+1},\tilde \nu_{F+2};\tilde \nu_{F+2},\tilde \mu_{F+1};\tau_y,\tau_y;\tilde \tau_a,\tilde \tau_{\tilde a};\Lambda,-\Lambda)} 
 \nonumber \\
&& \times \;
\color{red}{ \prod_{\ell=0}^{k-1}\Gamma_h(\tau_{\tilde A}\!+\!\mu_{F+1}\!+\!\mu_{F+2} +\ell\tau_Y,\tau_A\!+\!\nu_{F+1}\!+\!\nu_{F+2}+\ell\tau_Y) }\!\!  \nonumber \\
&&\times\textcolor{red}{\prod_{q=0}^{K-1}\prod_{b=1}^{F+1}
\Gamma_h^2(\tau_{\tilde A} +\mu_b +\mu_b+(2q+1)\tau_Y,\tau_A +\nu_b +\nu_b+(2q+1)\tau_Y)}\; ,
\end{eqnarray}
where we have introduced once more $\tau_k=\frac{\omega}{k+1}$.
Observe that the mass parameters for the two-index tensors $A$ and $\tilde A$, denoted as $\tau_A$ and $\tau_{\tilde A}$ are constrained by the cubic superpotential term $ A Y \tilde A\subset W$.
In this case we have $\tau_A = k \tau_k +x$ and $\tau_{\tilde A} = k \tau_k -x$, in addition to $\tau_Y = 2 \tau_k$.
Again, we have highlighted in red the contributions coming from the $\UU^2(3k)$ sector.
We further defined $K \equiv \frac{k-1}{2}$, which is an integer because we restrict to odd $k$.
The magnetic and electric fugacities are related by the following dictionary:
\begin{eqnarray}
\label{dictionaryADk}
&&
\tilde \mu_{b} = \tau_k -\mu_b + \frac{k(F+1)-1}{\tilde N} x , \quad 
\tilde \nu_{b} = \tau_k -\nu_b - \frac{k(F+1)-1}{\tilde N} x, \\
&&
\tilde \tau_a = \omega-\tau_k + \frac{N-k (F+1)}{\tilde N} x, \quad 
\tilde \tau_{\tilde a} =  \omega-\tau_k - \frac{N-k (F+1)}{\tilde N} x, \nonumber
\end{eqnarray}
while $\tilde \tau_y =\tau_Y= 2\tau_k$. Additionally, the balancing condition reads:
\begin{equation}
\label{bcantisymmAk}
\sum_{b=1}^F(\mu_b+\nu_b)+\frac{1}{2}(\mu_{F+1}+\mu_{F+2}+\nu_{F+1}+\nu_{F+2})=2 ((F+1)\omega- \tau_k(N+1))\,.
\end{equation}
The sector $\UU(3k)\times \UU(3k)$ has been studied explicitly in Section \ref{sec:Dconf}.
It consists of a quiver gauge theory with two gauge groups connected by a conjugated pair of bidundamentals. Each gauge group has a further pair of conjugated fundamentals and an adjoint. The FI of the two gauge groups are opposite.

We then proceed by further dualizing the sector $\UU(3k)\times \UU(3k)$ using the procedure spelled out in Section \ref{sec:Dconf} with opposite FI. 
The sector confines giving rise to towers of dressed mesons and monopoles. 
The mesons arising from this tower simplify against the ones associated with the hyperbolic Gamma functions appearing in red in  formula (\ref{quasiquadratoperfettoADk}). This simplification occurs at the level of the partition function thanks to the inversion relation for the hyperbolic Gamma functions and it is interpreted as an holomorphic mass term at the level of the superpotential.
The monopoles give rise to the contribution
\begin{equation}
\sum_{\ell=0}^{k-1}
\Gamma_h^2\Big(\pm\frac{\eta}{2}-\frac{1}{2}(\tilde\mu_{F+1}+\tilde\nu_{F+1})-(2\ell-1)\tau_k\Big)\Gamma_h^2\Big(-(\tilde\mu_{F+1}+\tilde\nu_{F+1})-2(\omega+(\ell-2)\tau_k)\Big).
\end{equation}
We can now plug this back into \eqref{quasiquadratoperfettoADk}, and after extracting the square root and substituting in the balancing condition \eqref{bcantisymmAk}, we obtain the final identity
\begin{equation}
\label{antisymmetricidDk}
\begin{aligned}
&
Z_{\UU(N)}^{(F\square; F \overline \square; 1 \overline{A};1 A;1 \text{Adj})}(\vec \mu;\vec \nu; \tau_{\tilde A};\tau_A;\tau_Y;\Lambda)
=Z_{\UU(3kF-N-2)}^{(F \square; F \overline \square; 1 \overline{A} ;1 A;1 \text{Adj})}
\left(\vec {\tilde \mu};\vec {\tilde \nu}; \tilde  \tau_{\tilde a};\tilde  \tau_a;\tilde\tau_y;-\Lambda
\right)\times\\
&\times \prod_{\ell=0}^{k-1}\prod_{1\leq b,c \leq F}
\Gamma_h(\mu_b+ \nu_c+\ell\tau_Y,\mu_b+ \nu_c+\tau_A +\tau_{\tilde A}+\ell\tau_Y) \\
&\times\prod_{\ell=0}^{k-1}\prod_{1\leq b<c \leq F}
\Gamma_h(\tau_{\tilde A} +\mu_b +\mu_c+\ell\tau_Y,\tau_A +\nu_b +\nu_c+\ell\tau_Y)
 \\
 &\times\prod_{q=0}^{K-1}\prod_{b=1}^F
\Gamma_h(\tau_{\tilde A} +\mu_b +\mu_b+(2q+1)\tau_Y,\tau_A +\nu_b +\nu_b+(2q+1)\tau_Y)\\
&\times\prod_{\ell=0}^{k-1}\Gamma_h\Big(\pm \frac{\eta}{2} +(F-1)\omega-\tau_k(N+2\ell+1 )-\frac{1}{2}\sum_{b=1}^{F} (\mu_b+\nu_b)\Big)
\\
& \times\;\prod_{\ell=0}^{k-1}\Gamma_h\Big(2\left(F\omega-\tau_k (N+\ell)\right)-\sum_{b=1}^{F} (\mu_b+\nu_b)\Big)
\,.
\end{aligned}
\end{equation}

The final duality can be summarized as follows
\begin{equation}
\label{MDkApure3d2}
\begin{array}{|cc|}
\hline
\UU(N) & \UU(3kF-N-2) \\
Y \oplus (A,\tilde A) \oplus F (Q,\tilde Q)& y \oplus (a,\tilde a) \oplus  F (q,\tilde q)   \\
W = \eqref{eq:WeleDkAAt} & W = \eqref{eq:WmagDkAAt} + Flip[\mathcal{M}^\pm_\ell,\mathcal{M}^{(0)}_\ell ] \\ 
\hline
\end{array}
\end{equation}
where $\mathcal{M}^{(0)}_\ell = \mathcal{M}^{(0)}\, Tr y^{\ell}$ and $\mathcal{M}^\pm_\ell = \mathcal{M}^\pm \, Tr y^{\ell}$ .
The flippers are the duals of the relative monopole operators acting as singlets in the magnetic phase.

Similarly to the case of the $A_k^{(A)}$ mezzanine model we can study this case through the original ARSW prescription.
Indeed, by choosing $N_c$ D4 branes and $N_f+1$ D6$_{\pm \theta}$ branes in the original setup we can remove the constraint from the KK monopole by assigning a large mass to one flavor. In the dual side, the $\UU(3k (N_f+1)-N_c-4)$ gauge groups is broken to 
$\UU(3k N_f-N_c-2)  \times \UU(3k-2)$. 
The $\UU(3k-2)$ is confining, because it is the limiting case of the duality that we are looking for, and we just proved it through the double scaling procedure. By removing this sector we obtain the expected duality.
Observe that the derivation of the duality  through the ARSW prescription is weaker here than in the $A_k^{(A)}$ mezzanine case, because  we need as an input the confining limit of the duality itself. It would be interesting to have a direct proof of this limiting case through tensor deconfinement.

\subsection{Mezzanine $D_{k+2}^{(S)}$}
\label{MDS3d}

This duality was  discussed in {\bf Section 6.3 of \cite{Brodie:1996xm}}. In addition to the adjoint $Y$, the field content consists of a pair of conjugated symmetric two-index tensors $S$ and $\tilde S$ and of $N_f$ pairs of fundamentals and antifundamentals $Q$ and $\tilde Q$. In the  following, we refer to this model as mezzanine $D_{k+2}^{(S)}$. On the electric side, the superpotential is
\begin{equation}
\label{eq:WeleDkSSt}
W = Y^{k+1} + S Y \tilde S.
\end{equation}
This model is dual to an $\SU(\tilde N_c =3kN_f -N_c+4)$ gauge theory with a dual adjoint $y$, a
pair of conjugated symmetric two-index tensors $s$ and $\tilde s$ and $N_f$ pairs of fundamentals and antifundamentals $q$ and $\tilde q$.
There are bifundamental mesons  $M_j = Q  Y^j S \tilde S \tilde Q$ and $N_j = Q Y^j \tilde Q$
with $j=0,\dots,k-1$ and,  in addition, there are also mesons of the type
$P_j = Q Y^j \tilde S Q$ and $\tilde P_j = \tilde Q Y^j  S \tilde Q$
that are
(anti)-symmetric for (odd) even   $j=0,\dots,k-1$. 
The chiral ring is truncated classically for odd values of $k$, while for even $k$ a quantum constraint was advocated in \cite{Brodie:1996xm}.  
The dual superpotential is
 \begin{equation}
\label{eq:WmagDkSSt}
W = y^{k+1} + s y \tilde s+ \sum_{j=0}^{k-1} (N_j\, q y^{k-j-1} \tilde q
+M_j\, q y^{k-j-1} \tilde s s \tilde q+ P_j\, q y^{k-j-1} \tilde s q+ \tilde P_j\, \tilde q y^{k-j-1} s \tilde q).
\end{equation}
 The duality discussed here reduces to the one considered in \cite{Amariti:2025gca}  for the case $k=1$. 
The global charges of the electric side are
\begin{equation}
\begin{array}{c|c|ccccc}
               &\SU(N_c)&\SU(N_f)   & \SU(N_f) & \UU(1)_B & \UU(1)_X & \UU(1)_R\\
               \hline
Q            & \square &  \square &1 &   \frac{1}{N_c}&0 &1- \frac{N_c-2}{(k+1)N_f} \\
\tilde Q    & \overline \square & 1  &  \square &\! \! \! \! \! -\frac{1}{N_c} &0  & 1- \frac{N_c-2}{(k+1)N_f} \\
Y & Adj.&1&1&0&0&\frac{2}{k+1}\\
\tilde S    & \overline {\square \! \square}&1   & 1&\! \! \! \! \!   -\frac{2}{N_c}&\! \! \! \! \! -1  &\frac{k}{k+1} \\
S           & \square \! \square&1    & 1&\frac{2}{N_c} & 1 &\frac{k}{k+1}  \\
\end{array}
\end{equation}
and the global charges of the magnetic side are
\begin{equation}
\begin{array}{c|c|cccccc}
&\SU(\tilde N_c)&\SU(N_f)   & \SU(N_f) & \UU(1)_B & \UU(1)_X & \UU(1)_R\\
\hline
q            & \square &  \overline \square &1 &   \frac{1}{\tilde N_c}& \frac{kN_f+2}{\tilde N_c} &1- \frac{\tilde N_c-2}{(k+1)N_f} \\
\tilde q    & \overline \square & 1  &  \overline\square&\! \! \! \! \! -\frac{1}{\tilde N_c} &\! \! \! \! \!   -\frac{kN_f+2}{\tilde N_c} & 1- \frac{\tilde N_c-2}{(k+1)N_f} \\
y & Adj.&1&1&0&0&\frac{2}{k+1}\\
\tilde s   & \overline {\square \! \square}  & 1& 1&\! \! \! \! \!  -\frac{2}{\tilde N_c}&\! \! \! \! \!  - \frac{N_c-kN_f}{\tilde N_c}&\frac{k}{k+1} \\
s            & \square \! \square &1 & 1&\frac{2}{\tilde N_c} &  \frac{N_c-kN_f}{\tilde N_c}&\frac{k}{k+1}  \\
\hline 
P_{2q+1}            & 1  & \begin{array}{c} \square \vspace{-2.9mm} \\  \square  \end{array}  &1& 0  & \! \! \! \! \!  -1  &2\frac{2q+k+1}{k+1}+2-2\frac{N_c-2}{(k+1)N_f}\\
P_{2p}            & 1  &  \square \! \square    &1& 0  & \! \! \! \! \!  -1  &2\frac{2p+k}{k+1}+2-2\frac{N_c-2}{(k+1)N_f}\\
\tilde P_{2q+1}            & 1  &  1 &\begin{array}{c} \square \vspace{-2.9mm} \\  \square  \end{array}  & 0  & 1  &2\frac{2q+k+1}{k+1}+2-2\frac{N_c-2}{(k+1)N_f}\\
\tilde P_{2p}            & 1  &  1&{\square \! \square}& 0  & 1  &2\frac{2p+k}{k+1}+2-2\frac{N_c-2}{(k+1)N_f}\\
M_j            & 1  & \square  &\square  & 0 &0&\frac{2(k+j)}{k+1}+2-2\frac{N_c-2}{(k+1)N_f}\\
N_j            & 1  & \square  &\square  & 0 &0&\frac{2j}{k+1}+2-2\frac{N_c-2}{(k+1)N_f}
\end{array}
\end{equation}
The integral formulas matching the superconformal index of the electric and of the dual phase for the $\UU(N)$ duality are obtained by integrating over the fugacities of the $\UU(1)_B$ symmetry from the formulas given in {\bf Section 9.4 of  \cite{Spiridonov:2009za}}. We refer the reader to \cite{Amariti:2025gca} for the detailed results for the case $k=1$. The case of higher $k$ is obtained by modifying the contributions of the $\UU(1)_X$ and $\UU(1)_R$-charges as read from the tables of charges above.
 Observe that, in this case, when increasing $k$ we need to consider the different parity for the mesons $P_\ell$ and $\tilde P_\ell$.

In the following, we study the reduction of this duality to 3d. In this case, as in the $A_k^{(S)}$ case we have two ways of proceeding. We  start first by employing again the double scaling method. We consider $N_c=2N$ and $N_f=2F+2$ and displace yet again $F$ flavor fugacities at the origin and at $1/(2r)$ and one both at $1/(4r)$ and $3/(4r)$. The dual gauge group is then broken to $\UU(\tilde N =3kF-N+2)^2\times\times\UU(3k)^2$. Making use of the double scaling procedure, we obtain again the following identity between partition functions:
\begin{eqnarray}
\label{quasiquadratoperfettoSDk}
&&
\left(
Z_{\UU(N)}^{(F\square; F \overline \square; 1 \overline{S};1 S;1 \text{Adj})}(\vec \mu;\vec \nu; \tau_{\tilde S};\tau_S;\tau_Y;\Lambda) \right)^2
\! \!=\!\! \left(Z_{\UU(3kF-N+2)}^{(F \square; F \overline \square; 1 \overline{S} ;1 S;1 \text{Adj})}
\left(\vec {\tilde \mu};\vec {\tilde \nu}; \tilde  \tau_{\tilde s};\tilde  \tau_s;\tilde\tau_y;-\Lambda
\right)\right)^2
\nonumber
\\
&&\times \prod_{\ell=0}^{k-1}\prod_{1\leq b,c \leq F}
\Gamma_h^2(\mu_b+ \nu_c+\ell\tau_Y,\mu_b+ \nu_c+\tau_S +\tau_{\tilde S}+\ell\tau_Y) 
\nonumber
\\
&& \times\prod_{\ell=0}^{k-1}\prod_{1\leq b<c \leq F}
\Gamma_h^2(\tau_{\tilde S} +\mu_b +\mu_c+\ell\tau_Y,\tau_S +\nu_b +\nu_c+\ell\tau_Y)
\nonumber \\
 &&\times\prod_{p=0}^{K}\prod_{b=1}^F
\Gamma_h^2(\tau_{\tilde S} +\mu_b +\mu_b+2p\tau_Y,\tau_S +\nu_b +\nu_b+2p\tau_Y)
\nonumber \\
&&\times \color{red}{ \prod_{\ell=0}^{k-1}\prod_{b=F+1}^{F+2}\!\! \Gamma_h(\mu_b\!+\! \nu_b+\ell\tau_Y,\mu_b\!+\! \nu_b+\tau_{S}+\tau_{\tilde S}+\ell\tau_Y)}
\nonumber
\\
&& \times
{\color{red} Z_{\UU^2(3k)}( \tilde \mu_{F+1},\tilde \nu_{F+2};\tilde \nu_{F+2},\tilde \mu_{F+1};\tau_y,\tau_y;\tilde \tau_s,\tilde \tau_{\tilde s};\Lambda,-\Lambda)}
\nonumber\\
&& \times \;
\textcolor{red}{ \prod_{\ell=0}^{k-1}\Gamma_h(\tau_{\tilde S}\!+\!\mu_{F+1}\!+\!\mu_{F+2} +\ell\tau_Y,\tau_S\!+\!\nu_{F+1}\!+\!\nu_{F+2}+\ell\tau_Y) }\; ,
\end{eqnarray}
where we have highlighted in red the contributions coming from the $\UU(3k)\times\UU(3k)$ sector and $K\equiv \frac{k-1}{2}$ as above. The requirement of $k$ being odd ensures that $K$ is an integer. The electric and magnetic fugacities are related by the following dictionary:
\begin{equation}
\label{dictionarySDk}
\begin{aligned}
\tilde \mu_{b} &= \tau_k -\mu_b + \frac{k(F+1)+1}{\tilde N} x,\quad 
\tilde \nu_{b} = \tau_k -\nu_b - \frac{k(F+1)+1}{\tilde N} x,  \\
\tilde \tau_a &= \omega-\tau_k + \frac{N-k (F+1)}{\tilde N} x,\quad
\tilde \tau_{\tilde a} =  \omega-\tau_k - \frac{N-k (F+1)}{\tilde N} x,\\
\end{aligned}
\end{equation}
and $\tilde \tau_y= \tau_Y = 2\tau_k$.
Additionally, the balancing condition reads:
\begin{equation}
\label{bcsymmD}
\sum_{b=1}^F(\mu_b+\nu_b)+\frac{1}{2}(\mu_{F+1}+\mu_{F+2}+\nu_{F+1}+\nu_{F+2})=2 ((F+1)\omega- \tau_k(N-1))\,.
\end{equation}
We then proceed by dualizing the $\UU(3k)^2$ sector as in Section \ref{sec:Dconf}.
In this way, we simplify the identity between partition functions, obtaining a relation between perfect squares.
By taking the square root, we obtain the final identity (up to an irrelevant phase):
\begin{equation}
\label{symmetricidDk}
\begin{aligned}
&
Z_{\UU(N)}^{(F\square; F \overline \square; 1 \overline{S};1 S;1 \text{Adj})}(\vec \mu;\vec \nu; \tau_{\tilde S};\tau_S;\tau_Y;\Lambda) 
=Z_{\UU(3kF-N+2)}^{(F \square; F \overline \square; 1 \overline{S} ;1 S;1 \text{Adj})}
\left(\vec {\tilde \mu};\vec {\tilde \nu}; \tilde  \tau_{\tilde s};\tilde  \tau_s;\tilde\tau_y;-\Lambda
\right)\times
\\
&\times \prod_{\ell=0}^{k-1}\prod_{1\leq b,c \leq F}
\Gamma_h(\mu_b+ \nu_c+\ell\tau_Y,\mu_b+ \nu_c+\tau_S +\tau_{\tilde S}+\ell\tau_Y) \\
&\times\prod_{\ell=0}^{k-1}\prod_{1\leq b<c \leq F}
\Gamma_h(\tau_{\tilde S} +\mu_b +\mu_c+\ell\tau_Y,\tau_S +\nu_b +\nu_c+\ell\tau_Y)
 \\
 &\times\prod_{p=0}^{K}\prod_{b=1}^F
\Gamma_h(\tau_{\tilde S} +\mu_b +\mu_b+2p\tau_Y,\tau_S +\nu_b +\nu_b+2p\tau_Y)\\
&\times\prod_{\ell=0}^{k-1}\Gamma_h\left(\pm \frac{\eta}{2} +(F+1)\omega-\tau_k(N+2\ell-1 )-\frac{1}{2}\sum_{b=1}^{F} (\mu_b+\nu_b)\right)
\\
& \times\;\prod_{\ell=0}^{k-1}\Gamma_h\left(2\left(F\omega-\tau_k (N+\ell+2)\right)-\sum_{b=1}^{F} (\mu_b+\nu_b)\right)
\,.
 \end{aligned}
\end{equation}

The final duality can be summarized as follows
\begin{equation}
\label{MDkSpure3d2}
\begin{array}{|cc|}
\hline
\UU(N) & \UU(3kF-N+2) \\
Y \oplus (S,\tilde S) \oplus F (Q,\tilde Q)& y \oplus (s,\tilde s) \oplus  F (q,\tilde q)   \\
W = \eqref{eq:WeleDkSSt} & W = \eqref{eq:WmagDkSSt} + Flip[\mathcal{M}^\pm_\ell ,\mathcal{M}^{(0)}_\ell ] \\ 
\hline
\end{array}
\end{equation}
where $\mathcal{M}^{(0)}_\ell = \mathcal{M}^{(0)}\, Tr y^{\ell}$ and $\mathcal{M}^\pm_\ell = \mathcal{M}^\pm \, Tr y^{\ell}$ , with $\ell+0,\dots,k-1$.
The flippers are the duals of the relative monopole operators acting as singlets in the magnetic phase.

In this case, we have an alternative way of obtaining the same result, similarly to what we did in the $A_k^{(S)}$ case. We take $N_c=N+1$ and $N_f=F$ and then we consider a background with a unbroken  $\UU(N)$ gauge group at the origin and a $\UU(1)$ at $1/(2r)$. Here, we do not need to take the square root, but we are able to reach the same identity \eqref{quasiquadratoperfettoSDk} after a shift in the FI.
The electric configuration is dual to a $\UU(\tilde N)\times \UU(1)$, where $\tilde N= 3k F-N+2$. Then, after a proper identification  of the $\UU(1)$ factor on the electric and magnetic side, we find the following identity between partition functions:
\begin{equation}
\label{quasiquadratoperfettoSDkU1}
\begin{aligned}
&
Z_{\UU(N)}^{(F\square; F \overline \square; 1 \overline{S};1 S;1 \text{Adj})}(\vec \mu;\vec \nu; \tau_{\tilde S};\tau_S;\tau_Y;\Lambda)
=Z_{\UU(3kF-N+2)}^{(F \square; F \overline \square; 1 \overline{S} ;1 S;1 \text{Adj})}
\left(\vec {\tilde \mu};\vec {\tilde \nu}; \tilde  \tau_{\tilde s};\tilde  \tau_s;\tilde\tau_y;-\Lambda
\right)\times
\\
&\times \prod_{1\leq b,c \leq F}
\prod_{\ell=0}^{k-1} \Gamma_h(\mu_b\!+\! \nu_c\!+\!\ell\tau_Y,\mu_b\!+\! \nu_c\!+\!(\ell+2)\tau_Y,\tau_{\tilde S} \!+\!\mu_b \! +\!\mu_c\!+\!\ell\tau_Y,\tau_S\! +\!\nu_b \!+\!\nu_c\!+\!\ell\tau_Y)
 \\
 &\times\prod_{p=0}^{K}\prod_{b=1}^F
\Gamma_h(\tau_{\tilde S} +\mu_b +\mu_b+2p\tau_Y,\tau_S +\nu_b +\nu_b+2p\tau_Y)\;.
\end{aligned}
\end{equation}
Here, the magnetic and electric quantity are related by the following dictionary:
\begin{equation}
\label{dictionarySDkU1}
\begin{aligned}
\tilde{\mu}_{b} &=
    3 \tau_k - \omega - \mu_b
    + \frac{kF+2}{3kF - N + 3}\, x,
&\qquad
\tilde{\nu}_{b} &=
    3 \tau_k - \omega - \nu_b
    - \frac{kF+2}{3kF - N + 3}\, x,
\\[6pt]
\tilde{\tau}_{\tilde a} &=
    \omega - \tau_k
    - \frac{N+1-kF}{3kF - N + 3}\, x,
&\qquad
\tilde{\tau}_{a} &=
    \omega - \tau_k
    + \frac{N+1-kF}{3kF - N + 3}\, x,
\end{aligned}
\end{equation}
and $\tilde \tau_y =\tau_Y= 2\tau_k$.
Additionally, we have the following constraint:
\begin{equation}
\label{constraintD}
\sum_{b=1}^F(\mu_b+\nu_b)=2 (F\omega- \tau_k(N-1))\,,
\end{equation}
that is enforced by a linear monopole superpotential.

Then, we need to remove this constraint by triggering a real mass flow. For this reason we 
consider  $F+2$ flavors and we 
 perform a real mass flow, with a large positive mass for one flavor and a large negative mass for another one.
On the electric side, we are left with a $\UU(N)$ theory with $F$ flavors, in addition to the two-index tensors $\Phi$, $A$ and $\tilde S$. On the dual side, the duality is preserved if in addition to the real mass flow we consider also a Higgs flow that breaks the dual gauge group to $\UU(3kF-N+2)\times\UU(3k)^2$. This flow automatically  eliminates the constraint \eqref{constraintD} and we obtain the following identity between partition function:
\begin{eqnarray}
\label{quasiquadratoperfettoSDkU12F}
&&
Z_{\UU(N)}^{(F\square; F \overline \square; 1 \overline{S};1 S;1 \text{Adj})}(\vec \mu;\vec \nu; \tau_{\tilde S};\tau_S;\tau_Y;\Lambda)
=Z_{\UU(3kF-N+2)}^{(F \square; F \overline \square; 1 \overline{S} ;1 S;1 \text{Adj})}
\left(\vec {\tilde \mu};\vec {\tilde \nu}; \tilde  \tau_{\tilde s};\tilde  \tau_s;\tilde\tau_y;-\Lambda
\right)
\nonumber \\
&&\times
Z_{\UU^2(3k)}( \tilde \mu_{F+1},\tilde \nu_{F+2};\tilde \nu_{F+2},\tilde \mu_{F+1};\tau_y,\tau_y;\tilde \tau_s,\tilde \tau_{\tilde s};\Lambda_{eff},-\Lambda_{eff}) \nonumber
\\
&&\times \prod_{\ell=0}^{k-1}\prod_{1\leq b,c \leq F}
\Gamma_h(\mu_b+ \nu_c+\ell\tau_Y,\mu_b+ \nu_c+\tau_S +\tau_{\tilde S}+\ell\tau_Y) \nonumber  \\
&&\times \prod_{\ell=0}^{k-1}\prod_{b,c=F+1}^{F+2}\!\! \Gamma_h(\mu_b\!+\! \nu_c+\ell\tau_Y,\mu_b\!+\! \nu_c+\tau_{S}+\tau_{\tilde S}+\ell\tau_Y) \nonumber \\
&&\times\prod_{\ell=0}^{k-1}\prod_{1\leq b<c \leq F}
\Gamma_h(\tau_{\tilde S} +\mu_b +\mu_c+\ell\tau_Y,\tau_S +\nu_b +\nu_c+\ell\tau_Y) \nonumber
 \\
&&\times\prod_{p=0}^{K}\prod_{b=1}^F
\Gamma_h(\tau_{\tilde S} +\mu_b +\mu_b+2p\tau_Y,\tau_S +\nu_b +\nu_b+2p\tau_Y) \nonumber \\
&&\times \prod_{p=0}^{K}\prod_{b=1}^{F+1}
\Gamma_h(\tau_{\tilde S} +\mu_b +\mu_b+2p\tau_Y,\tau_S +\nu_b +\nu_b+2p\tau_Y) \nonumber
 \\
&&\times \;
\prod_{\ell=0}^{k-1}\Gamma_h(\tau_{\tilde S}\!+\!\mu_{F+1}\!+\!\mu_{F+2} +\ell\tau_Y,\tau_S\!+\!\nu_{F+1}\!+\!\nu_{F+2}+\ell\tau_Y) \, ,
\end{eqnarray}
with $\Lambda_{eff} = \Lambda-\frac{1}{2}(\mu_{F+1}+\mu_{F+2}+\nu_{F+1}+\nu_{F+2})+2\omega$.
Here, the electric and magnetic fugacities are related by the following dictionary:
\begin{equation}
\label{dictionarySDkU12F}
\begin{aligned}
\tilde \mu_{b} &= 3 \tau_k -\omega-\mu_b + \frac{k(F+2)+2}{3 k (F+2) - N+3} x, \quad
\tilde \tau_a = \omega-\tau_k + \frac{N+1-k (F+2)}{3 k (F+2) - N+3} x,  \\
\tilde \nu_{b} &= 3 \tau_k -\omega-\nu_b -\frac{k(F+2)+2}{3 k (F+2) - N+3} x,\quad
\tilde \tau_{\tilde a} =  \omega-\tau_k - \frac{N+1-k (F+2)}{3 k (F+2) - N+3} x,
\end{aligned}
\end{equation}
and
$\tilde \tau_y =\tau_Y= 2\tau_k$, and the balancing condition reads:
\begin{equation}
\label{newbb}
\sum_{b=1}^F(\mu_b+\nu_b)+(\mu_{F+1}+\mu_{F+2}+\nu_{F+1}+\nu_{F+2})=2 ((F+2)\omega- \tau_k(N-1))\,.
\end{equation}
If we compare this balancing condition to \eqref{bcsymmD}, we notice that they differ by $-\frac{1}{2}(\mu_{F+1}+\mu_{F+2}+\nu_{F+1}+\nu_{F+2})+2\omega$, which is precisely the difference in the FI in $\Lambda_{eff}$. We can thus use again the results of Section \ref{sec:Dconf} to dualize the $\UU(3k)^2$ and find once more \eqref{symmetricidDk},  after substituting the condition \eqref{newbb}.

\subsection{Balcony $D_{k+2}$ }
\label{BD3d}

The last model under investigation has been  studied in {\bf Section 7 of \cite{Brodie:1996xm}}.
where it was referred to as  $D_{k+2}$ balcony model.
Observe that, in this case, both parities of $k$ are allowed.
The field content consists of an adjoint $Y$, a two-index antisymmetric tensor $A$ and a two-index conjugated symmetric  tensor  $\tilde S$ in addition to $N_f$  fundamentals $Q$  and $N_f-8$ antifundamentals  $\tilde Q$. On the electric side the superpotential is
\begin{equation}
\label{eq:WeleDkASt}
W = Y^{k+1} + A Y \tilde S\, .
\end{equation}
This model is dual to an $\SU(\tilde N_c = 3k (N_f-4) -N_c)$ gauge theory with 
an adjoint $y$, a two-index antisymmetric tensor $a$ and a two-index conjugated symmetric  tensor  $\tilde s$ in addition to $N_f$  fundamentals $q$  and $N_f-8$ antifundamentals  $\tilde q$. 
In this case the chiral rings truncates classically for either even or odd $k$. 
There are bifundamental mesons  $M_j = Q Y^j \tilde Q$ and  $N_j = Q Y^j \tilde  S A \tilde Q$  with $j=0,\dots,k-1$. 
The other mesons 
$P_j = Q Y^j \tilde S Q$ and $\tilde P_j = \tilde Q A Y^j   \tilde Q$ with $j=0,\dots,k-1$ are in the two-index symmetric and antisymmetric representation of the flavor symmetry group  respectively. The dual superpotential is
\begin{equation}
\label{eq:WmagDkASt}
W =y^{k+1} + y a \tilde{s}
+ \sum_{j=0}^{k-1}( 
    N_j\, \tilde q\, y^{k-j-1} q 
+ P_j\, q\, y^{k-j-1} a q
+ \tilde P_j\, \tilde q\, \tilde{s}\, y^{k-j-1} \tilde q
+ M_j\, \tilde q\, y^{k-j-1} \tilde{s} a q)
.
\end{equation}

The global charges of the electric side 
\begin{equation}
\begin{array}{c|c|ccccc}
               &\SU(N_c)&\SU(N_f)   & \SU(N_f-8) & \UU(1)_B & \UU(1)_X & \UU(1)_R\\
               \hline
Q            & \square &  \square &1 &   \frac{1}{N_c}&-1+\frac{6}{N_f}&1- \frac{N_c+6k}{(k+1)N_f} \\
\tilde Q    & \overline \square & 1  &  \square &\! \! \! \! \! -\frac{1}{N_c} &1+\frac{6}{N_f-8}  & 1- \frac{N_c-6k}{(k+1)(N_f-8)} \\
Y & Adj.&1&1&0&0&\frac{2}{k+1}\\
\tilde S    & \overline{ \square \! \square} &1   & 1&\! \! \! \! \!   -\frac{2}{N_c}&\! \! \! \! \! -1  &\frac{k}{k+1} \\
A            & \begin{array}{c} \square \vspace{-2.9mm} \\  \square  \end{array} &1    & 1&\frac{2}{N_c} & 1 &\frac{k}{k+1}  \\
\end{array}
\end{equation}
and of the magnetic side
\begin{equation}
\begin{array}{c|c|cccccc}
&\SU(\tilde N_c)&\SU(N_f)   & \SU(N_f-8) & \UU(1)_B & \UU(1)_X & \UU(1)_R\\
\hline
q            & \square &  \overline \square &1 &   \frac{1}{\tilde N_c}& 1-\frac{6}{N_f} &1- \frac{\tilde N_c+6k}{(k+1)N_f} \\
\tilde q    & \overline \square & 1  &  \overline\square&\! \! \! \! \! -\frac{1}{\tilde N_c} & -1-\frac{6}{N_f-8} & 1- \frac{\tilde N_c-6k}{(k+1)(N_f-8)} \\
y & Adj.&1&1&0&0&\frac{2}{k+1}\\
\tilde s    &\overline{\square\!\square}  & 1& 1&\! \! \! \! \!  -\frac{2}{\tilde N_c}&1&\frac{k}{k+1} \\
a            & \begin{array}{c} \square \vspace{-2.9mm} \\  \square  \end{array}    &1 & 1&\frac{2}{\tilde N_c} &\! \! \! \! \!-1&\frac{k}{k+1}  \\
\hline 
P_{j}            & 1  &  \begin{array}{c} \square \vspace{-2.9mm} \\  \square  \end{array}   &1& 0  &-\frac{3(N_f-4)}{N_f} &\frac{2j+k}{k+1}+2-2\frac{N_c+6k}{(k+1)N_f}\\
\tilde P_{j}            & 1  & 1 &\begin{array}{c} \square \vspace{-2.9mm} \\  \square  \end{array}   & 0  & \frac{3(N_f-4)}{N_f-8}  &\frac{2j+k}{k+1}+2-2\frac{N_c-6k}{(k+1)(N_f-8)}\\
M_j            & 1  & \square  &\square  & 0 &\frac{12(N_f-4)}{N_f(N_f-8)}&\frac{2(k+j)}{k+1}+2-\frac{2N_c(N_f-4)+48k}{(k+1)N_f(N_f-8)}
\\
N_j            & 1  & \square  &\square  & 0 &\frac{12(N_f-4)}{N_f(N_f-8)}&\frac{2j}{k+1}+2-\frac{2N_c(N_f-4)+48k}{(k+1)N_f(N_f-8)}
\end{array}
\end{equation}

As in the $A_k$ balcony duality, here, we study the reduction to 3d of  two different models, connected by an RG flow deformation. First, we consider a deformed version of the balcony model with an additional cubic superpotential term for the conjugated symmetric and eight fundamentals. This is the model that can be read from the brane picture. Then, we study the undeformed case, corresponding to the original balcony duality of \cite{Brodie:1996xm}, by consistently adapting the results obtained from the application of the HW transition in the deformed case.

\subsubsection*{The deformed case}
We start by considering the brane picture discussed in \cite{Brunner:1998jr} and reviewed in subsection \ref{branepicture}. We consider 
$N_c=2N$ D4 branes and $2F+2$ pairs of D6$_{\pm \theta}$. In the electric phase, upon T-duality, we focus on the configuration with $N$ D3 at the origin and $N$ D3 at $1/(2r)$.
We further displace $F$ flavors at the origin and at $1/(2r)$ and one both at $1/(4r)$ and $3/(4r)$. 
Then, we consider the HW transition for this brane picture and we obtain a dual configuration with $3kF-N+2k$ D3 at the origin and  at $1/(2r)$ in addition to $3k$ D3 both at $1/(4r)$ and at $3/(4r)$.
We then translate this picture in the reduction of the identity between the superconformal 
indices and we indeed find that the divergent phase cancels between the dual phase, leaving us with the following identity 
\begin{eqnarray}
\label{quasiquadratoperfettobalconyDkbroken}
&&
\left(
Z_{\UU(N)}^{((F+4)\square; F \overline \square; 1 \overline{S};1 A;1 \text{Adj})}(\vec \mu,\vec \zeta;\vec \nu; \tau_{\tilde S};\tau_A;\tau_Y;\Lambda) \right)^2=
 \nonumber
\\
&& 
\left(Z_{\UU(3kF-N+2k)}^{((F+4) \square; F \overline \square; 1 \overline{S} ;1 A;1 \text{Adj})}
\left(\vec {\tilde \mu},\vec {\tilde \zeta};\vec {\tilde \nu}; \tilde  \tau_{\tilde s};\tilde  \tau_a;\tilde\tau_y;-\Lambda
\right)\right)^2\times \nonumber
\\ &&\times\prod_{\ell=0}^{k-1}\left(\prod_{1\leq b,c \leq F}
\Gamma_h^2(\tau_{\tilde S} +\mu_b +\mu_c+\ell\tau_Y)\prod_{1\leq b<c \leq F}\Gamma_h^2(\tau_A +\nu_b +\nu_c+\ell\tau_Y)\right) \nonumber \\
&&\times \prod_{\ell=0}^{k-1}\left(\prod_{1\leq b,c \leq F}
\Gamma_h^2(\mu_b+ \nu_c+\ell\tau_Y,\mu_b+ \nu_c+\tau_A +\tau_{\tilde S}+\ell\tau_Y) \prod_{b=1}^{F}\prod_{d=1}^4
\Gamma_h^2(\mu_b+ \zeta_d+\ell\tau_Y)\right)\nonumber  \\
&&\times \color{red}{ \prod_{\ell=0}^{k-1}\left(\prod_{b,c=F+1}^{F+2}\Gamma_h(\mu_b\!+\! \nu_c+\ell\tau_Y,\mu_b\!+\! \nu_c+\tau_{A}+\tau_{\tilde S}+\ell\tau_Y)\prod_{b=F+1}^{F+2}\prod_{d=1}^4\Gamma_h(\mu_b\!+\! \zeta_d+\ell\tau_Y)\right)}\nonumber \\
&& \times
{\color{red} Z_{\UU^2(3k)}( \tilde \mu_{F+1},\tilde \nu_{F+2};\tilde \nu_{F+2},\tilde \mu_{F+1};\tau_y,\tau_y;\tilde \tau_a,\tilde \tau_{\tilde s};\Lambda,-\Lambda)} \nonumber \\
&& \times \;
\color{red}{ \prod_{\ell=0}^{k-1}\Gamma_h(\tau_{\tilde S}\!+\!\mu_{F+1}\!+\!\mu_{F+2} +\ell\tau_Y,\tau_A\!+\!\nu_{F+1}\!+\!\nu_{F+2}+\ell\tau_Y) } \, , 
\end{eqnarray}
where the dictionary between the electric and the magnetic variables is the following: 
\begin{eqnarray}
\label{dictionarybalconyDkbroken}
\tilde \mu_{b} &=& \tau_k -\mu_b + \frac{k(F+1)+1}{\tilde N} x \,, \nonumber \\
\tilde \nu_{b} &=& \tau_k -\nu_b - \frac{k(F+1)+1}{\tilde N} x\,, \nonumber\\
\tilde \zeta_{d} &=& \tau_k -\zeta_d + \left(\frac{N-k (F+1)}{2\tilde N}-\frac{1}{2}\right) x\,, \\
\tilde \tau_a &=& \omega-\tau_k + \frac{N-k (F+1)}{\tilde N} x\, ,\nonumber \\
\tilde \tau_{\tilde s} &=&  \omega-\tau_k - \frac{N-k (F+1)}{\tilde N} x\, ,\nonumber
\end{eqnarray}
and $\tilde \tau_y =\tau_Y =  2\tau_k$.
The mass parameters in this case are constrained by a balancing condition that can be read from the 4d one, and it signals the presence of a monopole 
superpotential. We have 
\begin{equation}
\sum_{b=1}^F(\mu_b+\nu_b)+\frac{1}{2}(\mu_{F+1}+\mu_{F+2}+\nu_{F+1}+\nu_{F+2})=2 ((F+2)\omega- \tau_k(N+1))\,.
\end{equation}
Again, the red quantities in \eqref{quasiquadratoperfettobalconyDkbroken} can be eliminated by using the results in Section \ref{sec:Dconf}.
Indeed the $\UU(3k)^2$ sector confines, giving rise to towers of mesons and monopole.
The mesons simplify with the ones represented in red in 
\eqref{quasiquadratoperfettobalconyDkbroken} while the monopoles contribute to the final partition function, giving rise to an identity between two perfect squares.
The final relation, up to an irrelevant sign, becomes
\begin{equation}
\label{balconyidDkbroken}
\begin{aligned}
&
Z_{\UU(N)}^{((F+4)\square; F \overline \square; 1 \overline{S};1 A;1 \text{Adj})}(\vec \mu,\vec \zeta;\vec \nu; \tau_{\tilde S};\tau_A;\tau_Y;\Lambda)
=Z_{\UU(3kF-N+2k)}^{((F+4) \square; F \overline \square; 1 \overline{S} ;1 A;1 \text{Adj})}
\left(\vec {\tilde \mu},\vec {\tilde \zeta};\vec {\tilde \nu}; \tilde  \tau_{\tilde s};\tilde  \tau_a;\tilde\tau_y;-\Lambda
\right)\times
\\
&\quad\times \prod_{\ell=0}^{k-1}\left(\prod_{1\leq b,c \leq F}
\Gamma_h(\mu_b+ \nu_c+\ell\tau_Y,\mu_b+ \nu_c+\tau_A +\tau_{\tilde S}+\ell\tau_Y) \prod_{b=1}^{F}\prod_{d=1}^4
\Gamma_h(\mu_b+ \zeta_d+\ell\tau_Y)\right) \\
&\quad\times\prod_{\ell=0}^{k-1}\left(\prod_{1\leq b,c \leq F}
\Gamma_h(\tau_{\tilde S} +\mu_b +\mu_c+\ell\tau_Y)\prod_{1\leq b<c \leq F}\Gamma_h(\tau_A +\nu_b +\nu_c+\ell\tau_Y)\right)\\
&\quad\times
\prod_{\ell=0}^{k-1}\Gamma_h\left(\pm \frac{\eta}{2} +(F+2)\omega-\tau_k(N+2\ell+1 )-\frac{1}{2}\sum_{b=1}^{F} (\mu_b+\nu_b)\right)
\\
&\quad \times\;\prod_{\ell=0}^{k-1}\Gamma_h\left(2\left((F+1)\omega-\tau_k (N+\ell)\right)-\sum_{b=1}^{F} (\mu_b+\nu_b)\right)
\,.
\end{aligned}
\end{equation}

The final duality can be summarized as follows
\begin{equation}
\label{BDkpure3d}
\begin{array}{|cc|}
\hline
\UU(N) & \UU(3kF-N+2k) \\
Y \oplus (A,\tilde S) \oplus F (Q,\tilde Q) \oplus 4 T & y \oplus (a,\tilde s) \oplus  F (q,\tilde q) \oplus 4 t   \\
W = \eqref{eq:WeleDkASt} + \tilde S T^2 & W = \eqref{eq:WmagDkASt}+ \tilde s t^2 + Flip[\mathcal{M}^\pm_\ell ,\mathcal{M}^{(0)}_\ell ] \\ 
\hline
\end{array}
\end{equation}
where $\mathcal{M}^{(0)}_\ell = \mathcal{M}^{(0)}\, Tr y^{\ell}$ and $\mathcal{M}^\pm_\ell = \mathcal{M}^\pm \, Tr y^{\ell}$ , with $\ell=0,\dots,k-1$.
The flippers are the duals of the relative monopole operators acting as singlets in the magnetic phase.

\subsubsection*{The undeformed case}

We conclude the analysis by discussing the reduction of the original $D_{k+2}$ balcony duality proposed in \cite{Brodie:1996xm}.
This duality cannot be obtained from the brane picture, and it has been reviewed above.
In the case at hand we take $N_c=2N$ and we consider $2F+2$ fundamental flavors and $2F-6$ antifundamental flavors. Inspired by the discussion in the deformed case, we consider we do not turn on a background for  $F$ fundamentals and $F-4$ antifundamentals, corresponding in the brane picture of the deformed case to the D5${\pm \theta}$ left at the origin.  Then, we turn on a background proportional to $1/(2r)$ for $F$ fundamentals and $F-4$ antifundamentals, corresponding  in the brane picture of the deformed case to the D5${\pm \theta}$ displaced at $1/(2r)$ in the compact T-dual direction. 
Furthermore we consider a mass proportional to $1/(4r)$  for the $(F+1)$-th flavor 
and a mass proportional to $3/(4r)$  for the $(F+2)$-th flavor.
The dual gauge group then gets broken to $\SU(3k(F-2)-N+2k)^2\times\UU(3k)^2$.
If we further identify the masses symmetrically between the specular sectors we arrive to the following identity between partition functions:
\begin{eqnarray}
\label{quasiquadratoperfettobalconyDk}
&&
\left(
Z_{\UU(N)}^{(F\square; (F-4) \overline \square; 1 \overline{S};1 A;1 \text{Adj})}(\vec \mu,\vec \zeta;\vec \nu; \tau_{\tilde S};\tau_A;\tau_Y;\Lambda) \right)^2
=\nonumber \\
&&
\left(Z_{\UU(3k(F-2)-N+2k)}^{(F \square; (F-4) \overline \square; 1 \overline{S} ;1 A;1 \text{Adj})}
\left(\vec {\tilde \mu},\vec {\tilde \zeta};\vec {\tilde \nu}; \tilde  \tau_{\tilde s};\tilde  \tau_a;\tilde\tau_y;-\Lambda
\right)\right)^2\times
\nonumber \\
&&
\times\prod_{\ell=0}^{k-1}\left(\prod_{1\leq b,c \leq F}
\Gamma_h^2(\tau_{\tilde S} +\mu_b +\mu_c+\ell\tau_Y)\prod_{1\leq b<c \leq F-4}\!\!\!\!\!\!\Gamma_h^2(\tau_A +\nu_b +\nu_c+\ell\tau_Y)\right)
\nonumber \\
&&
\times \prod_{\ell=0}^{k-1}\prod_{b=1}^F\prod_{c=1}^{F-4}
\Gamma_h^2(\mu_b+ \nu_c+\ell\tau_Y,\mu_b\!+\! \nu_c\!+\!\tau_A \!+\!\tau_{\tilde S}\!+\!\ell\tau_Y) 
  \\
&& 
  \times {\color{red}  Z_{\UU^2(3k)}( \tilde \mu_{F+1},\tilde \nu_{F+2};\tilde \nu_{F+2},\tilde \mu_{F+1};\tau_y,\tau_y;\tilde \tau_a,\tilde \tau_{\tilde s};\Lambda,-\Lambda)}
\nonumber \\
&&
\times \color{red}{ \prod_{\ell=0}^{k-1}\prod_{b=F+1}^{F+2}\prod_{c=F-3}^{F-2}\Gamma_h(\mu_b\!+\! \nu_c+\ell\tau_Y,\mu_b\!+\! \nu_c+\tau_{A}+\tau_{\tilde S}+\ell\tau_Y)}
 \nonumber
 \\
&&
 \times \;
\color{red}{ \prod_{\ell=0}^{k-1}\bigg(\Gamma_h(\tau_{\tilde S}\!+\!\mu_{F+1}\!+\!\mu_{F+2} +\ell\tau_Y,\tau_A\!+\!\nu_{F-3}\!+\!\nu_{F-2}+\ell\tau_Y)\prod_{b=F+1}^{F+2} \Gamma_h(\tau_{\tilde S}\!+\!\mu_{b}\!+\!\mu_{b} +\ell\tau_Y)\bigg)}. \nonumber
\end{eqnarray}
Here the balancing condition reads as 
\begin{equation}
\sum_{b=1}^F(\mu_b+\nu_b)+\frac{1}{2}(\mu_{F+1}+\mu_{F+2}+\nu_{F+1}+\nu_{F+2})=2 ((F-1)\omega- \tau_k N+x)\,,
\end{equation}
and the following dictionary between the electric and the magnetic parameters holds
\begin{equation}
\label{dictionarybalconyDk}
\begin{aligned}
\tilde \mu_{b} &= \tau_k -\mu_b - \frac{F-2}{F+1}x, \quad
\tilde \tau_a = \omega-\tau_k + x, \\
\tilde \nu_{b}& = \tau_k -\nu_b +\frac{F+1}{F-2}x, \quad
\tilde \tau_{\tilde s} =  \omega-\tau_k - x,
\end{aligned}
\end{equation}
and $\tilde \tau_y =\tau_Y= 2\tau_k$.
Again, we can get rid of the confining $\UU(3k)^2$ sector and we arrive to the following identity
\begin{equation}
\label{balconyidDk}
\begin{aligned}
&
Z_{\UU(N)}^{(F\square; (F-4) \overline \square; 1 \overline{S};1 A;1 \text{Adj})}(\vec \mu,\vec \zeta;\vec \nu; \tau_{\tilde S};\tau_A;\tau_Y;\Lambda) \!\!=\!\!Z_{\UU(3k(F-2)-N+2k)}^{(F \square; (F-4) \overline \square; 1 \overline{S} ;1 A;1 \text{Adj})}
\left(\vec {\tilde \mu},\vec {\tilde \zeta};\vec {\tilde \nu}; \tilde  \tau_{\tilde s};\tilde  \tau_a;\tilde\tau_y;-\Lambda
\right)
\\
&\quad\times \prod_{\ell=0}^{k-1}\prod_{b=1}^F\prod_{c=1}^{F-4}
\Gamma_h(\mu_b+ \nu_c+\ell\tau_Y,\mu_b+ \nu_c+\tau_A +\tau_{\tilde S}+\ell\tau_Y)  \\
 &\quad\times\prod_{\ell=0}^{k-1}\left(\prod_{1\leq b,c \leq F}
\Gamma_h(\tau_{\tilde S} +\mu_b +\mu_c+\ell\tau_Y)\prod_{1\leq b<c \leq F-4}\!\!\!\!\!\!\Gamma_h(\tau_A +\nu_b +\nu_c+\ell\tau_Y)\right)\\
&\quad\times
\prod_{\ell=0}^{k-1}\Gamma_h\left(\pm \frac{\eta}{2} +(F-1)\omega-\tau_k(N+2\ell)+x-\frac{1}{2}\sum_{b=1}^{F} (\mu_b+\nu_b)\right)
\\
&\quad \times\;\prod_{\ell=0}^{k-1}\Gamma_h\left(2\left((F-2)\omega-\tau_k (N+\ell-1)+x\right)-\sum_{b=1}^{F} (\mu_b+\nu_b)\right)
\,,
\end{aligned}
\end{equation}
representing the pure identity for the pure 3d $\mathcal{N}=2$ $D_{k+2}$ balcony  duality.

The final duality can be summarized as follows
\begin{equation}
\label{BDkpure3d2}
\begin{array}{|cc|}
\hline
\UU(N) & \UU(3k(F-2)-N+2k) \\
Y \oplus (A,\tilde S) \oplus F Q \oplus (F-4) \tilde Q \quad & y \oplus (a,\tilde s)\oplus F q \oplus (F-4) \tilde q   \\
W = \eqref{eq:WeleDkASt}  & W = \eqref{eq:WmagDkASt} + Flip[\mathcal{M}^\pm_\ell ,\mathcal{M}^{(0)}_\ell ] \\ 
\hline
\end{array}
\end{equation}
where $\mathcal{M}^{(0)}_\ell = \mathcal{M}^{(0)}\, Tr y^{\ell}$ and $\mathcal{M}^\pm_\ell = \mathcal{M}^\pm \, Tr y^{\ell}$ , with $\ell=0,\dots,k-1$.
The flippers are the duals of the relative monopole operators acting as singlets in the magnetic phase.

%
%
%
%
%
\section{Comments on CS and $\SU(N)$ dualities}
\label{CSSU}

In this section, we  discuss two generalizations of the 3d dualities studied here.
The first generalization concerns the generation of CS terms in the action, while the second one consist of the un-gauging of the baryonic symmetry, giving rise to dualities involving $\SU(N)$ factors.
We will be rather sketchy in our analysis leaving many technical details to the interested reader.

%
\subsection{Dualities with CS terms }
\label{DualCS}
%

The construction is rather canonical, because the CS terms are generated by real mass flows on the fundamentals. Namely we consider $F+\kappa$ pairs of conjugated fundamental and assign them a large real mass. This procedure generates the CS terms on the electric side, and once we map the  real mass flow on the dual side, we obtain the final duality.
It is important to observe that these final dualities hold if opportune CS contact terms for the global symmetries are turned on. Such CS contact terms reflect into a complex phase in the identity between the three sphere partition functions.

We start by considering on the electric side $\UU(N)$ gauge theories with $F+\kappa$ pairs of fundamentals and antifundamentals in the mezzanine models. In the balcony models we  have $F +\kappa-4$ fundamentals and $F+\kappa$ antifundamentals.
Assigning large real masses to $\kappa$ pairs of fundamentals and antifundamentals triggers a  real mass flow, and we end up with $\UU(N)$ gauge theories at CS level $\kappa$.
Depending on the duality we have a different dual gauge group.
Below, we  summarize the dual group arising for each model studied here
\begin{center}
\begin{tabular}{|c|c|}
\hline
Model & Dual gauge group \\
\hline
Mezzanine $A_k^{(A)}$ &  $\UU((2k+1)(F+|\kappa|)-N-2k)_{-\kappa}$   \\
Mezzanine $A_k^{(S)}$ & $\UU((2k+1)(F+|\kappa|)-N+2k)_{-\kappa}$  \\
Deformed Balcony
\footnote{Observe that in this case we  restricted the power of the adjoint to be even, and strictly speaking this boils down to consider the $A_{2k+1}$ case.  While in the body of the paper we kept on referring 
to the $A_k$ case, keeping the notation of \cite{Brodie:1996xm}, here  we have explicitly denoted this case as $A_{2k+1}$}
 $A_{2k+1}$& $\UU((4k+3)(F+|\kappa|)-N+2)_{-\kappa}$   \\
Undeformed Balcony $A_{2k+1}$& $\UU((4k+3)(F+|\kappa|-2)-N+2)_{-\kappa}$ \\
\hline
Mezzanine $D_{k+2}^{(A)}$ &  $\UU(3k(F+|\kappa|)-N-2)_{-\kappa}$  \\  
Mezzanine $D_{k+2}^{(S)}$ & $\UU(3k(F+|\kappa|)-N+2)_{-\kappa}$  \\
Deformed Balcony $D_{k+2}$ &$\UU(3k(F+|\kappa|)-N+2k)_{-\kappa}$  \\
Undeformed Balcony $D_{k+2}$ &$\UU(3k(F+|\kappa|-2)-N+2k)_{-\kappa}$ \\
\hline
\end{tabular}
\end{center}
Such real mass flows can be engineered at the level of the three sphere partition function, by taking a limit on the associated mass parameters. It turns out that the divergent phases cancel in the identity. The massless spectrum is identical to the one of the 4d duality, indeed, the monopoles are all massive. There is  a complex phase in the final identities, which indicates the presence of non trivial CS contact terms, that needs to be matched as explained in \cite{Closset:2012vp}. We leave a detailed analysis of the matching of the partition functions and the relative CS contact terms to the interested reader. 

Rather, here, we compare the results obtained above with similar dualities proposed in the literature.
The mezzanine $A_k^{(A)}$ and $A_k^{(S)}$ have been proposed in \cite{Kapustin:2011vz} (see {\bf Section 3.4} and {\bf 3.5} of such a reference) and they coincide with our findings for positive $\kappa$ (here we did not restrict ourselves to $\kappa>0$ as done in \cite{Kapustin:2011vz}).

In general, the brane setup discussed above can be used to engineered the real mass flow that gives rise to the duality between CS theories. The  mezzanine $A_k^{(A)}$ was explicitly discuss in \cite{Amariti:2015mva} (see also \cite{Amariti:2016kat}) while the other cases can be studied analogously. The flows are triggered by breaking a D5$_{\pm \theta}$  brane 
on an NS$_{\pm \theta}$ brane, creating two semi-infinte half D5 branes that can slide along the NS branes in the direction $x_3$. The opposite  motion along $x_3$ of $k$ semi-infinte pairs of D5 branes 
forces one to generate a $(1,k)$ fivebrane in the plane parameterized by $x_3$ and $x_7$, and such effects reflects in the generation of a CS term \cite{Aharony:1997ju,Kitao:1998mf} in the action.
The dualities discussed above in presence of CS terms can be all obtained by HW transition on the associated brane setup.

%
%
%
%
%
\subsection{$\SU(N)$ dualities}
\label{sec:sun}
%
%
%
%
%
We can also un-gauge the baryonic symmetry, giving rise on the electric side to an $\SU(N)$ gauge theory.
Such an un-gauging is performed by gauging the topological symmetry $\UU(1)_J$. There is a  mixed CS term between the overall $\UU(1) \subset \UU(N)$ and the gauged topological symmetry. By integrating out the massive vector multiplet, one is left with an $\SU(N)$ gauge symmetry. The baryonic symmetry is the topological symmetry of the gauged $\UU(1)_J$.

While this procedure converts the electric  $\UU(N)$ into $\SU(N)$ on the electric side, the situation is more involved in the dual theories. Indeed, in this case we have monopoles charged under the topological symmetry, and in addition monopoles that are uncharged under such a symmetry.

Let us focus on a simple example in order to have a better understanding.
For simplicity we focus on the $A_1$ cases, corresponding to the cases studied in \cite{Amariti:2025gca}, and we look at the gauging from the perspective of the three sphere partition function. 
This boils down to integrate over 
\begin{equation}
\frac{1}{2} \int d\Lambda \,e^{2i \pi N m_B \Lambda}\, .
\end{equation}

The integral over $\Lambda$ on the electric side gives rise to the delta function that reduces $\UU(N)$ into $\SU(N)$. On the dual side we have a SQED sector, and, after performing this integral, we are left with  
a pair of baryon and antibaryon, in addition to a singlet that we denote as $V_1$. Such a singlet is a dressed\footnote{Dresse by a quadratic gauge invariant combination of two-index tensors} monopole of the electric $\SU(N)$ theory, that acts a singlet in the dual phase, flipping the product of baryon and antibaryon.
There is a second singlet, the uncharged original monopole of the $\UU(N)$ theory (denoted as $V_0$ in 
 \cite{Amariti:2025gca}) that still flips the dual monopole $v_0$ in the dual side.
 
 In formulas, the two monopoles   $V_{\ell=0,1}$ and the baryons contribute to the dual partition function
in the mezzanine $A_1^{(A)}$ case as 
 \begin{equation}
\prod_{\ell=0,1} \Gamma_h((2F-N+\ell)\omega-\sum_{a=1}^F(\mu_a+\nu_a))
 \Gamma_h \Big(\pm \sum_{i=1}^{\tilde N}\sigma_i +\frac{N-2F+1}{2} \omega+\frac{1}{2}\sum_{a=1}^F(\mu_a+\nu_a)  \Big) \, ,
  \end{equation}
while  in the mezzanine $A_1^{(S)}$  and  in the deformed balcony  $A_1$  we have
 \begin{equation}
\prod_{\ell=0,1} 
 \Gamma_h((2F-N+\ell+2)\omega-\sum_{a=1}^F(\mu_a+\nu_a))
 \Gamma_h \Big(\pm \sum_{i=1}^{\tilde N}\sigma_i +\frac{N-2F-1}{2} \omega+\frac{1}{2}\sum_{a=1}^F(\mu_a+\nu_a)  \Big) .
    \end{equation}
These results can be generalized to the $A_k$ family as well, indeed, in these cases, the $U(1)$ sector in the dual phase can always been integrated out using the SQED/XYZ duality.
The situation requires a more sophisticated analysis in the $D_{k+2}$ dualities, where there are towers of monopoles and anti-monopoles charged under the topological symmetry. In such cases it is not possible in general to integrate over the abelian factor and the dual theories correspond to $\UU(N) \times \UU(1)$ gauge theories.

\section{Conclusions}
\label{sec:conc}

In this paper, we have extended and generalized the analysis of \cite{Amariti:2025gca} in two main directions. 

In the first part we considered  the 4d $A_k$ and $D_{k+2}$ families, which are realised in brane setups by replacing the single central NS or NS$'$ fivebrane with a stack of $k$ such fivebranes. 
For each of the six models listed in the table of Section \ref{branepicture}, we studied the 4d/3d reduction both from the brane and the field-theory perspectives, and we established the matching of the squashed three-sphere partition functions. The reduction relies on a double scaling limit, whose structure is dictated by the brane picture, and on the validity of the parent 4d superconformal index identities. A key technical ingredient is the confinement of two-node quiver sectors: these are new confining dualities with $\UU(2k+1)^2$ and $\UU(3k)^2$ gauge groups, which we proved via the tensor deconfinement technique in section \ref{newconf}.

In the second part, we derived new families of 3d $\mathcal{N}=2$ dualities with $\UU(n)$ gauge groups and pairs of two-index tensors from the parent theories already studied there. We obtained dualities with Chern-Simons terms by triggering real mass flows on the fundamentals, and $\SU(n)$ dualities by gauging the topological symmetry.

Several open problems and natural continuations of this work deserve to be mentioned.
 
\paragraph{Chiral-ring issues in the $D$-type models.}
A subtle point arises in the confining $D_k$ quiver of Section \ref{sec:Dconf}: the classical chiral ring is truncated only for a specific sign of the higher-degree term in the superpotential of the second adjoint (the coefficient $(-1)^k$). In \cite{Brodie:1996xm} it was observed that \emph{it would be surprising if the duality depended critically on the numerical coefficients in the superpotential.} The brane picture does not shed immediate light on this sign ambiguity, since it does not distinguish the even and odd $k$ cases for the $D$-type models, even in the single-group version obtained by including an orientifold. We have not resolved this issue here; our approach has been to import the relevant constraints (anomaly cancellation, chiral ring truncation) from the field theory analysis, leaving a full geometric understanding of these constraints to future work.

\paragraph{Towards a direct proof via tensor deconfinement.}
The confinement results in the appendices suggest the possibility of  a fully self-contained proof of the 3d dualities, without assuming the 4d index identities as an input. Along the lines of \cite{Benvenuti:2024glr,Hwang:2024hhy}, one could attempt to prove the main identities of Sections \ref{sec:Aconf} and \ref{sec:Dconf} by a direct application of the tensor deconfinement technique to the full two-node quivers studied there. 
 
\paragraph{Stage models and product gauge groups.}
\begin{itemize}
\item The natural next step is to consider brane configurations with a larger number of stacks of NS fivebranes, engineering 4d theories with longer quiver structures, denoted as  \emph{stage} models in  \cite{Brodie:1996xm}. The double scaling prescription analysed here should generalize to those cases, producing an extended family of 3d dualities with product gauge groups. Moreover, it would be interesting to analyze the structure of monopole superpotential for the effective description on the circle along the lines of the analysis of 
\cite{Csaki:2017cqm}.
\item A related question is whether one can generalize the chiral dualities  constructed in \cite{Benini:2011mf} to our setting. The original $A_k$ and $D_{k+2}$ cases were addressed in \cite{Amariti:2020xqm,Amariti:2022lbw}, and an extension involving two two-index tensors is expected.
\end{itemize}
 
\paragraph{The $D_{k+2}$ orchestra and nilpotent vacua.}
\begin{itemize}
\item For certain $D$-type models, defined in \cite{Brodie:1996xm} as orchestra, the presence of nilpotent flat directions may obstruct the duality  \cite{Bryan:2025fxk}. It would be interesting to examine whether such obstructions are visible at the level of the superconformal index or the three-sphere partition function
\footnote{See also \cite{Kutasov:2014wwa,Intriligator:2016sgx} for further discussions on the chiral ring truncation of the orchestra models with $\SU(N)$ gauge groups.}. A direct proof via tensor deconfinement does not appear straightforward in those cases, which are also the models lacking a  brane interpretation in the literature. Understanding the fate of the other $D$-type models under these constraints is left for future investigation.
\end{itemize}

\section*{Acknowledgments}
This work  has been supported in part by the Italian Ministero dell'Istruzione, Università e Ricerca (MIUR), in part by Istituto Nazionale di Fisica Nucleare (INFN) through the “Gauge Theories, Strings, Supergravity” (GSS) research project.

\bibliographystyle{JHEP}
\bibliography{reffollowup.bib}
\end{document}